\pdfoutput=1
\documentclass{article}

\PassOptionsToPackage{sort&compress}{natbib}
\usepackage{iclr2027_conference,times}
\setcitestyle{numbers,square}

\usepackage[utf8]{inputenc} \usepackage[T1]{fontenc} \usepackage{hyperref} \usepackage{url} \usepackage{booktabs} \usepackage{amsfonts} \usepackage{nicefrac} \usepackage{microtype}
\usepackage{enumitem}   %
\usepackage{graphicx} \usepackage{xcolor}
\usepackage{rotating}
\makeatletter
\newcommand\rotpage[1]{\expandafter\gdef\csname rot@p@#1\endcsname{}}
\AddToHook{env/sidewaystable/end}{\if@filesw\write\@auxout{\string\rotpage{\the\c@page}}\fi}
\AtBeginDocument{\edef\rot@attr{\the\pdfpageattr}}
\AddToHook{shipout/before}{\ifcsname rot@p@\the\c@page\endcsname
  \global\pdfpageattr\expandafter{\rot@attr/Rotate 90}\else\global\pdfpageattr\expandafter{\rot@attr}\fi}
\makeatother
\hypersetup{colorlinks=true, linkcolor=blue!45!black, citecolor=blue!45!black, urlcolor=blue!45!black, filecolor=blue!45!black} \usepackage{amsmath} \usepackage{multirow} \usepackage{makecell} \usepackage{subcaption} \usepackage{tikz} \usepackage{pgfplots}
\usepackage{arydshln}  %

\usepackage{placeins} %
\pgfplotsset{compat=1.18} \usepgfplotslibrary{groupplots} \usetikzlibrary{positioning, calc, fit}

\newcommand{\PRIORRHOMMSU}{.67}

\title{Audio Token Attention Is Predictable\\Before the Language Model Runs}

\newif\ifpreprint \preprinttrue
\ifpreprint \iclrfinalcopy \fi
\author{Kyoungjun Park, Yunzhe Li, Lili Qiu \\
The University of Texas at Austin \\
\texttt{\{kjpark,lili\}@cs.utexas.edu}}

\DeclareUnicodeCharacter{03C1}{\ensuremath{\rho}}\DeclareUnicodeCharacter{2265}{\ensuremath{\ge}}

\usepackage[font=small]{caption}
\begin{document}

\maketitle
\ifpreprint \lhead{Preprint} \fi

\begin{abstract}
A large audio language model (LALM) turns a minute of speech into $750$--$1{,}500$ tokens and prefills every one. Image-token pruning often cuts after the language model's first layers, where image tokens draw little attention. Audio tokens draw much more attention there, and their ranking is still far from final, so audio needs a ranking before the language model runs. Surprisingly, the attention an audio token will receive across the language model is already linearly predictable from its encoder output, before the language model runs. A linear map, fitted in closed form without labels, predicts this all-layer attention ranking at
$\rho{\ge}.69$ on eleven of thirteen LALMs. Our method, \textbf{Triage}, cuts audio tokens by this prediction and, on multiple choice, cuts again at layer~$2$, correcting the prediction with the attention observed there. Triage sets its compression without labels, under two budgets that limit how far its output may differ from the model's own full-audio output. At the conservative budget, its word error rate and accuracy stay within $.04$ of full audio. At the aggressive budget, Triage beats every baseline in all twelve transcription cases. On multiple choice, at $2.2$--$5\times$ compression, it outperforms DART, the strongest baseline on average, by $.043$ in mean accuracy. Because it cuts before the language model, it raises the audio that fits in Qwen2.5-Omni-3B's context window from $21.8$ to about $62$~minutes. At its most compressive point, Triage lets one GPU serve $4\times$ as many concurrent $5$-minute streams of that model. Project page: \url{https://audio-triage.github.io}
\end{abstract}

\section{Introduction}
\label{sec:intro}

A large audio language model (LALM) answers questions about speech, sound and music directly from the audio~\cite{qwen25omni,qwen3omni,voxtral,phi4mm}. Its audio encoder and projector, which together we call the \emph{encoder}, turn the waveform into one vector per audio token, the \emph{encoder output} $\mathbf{e}_i$ that the language model receives for token~$i$. A minute of speech becomes $750$ to $1{,}500$ tokens, and each costs twice: the language model must prefill it, and it takes up space in the context window. The context window of Qwen2.5-Omni-3B (Qwen 3B) holds $21.8$ minutes of audio, so an hour-long lecture does not fit.

\begin{figure*}[t]
\centering
\begin{minipage}[t]{0.578\textwidth}\centering
{\sffamily\bfseries 1. \mbox{The prior predicts attention}}\\[4pt]
\includegraphics[width=\linewidth,trim=0 34pt 541pt 64pt,clip]{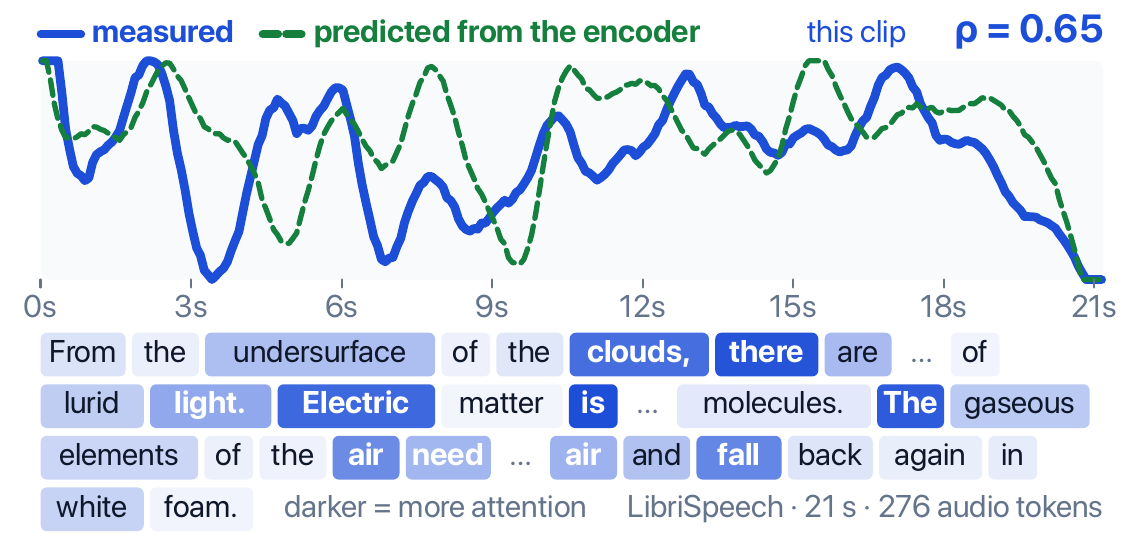}
\end{minipage}\hfill
\begin{minipage}[t]{0.392\textwidth}\centering
{\sffamily\bfseries 2. \mbox{Audio is not yet prunable}}\\[4pt]
\includegraphics[width=\linewidth,trim=0 0 0 48pt,clip]{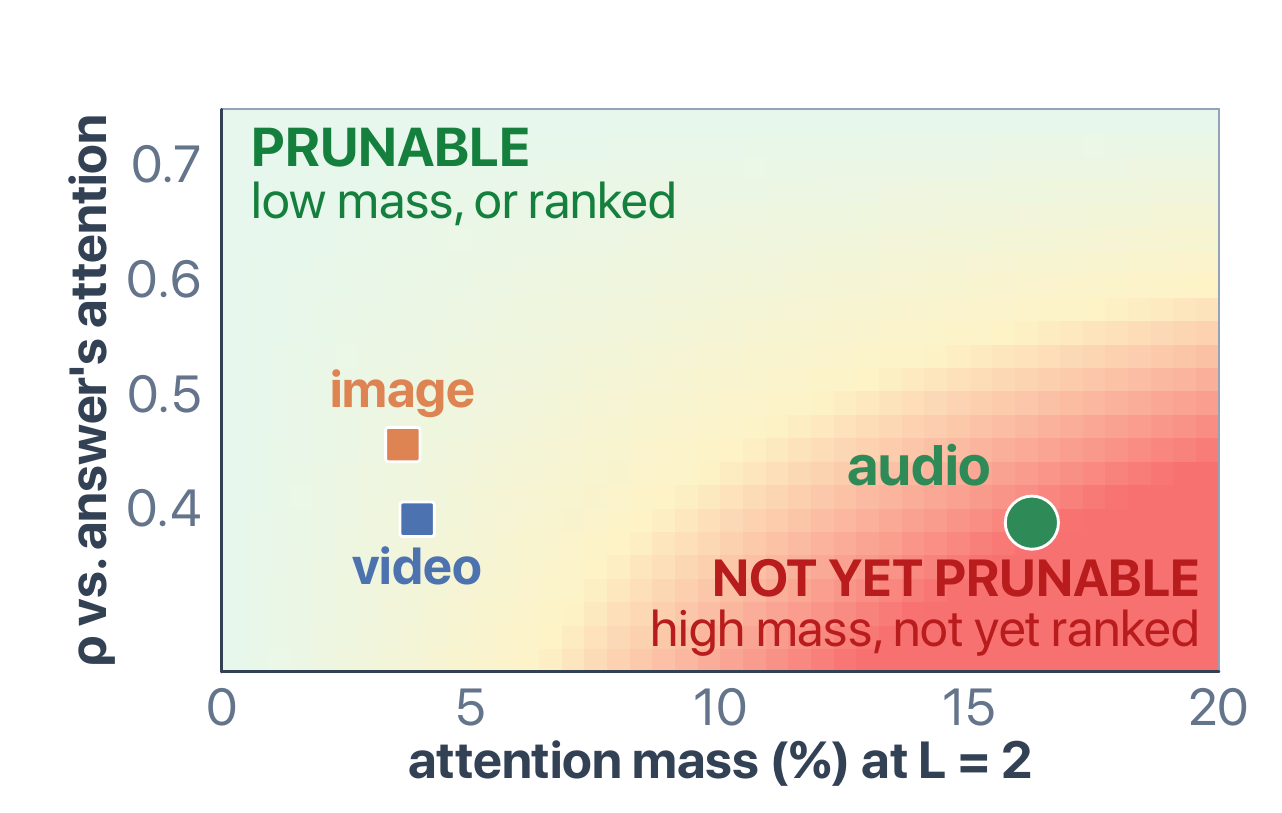}
\end{minipage}\\[6pt]
{\sffamily\bfseries 3. Other signals available before the language model predict attention weakly}\\[2pt]
\includegraphics[width=\textwidth,trim=0 6pt 0 0,clip]{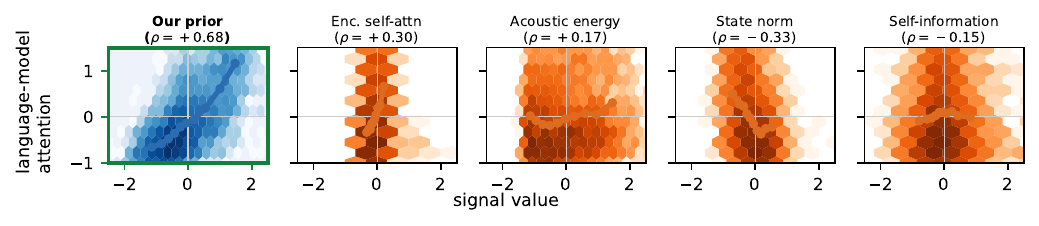}
\caption{\textbf{Before the language model runs, the prior predicts which audio tokens it will attend to, but other signals available at that point do so only weakly.} All panels: Qwen 30B. \textbf{(1)}~A held-out LibriSpeech clip with the median $\rho$: each audio token's language-model attention and the prior's prediction of it. \textbf{(2)}~At layer~2, each modality's share of the text positions' attention and how well the tokens' ranking by the answer's attention through that layer agrees with their all-layer ranking. The shading is schematic. \textbf{(3)}~Language-model attention against the prior (green) and four other signals available before the language model, over held-out LibriSpeech tokens ($\rho$ pooled over all tokens).}
\label{fig:teaser}\label{fig:xtower}\label{fig:layer-analysis}
\end{figure*}

Only a cut before the language model reduces both costs: a cut inside it saves only the later layers, and the tokens it drops already hold positions in the context window. Yet methods that rank by the language model's attention must run it first~\cite{fastv,fastav,headrouter,tasktoken,snapkv,h2o,pyramidkv,audiokv}, and methods that cut earlier rank by position, acoustic energy, self-information, similarity or the encoder's own attention~\cite{ibm,ltbm,intensityprune,llmlingua,segwise,vispruner}. The loudness, self-information and encoder-attention scores predict the language model's attention only weakly (Fig.~\ref{fig:teaser}, panel~3). Two vision methods train a network to predict the language model's attention before it runs~\cite{fast3d,map}. SpeechPrune~\cite{speechprune} needs the question and approximates only layer~1's attention.

Image-token pruning often cuts after the language model's first layers, which costs little if tokens are weakly attended by then or already ranked. Audio tokens meet neither condition (\S\ref{sec:doublebind}). When the same passages are given as speech and as rendered pages, with the same questions, audio draws over twice the page's attention at layer~2 on both Qwen3-Omni-30B (Qwen 30B) and Qwen 3B, and it draws more than the page on every passage and per token. At that layer, ranking the audio tokens by the attention the answer has paid them so far agrees with their all-layer ranking at only Spearman $\rho{=}.39$ (Fig.~\ref{fig:teaser}, panel~2). So audio needs a ranking before the language model forms its own.

Let $g_i$, the \emph{language-model attention} of audio token~$i$, be the attention it receives from the audio positions, summed over all layers and heads. In our prompts the question follows the audio, so under causal attention the audio positions never see it: $g$ does not depend on the question, and a cut made before the question arrives can use it. Surprisingly, $g$ is already linearly decodable from the encoder output, before the language model processes any audio tokens. A linear map, $s_i=\mathbf{w}^{\top}\mathbf{e}_i+b$, predicts $g_i$ (Fig.~\ref{fig:teaser}, panels~1 and~3), and we call it \emph{the prior}. Our method, \emph{Triage}, cuts in two stages (\S\ref{sec:method}): \emph{Stage~1} cuts on the prior before the language model runs, and \emph{Stage~2}, used on multiple choice, cuts again after the first two layers, correcting the prior with the attention observed there. A label-free calibration sets how much each stage cuts, under a conservative or an aggressive budget on how far the output may drift from the full-audio output. App.~\ref{app:map} lists where each main claim is measured. Our contributions:
\begin{itemize}[nosep,leftmargin=*]
\item \textbf{Audio needs a ranking before the language model runs.} Where image-token pruning cuts, at layer~2, audio tokens are still heavily attended and far from their final ranking (\S\ref{sec:doublebind}).
\item \textbf{Future language-model attention is predictable before the language model runs.} Fitted in closed form without labels, the prior predicts $g$ at $\rho{\ge}.69$ on eleven of thirteen LALMs. On seven of them, the loudness, self-information and encoder-attention scores used by other methods reach at most $|\rho|{=}.31$. With position, loudness and other waveform and word properties controlled, Qwen 3B's prior keeps $95$ to $97\%$ of its $\rho$. Even against the answer's attention, which depends on the question, the prior ranks tokens better than layer~2's attention does (\S\ref{sec:stage1}). Deleting its top tokens raises Voxtral's word error rate (WER) more than deleting energy- or position-matched sets (\S\ref{sec:causal-body}).
\item \textbf{Triage stays close to full audio at one budget and leads the baselines at the other.} At its conservative budget, WER and accuracy stay within $.04$ of full audio. At its aggressive budget, it has lower WER than every baseline, mid-prefill included, in all twelve transcription cases. On multiple choice, at $2.2$ to $5\times$ compression, it outperforms DART, the strongest baseline on average, by $.043$ in mean accuracy (\S\ref{sec:main-results}).
\item \textbf{Only a cut before the language model saves context.} On Qwen 3B, cutting the tokens $2.86\times$ raises the audio that fits in the context window from $21.8$ to about $62$ minutes. On natural recordings of $20$ to $43$ minutes, aggressive Triage outperforms truncating them to fit, voting over one-minute windows and DART at the same compression. At its most compressive point, Triage also lets one GPU serve $4\times$ as many concurrent streams (\S\ref{sec:efficiency}).
\end{itemize}

\section{Related Work}
\label{sec:related}

\paragraph{Where existing methods cut, and by what.} KV eviction ranks tokens by the language model's attention after the prefill, so it frees only the decode cache~\cite{snapkv,h2o,pyramidkv,audiokv}. Mid-prefill pruning ranks them by the attention paid so far and cuts inside the language model: at layer~2 in FastV~\cite{fastv} and the concurrent HeadRouter~\cite{headrouter}, at the middle layer in FastAV~\cite{fastav}, and after the first layers in IVTP's second stage~\cite{ivtp}. Lei et al.~\cite{tasktoken} cut at the input using layer~1's attention, and DART~\cite{dart} at layer~2 by dissimilarity to a few pivots. Methods that cut before the language model rank tokens by position~\cite{ibm}, by similarity to nearby tokens~\cite{ltbm}, by acoustic energy or the encoder's own attention~\cite{segwise,vispruner,ivtp,intensityprune}, or by self-information~\cite{llmlingua,longllmlingua}, which LLMLingua-2~\cite{llmlingua2} replaced with a distilled token classifier. In speech, one work learns an entropy-based compressed representation~\cite{entropyspeech} and a concurrent one pools acoustically similar tokens~\cite{affinitypool}. \S\ref{sec:stage1} tests these signals against language-model attention.

\paragraph{Predicting the language model's attention.} Closest to our work, Fast3D~\cite{fast3d} and the concurrent MAP~\cite{map} train a network to predict the language model's attention to visual tokens before it runs. Fast3D fits a $159$M-parameter network that also sees the prompt and cuts inside the language model from layer~2 on. MAP distils one middle layer's attention into a light predictor and cuts before the first layer. In speech, SpeechPrune~\cite{speechprune} also cuts before the language model, but it needs the question and approximates only the first layer's attention. The prior differs in its target and its fit: it predicts $g$, the attention summed over all layers, without seeing the question, and it is a linear map on the encoder output alone, fitted in closed form in seconds, without labels. The prior reaches $\rho{\ge}.69$ on eleven of thirteen LALMs, and on three models a trained MLP cuts no more accurately (\S\ref{sec:readnotspend}).

\section{Analysis: What Decides a Cut, and When}
\label{sec:analysis}

\subsection{At Layer~2, Audio Is Still Heavily Attended and Not Yet Ranked}
\label{sec:doublebind}

A cut at layer~2 of the language model is safe under either of two conditions. Under the \emph{attention condition}, the tokens it drops are already weakly attended, so losing them costs little. Under the \emph{ranking condition}, the model has already ranked the tokens, so the cut can pick the right ones. For images, FastV~\cite{fastv} shows the first: in LLaVA~\cite{llava}, each image token draws $0.21\%$ of the attention a system-prompt token draws after layer~2. We test both conditions on Qwen 30B with $100$ items per modality: audio from AudioMarathon-RACE~\cite{audiomarathon}, images from MMMU~\cite{mmmu} and token-dense video from MVBench~SSv2~\cite{mvbench}.

Only images and video meet the attention condition. At layer~2 the text positions send $16.2\%$ of their attention to audio, against $3.6\%$ to images and $3.9\%$ to video (Fig.~\ref{fig:teaser}, panel~2). A controlled run confirms the gap. Given the same RACE passage as speech and as a rendered page, with the same question, audio draws more attention at layer~2 on all $40$ passages: $2.3\times$ the page's share on Qwen 30B and $2.1\times$ on Qwen 3B. Per token, it also draws more on Qwen 30B, Qwen 3B and Phi-4. Token count and temporal structure do not explain the gap: video has more tokens than audio and is also temporal, yet from layer~2 onward it draws much less attention than audio, closer to images.

On Qwen 30B, no modality meets the ranking condition. At layer~2 the tokens' ranking by the attention the model's answer pays them agrees with their all-layer ranking at only Spearman $\rho{=}.39$ for audio, $.39$ for video and $.46$ for images. Images and video can afford an unsettled ranking, since they meet the attention condition, but audio meets neither condition. Deeper layers rank better (Fig.~\ref{fig:layerconv}), but a token dropped inside the language model has already taken a position in the context window. So audio needs a ranking before the language model runs.

\subsection{The Encoder Output Predicts Language-Model Attention}
\label{sec:stage1}

\paragraph{Target.} The target $g_i$ is the attention audio token~$i$ receives from the audio positions over the whole forward pass,
\begin{equation}
\label{eq:importance}
g_i \;=\; \sum_{\ell=1}^{N_{\mathrm{layer}}} \sum_{h=1}^{H} \sum_{q \in \mathcal{Q}} a_{\ell,h}(q, i),
\end{equation}
where $a_{\ell,h}(q,i)$ is the post-softmax attention from position~$q$ to token~$i$ at layer~$\ell$ and head~$h$ of $H$ (zero for $q{<}i$ under causal masking), and $\mathcal{Q}$ is the set of audio positions. We call $g$ \emph{language-model attention}. Unlike \S\ref{sec:doublebind}'s share, $g$ ranks the audio tokens against each other. We use the audio positions for two reasons. In our prompts the question follows the audio, so under causal attention the audio positions never see it, and $g$ does not depend on the question, as required for a score used before the question arrives. Even positions that do see the question attend mostly to the same tokens: on Qwen 3B, the top $25\%$ of tokens by attention from the last eight prompt rows overlap at Jaccard $.71$ across different questions on the same RACE passage, against $.14$ for random sets of that size. We chose $g$ once, on a separate sample on Qwen 3B (\S\ref{sec:readnotspend}), and use it unchanged on transcription, on every other model and on long recordings (\S\ref{sec:efficiency}). One full-audio forward pass gives $g$, so fitting against it needs no labels, while cutting on $g$ itself would cost the full prefill.

\paragraph{The prior.} A ridge regression on the encoder output predicts $g$ on all seven models of Tab.~\ref{tab:crossarch}, at Spearman $\rho{=}.69$ to $.83$ on $100$ held-out clips. It maps each encoder output $\mathbf{e}_i$, the vector the language model receives for audio token~$i$, to the per-clip $z$-normalised $g_i$ ($\lambda{=}10$, closed form) and is fitted in seconds on a CPU, without labels. Its score $s_i=\mathbf{w}^\top\mathbf{e}_i+b$ is \emph{the prior}, known before the language model runs and corrected by Stage~2 with the attention observed at layer~2. In effect, the prior ranks the audio tokens by how far each lies along one learned direction in the encoder's output space. In Tab.~\ref{tab:crossarch}, the four models that attach a Whisper or conformer~\cite{conformer} encoder to a pretrained language model (Ultravox, Phi-4, Voxtral and Granite) span the whole range, so the prior does not need the encoder and the language model to be trained together. Triage deploys one prior per model, fitted on clips from several benchmarks at once: on Qwen 3B, a prior pooled over six benchmarks, including sound and music, predicts attention on each at least as well as one fitted to that benchmark alone.

\begin{table}[!ht]
\centering
\caption{\textbf{The prior predicts language-model attention on all seven models, while three signals used by existing methods predict it weakly and the state norm changes sign across models.} Each entry is the per-clip Spearman $\rho$ against $g$, averaged over $100$ held-out LibriSpeech clips per model. Fig.~\ref{fig:readprops}b adds three more LALMs ($\rho{=}.82$ to $.88$), DiVA ($\rho{=}.91$) and the two Qwen-Audio models, where the prior fails. DiVA is left out here because its learned query tokens have no time order. \textbf{Bold}: each row's largest $|\rho|$.}
\label{tab:crossarch}
\footnotesize
\setlength{\tabcolsep}{3pt}
\begin{tabular}{@{}l c @{\hspace{13pt}} cccc@{}}
\toprule
$\rho$ against $g$ & \textbf{the prior} & enc.\ self-attn & acoustic energy & state norm $\|\mathbf{e}_i\|_2$ & self-information \\
\midrule
Ultravox, $1$B & $\mathbf{+.83}$ & $+.17$ & $+.28$ & $-.45$ & $-.26$ \\
Phi-4, $5.6$B & $\mathbf{+.79}$ & $+.23$ & $+.12$ & $+.03$ & $+.12$ \\
Voxtral, $3$B & $\mathbf{+.79}$ & $+.10$ & $+.15$ & $-.36$ & $-.13$ \\
Qwen2.5-Omni, $3$B & $\mathbf{+.75}$ & $+.17$ & $+.27$ & $-.49$ & $-.12$ \\
Qwen2.5-Omni, $7$B & $\mathbf{+.73}$ & $+.18$ & $+.29$ & $-.54$ & $-.27$ \\
Qwen3-Omni, $30$B & $\mathbf{+.70}$ & $+.31$ & $+.19$ & $-.33$ & $-.14$ \\
Granite, $2$B & $\mathbf{+.69}$ & $+.08$ & $+.18$ & $+.12$ & $+.25$ \\
\bottomrule
\end{tabular}
\end{table}

\paragraph{Simpler signals.} The signals used by existing methods (the encoder's self-attention, acoustic energy and self-information) reach at most $|\rho|{=}.31$ on the seven models of Tab.~\ref{tab:crossarch}, and the state norm, the strongest unfitted signal on five models, flips sign on Phi-4 and Granite. A regression fitted to the waveform stays well below the prior: on Qwen 3B, the same regression refit on seven waveform descriptors of each token, including its position and loudness, reaches only $\rho{=}.49$ on LibriSpeech, against the prior's $.76$. Position is the most informative of them, since under causal attention an early token is seen by more positions, but the prior correlates with it far less ($\rho{=}{-}.226$, against $-.485$ for $g$). Such properties do not explain the prior, even combined linearly: on Qwen 3B the prior keeps $95$ to $97\%$ of its $\rho$ on two benchmarks when eleven such properties of the waveform and its words, including each token's position, are controlled for at once.

\paragraph{Deep layers.} The prior predicts attention deep in the model, not just the layer-1 attention between audio positions, which is a function of $\mathbf{e}$ by construction. Refit against $g$ restricted to a range of layers, it does worst on layer~1 alone, and on the deepest quarter of layers it still reaches $\rho{=}.78$ on Voxtral and $.71$ on Qwen 3B. The prior also predicts more than the attention sink: with the sink's column dropped from $g$, its $\rho$ does not fall ($.82$ and $.77$).

\begin{figure}[t]
\centering
\includegraphics[width=\linewidth]{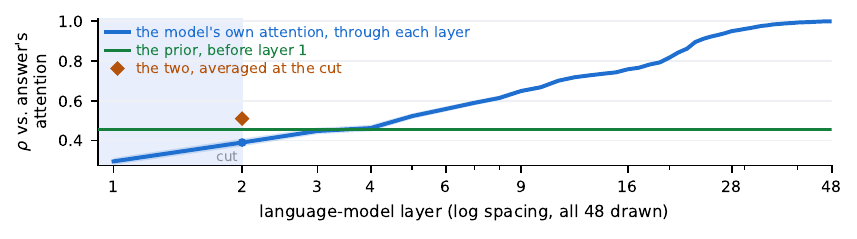}
\caption{\textbf{Where Stage~2 cuts, the prior ranks the audio tokens better than the language model's own attention, and the two combined rank them best.} Qwen 30B, AudioMarathon-RACE. Every ranking is scored against the answer's attention: the attention the answer pays each audio token over all $48$ layers. The blue curve ranks the tokens by that attention summed only through each layer, so it reaches $1$ at layer~48 by construction. The green line is the prior, fixed before layer~1. The diamond averages the two rankings at the cut. Past the strip marked \emph{cut}, the prefill over every token is paid.}
\label{fig:layerconv}
\end{figure}

\paragraph{Ranking at layer~2.} Where Stage~2 cuts, the prior already ranks the audio tokens better than the language model's own attention, and a token the prior drops costs no prefill. Fig.~\ref{fig:layerconv} scores both against a harder target than the $g$ of Tab.~\ref{tab:crossarch}: the answer's attention, which depends on a question the prior never sees. At layer~2 the prior scores $\rho{=}.46$ against the attention's $.39$, and averaging the two rankings beats both ($.51$). Stage~2 builds on this combination. On Qwen 3B, we cut at layer~2 to keep three different fractions of the audio tokens on three multiple-choice benchmarks. In eight of these nine settings, cutting on the prior is more accurate than cutting on the attention observed there. FastV's signal takes that attention from the last prompt rows only, and the prior beats it in all nine.

\subsection{The Prior's Top Tokens Matter, and One Pass Separates Its Failures}
\label{sec:causal-body}\label{sec:screen}

\begin{figure}[t]
\centering
\definecolor{ctlpos}{gray}{0.42}%
\definecolor{ctlnrg}{HTML}{AA3377}\definecolor{ctlinfo}{HTML}{0072B2}%
\begin{subfigure}[t]{0.435\linewidth}\centering
\begin{tikzpicture}
\begin{axis}[
  width=\linewidth, height=4.9cm,
  xmin=-3, xmax=53, ymin=0, ymax=1.02,
  xtick={0,10,20,35}, xticklabels={$0\%$,$10\%$,$20\%$,$35\%$},
  xlabel={audio tokens deleted}, ylabel={mean per-clip WER},
  xlabel near ticks, ylabel near ticks, ytick={0,0.5,1.0},
  tick label style={font=\scriptsize}, label style={font=\scriptsize},
  ymajorgrids, grid style={black!10},
]
\addplot[ctlpos, semithick, mark=o, mark size=1.7pt, mark indices={2,3,4}]       coordinates {(0,0.070) (10,0.152) (20,0.126) (35,0.074)};
\addplot[ctlnrg, semithick, mark=diamond, mark size=2.0pt, mark indices={2,3,4}] coordinates {(0,0.070) (10,0.045) (20,0.104) (35,0.060)};
\addplot[ctlinfo, semithick, mark=triangle, mark size=2.2pt, mark indices={2,3,4}] coordinates {(0,0.070) (10,0.181) (20,0.621) (35,0.927)};
\addplot[red!75!black, very thick, mark=*, mark size=2.2pt, mark indices={2,3,4}]
  coordinates {(0,0.070) (10,0.573) (20,0.880) (35,0.935)};
\addplot[only marks, mark=*, mark size=1.8pt, black!70] coordinates {(0,0.070)};
\coordinate (full) at (axis cs:0,0);
\node[font=\scriptsize\bfseries, text=red!75!black, anchor=west, inner sep=1pt] at (axis cs:36.3,0.955) {the prior};
\node[font=\scriptsize, text=ctlinfo, anchor=north west, inner sep=1pt] at (axis cs:27.4,0.735) {self-info};
\node[font=\scriptsize, text=ctlpos, anchor=west, inner sep=1pt] (lpos) at (axis cs:37.0,0.165) {position};
\draw[ctlpos, line width=0.4pt, shorten >=1.9pt] (lpos.west) -- (axis cs:35,0.074);
\node[font=\scriptsize, text=ctlnrg, anchor=west, inner sep=1pt] (lnrg) at (axis cs:37.0,0.030) {same energy};
\draw[ctlnrg, line width=0.4pt, shorten >=2.1pt] (lnrg.west) -- (axis cs:35,0.060);
\end{axis}
\node[font=\tiny, anchor=north, inner sep=0pt, overlay] at ([yshift=-10.5pt]full) {full audio};
\end{tikzpicture}
\caption{Deleting the prior's top tokens costs the most.}
\label{fig:causaldel}
\end{subfigure}\hfill
\begin{subfigure}[t]{0.545\linewidth}\centering
\begin{tikzpicture}[lead/.style={black!60, line width=0.4pt, shorten >=2.4pt}]
\begin{axis}[
  width=\linewidth, height=4.9cm,
  xmode=log, log basis x=10,
  xmin=0.115, xmax=4.4, ymin=0.16, ymax=1.06,
  xlabel={$\Delta_2$: drift, before fitting}, ylabel={$\rho$ of the prior, after fitting},
  xtick={0.15,0.25,0.5,1.0,2.0}, xticklabels={0.15,{\bfseries 0.25},0.5,1.0,2.0},
  ytick={0.2,0.4,0.6,0.8,1.0},
  xlabel near ticks, ylabel near ticks,
  tick label style={font=\scriptsize}, label style={font=\scriptsize},
]
\addplot[draw=none, fill=red!7,  forget plot] coordinates {(0.115,0.16) (0.25,0.16) (0.25,1.06) (0.115,1.06)} --cycle;
\addplot[draw=none, fill=blue!6, forget plot] coordinates {(0.25,0.16) (4.4,0.16) (4.4,1.06) (0.25,1.06)} --cycle;
\addplot[black!55, dashed, very thick, forget plot] coordinates {(0.25,0.16) (0.25,0.257)};
\addplot[black!55, dashed, very thick, forget plot] coordinates {(0.25,0.320) (0.25,0.535)};
\addplot[black!55, dashed, very thick, forget plot] coordinates {(0.25,0.598) (0.25,1.06)};
\addplot[only marks, mark=*, mark size=2.0pt, blue!55!black, forget plot] coordinates
  {(0.941,0.829) (0.858,0.911) (0.739,0.750) (0.906,0.794) (0.665,0.727) (0.624,0.689)
   (0.395,0.787) (2.100,0.701)};
\addplot[only marks, mark=*, mark size=2.0pt, red!70!black, forget plot] coordinates
  {(0.159,0.288) (0.144,0.566)};
\addplot[only marks, mark=*, mark size=2.0pt, blue!55!black, forget plot] coordinates
  {(1.342,0.819) (1.143,0.879) (0.747,0.843)};
\node[font=\scriptsize, anchor=south] at (axis cs:0.858,0.928) {DiVA};
\node[font=\scriptsize, anchor=south] at (axis cs:0.395,0.805) {Voxtral};
\node[font=\scriptsize, anchor=north east] at (axis cs:3.87,0.669) {Qwen3-Omni-30B};
\node[font=\scriptsize, anchor=west] at (axis cs:0.168,0.566) {Qwen2-Audio};
\node[font=\scriptsize, anchor=west] at (axis cs:0.183,0.288) {Qwen-Audio-Chat};
\node[font=\scriptsize, anchor=east] at (axis cs:0.590,0.689) {Granite};
\node[font=\scriptsize, anchor=south west] at (axis cs:1.19,0.900) {Aero};
\node[font=\scriptsize, anchor=west] at (axis cs:1.43,0.819) {SeaLLMs};
\node[font=\scriptsize, anchor=east] (mida) at (axis cs:0.603,0.975) {MiDashengLM};
\draw[lead] (mida.east) -- (axis cs:0.747,0.843);
\node[font=\scriptsize, anchor=west] (ultra) at (axis cs:1.035,0.728) {Ultravox};
\draw[lead] (ultra.west) -- (axis cs:0.941,0.829);
\node[font=\scriptsize, anchor=west] (phi) at (axis cs:0.93,0.555) {Phi-4};
\draw[lead] (phi.west) -- (axis cs:0.906,0.794);
\node[font=\scriptsize, anchor=west] (o3b) at (axis cs:0.80,0.470) {Qwen2.5-Omni-3B};
\draw[lead] (o3b.west) -- (axis cs:0.739,0.750);
\node[font=\scriptsize, anchor=west] (o7b) at (axis cs:0.80,0.385) {Qwen2.5-Omni-7B};
\draw[lead] (o7b.west) -- (axis cs:0.665,0.727);
\end{axis}
\end{tikzpicture}
\caption{One pass, before fitting, separates the prior's failures.}
\label{fig:screen}
\end{subfigure}
\caption{\textbf{Deleting the prior's top tokens costs the most, and one forward pass separates the prior's failures.} \textbf{(a)}~Voxtral, LibriSpeech. Each line deletes the same number of tokens by a different rule. \emph{Same energy} and \emph{position} delete other tokens that match the prior's top tokens in acoustic energy or in their spread over the clip. The three controls not drawn (random, the prior's lowest-ranked and the loudest tokens) cost less at every rate. \textbf{(b)}~Thirteen LALMs~\cite{qwen25omni,qwen3omni,voxtral,phi4mm,ultravox,diva,granitespeech,qwenaudio,qwen2audio,seallmsaudio,midashenglm,aero}. Each is screened with one forward pass over $20$ unlabelled clips.}
\label{fig:readprops}
\end{figure}
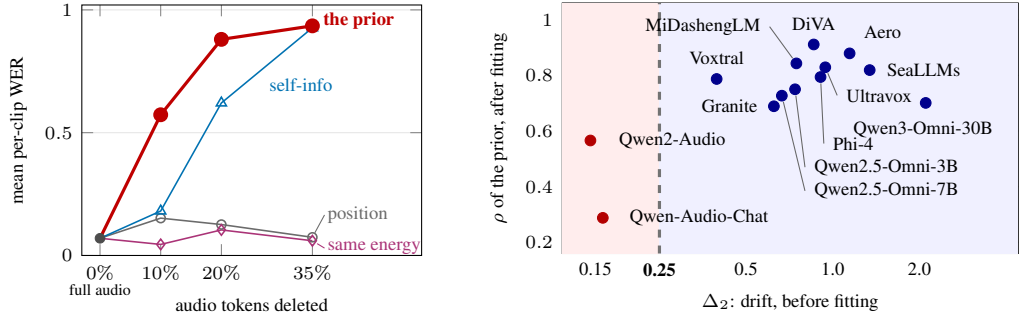

\textbf{Deleting the prior's top tokens costs the most.} If the model relies on the tokens the prior ranks highest, deleting them should hurt more than deleting other tokens. On Voxtral, this deletion raises WER more than deleting the same number of tokens by any of the three other rules in Fig.~\ref{fig:readprops}a, though self-information nearly matches it at the $35\%$ rate. Two of those rules match the prior's top tokens in acoustic energy or in position, and both cost far less, so the prior captures more than energy or position: at the $10\%$ rate its deletion costs $+.53$ WER more than the same-energy one. On Phi-4, and on Qwen 3B with multiple-choice questions, this deletion also costs more than the same-energy one.

\textbf{One unlabelled forward pass separates the LALMs the prior fails on.} The prior reaches $\rho{\ge}.69$ on eleven of the thirteen LALMs in Fig.~\ref{fig:readprops}b, from five language-model families, and fails on the two Qwen-Audio models ($\rho{=}.29$ and $.57$). A screen tells the two groups apart before anything is fitted: one forward pass over $20$ unlabelled clips measures how far the audio positions move through the language model's first two layers, where Stage~2 cuts. This drift, averaged over audio positions and clips, is $\Delta_2 = \lVert \mathbf{h}_2-\mathbf{h}_0\rVert/\lVert \mathbf{h}_0\rVert$, where $\mathbf{h}_0$ is the language model's input at an audio position and $\mathbf{h}_\ell$ its hidden state after layer~$\ell$. A threshold of $0.25$ splits the thirteen models. SeaLLMs-Audio and Aero-1-Audio attach Qwen2-Audio's encoder to Qwen2.5 language models, and the prior works on both, so the screen does not simply flag that encoder. The screen needs no prior, labels or task run and takes under two minutes on one A100.

\subsection{A Linear Map Is Enough}
\label{sec:readnotspend}

\textbf{Targets.} For targets taken from the language model, the better the prior predicts a target, the more accurate the cut. On Qwen 3B and DREAM, we fit the prior to each of thirteen such targets, including variants of $g$ and gradient saliency, and cut on each prediction, keeping $25\%$ of the tokens. A target's $\rho$ and its cut's accuracy have a rank correlation of $+.66$, and $g$'s cut is the most accurate of the thirteen ($.815$). Targets that also use the attention of the text positions, which see the question, do no better (at most $.812$). Acoustic energy, a property of the sound rather than of the model, is the exception: the prior predicts it best ($\rho{=}.88$), but its cut reaches only $.662$. A target must therefore come from the model itself, not merely be easy to predict.

\textbf{Predictors.} Training a network to predict attention, as Fast3D~\cite{fast3d} and MAP~\cite{map} do, does not help here. A $1.3$M-parameter MLP fits $g$ better than the prior, mostly among low-ranked tokens a cut discards, yet on Qwen 3B it cuts no more accurately: over nine paired settings (three benchmarks at three operating points), the median accuracy difference is $+.005$ in the prior's favour. On Qwen 30B and Voxtral, at the deployed operating points, its cut is never significantly more accurate than the prior's either. An $8.5$M-parameter Perceiver connector~\cite{perceiver}, trained against the frozen model's own answers, does not significantly beat the prior at the deployed compressions and is less accurate at a higher compression. It also leaves Stage~2 no per-token score to refine, since it replaces the audio tokens with new vectors. A closed-form linear map is therefore enough, and Triage uses it.

\section{Triage: Cutting with the Prior}
\label{sec:method}\label{sec:stage1cut}\label{sec:stage2}\label{sec:deployment}\label{sec:experiments}

\begin{figure}[t]
\centering
\includegraphics[width=\linewidth]{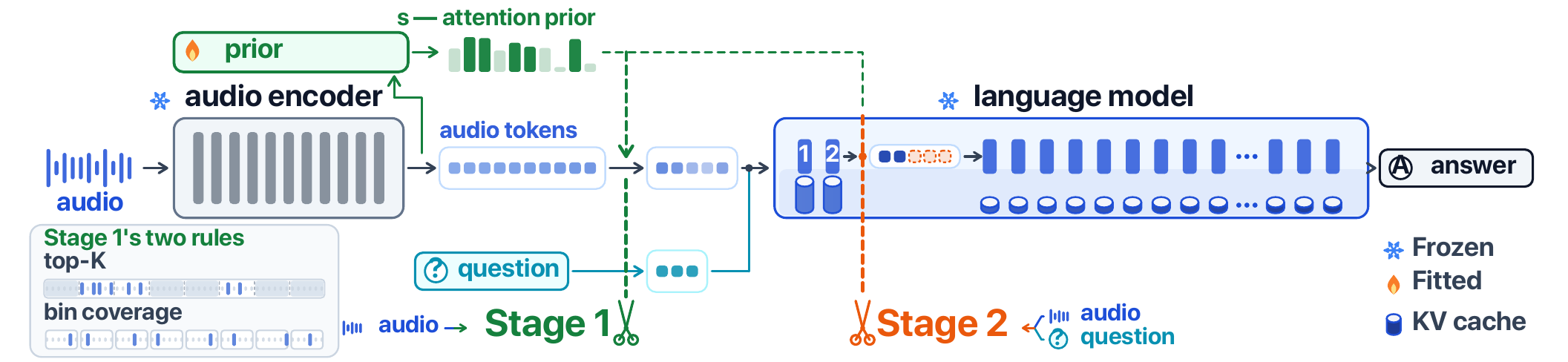}
\caption{\textbf{Triage: two cuts, one prior.} \textbf{Stage~1} cuts the encoder's $N$ audio tokens to $K_1$ before the language model runs. The bars above the first strip are the prior $s$, solid where Stage~1 keeps the token. \textbf{Stage~2} cuts $K_1{\to}K$ at layer~2, correcting the prior with the attention observed there. Transcription runs Stage~1 alone. Lower left (schematic): \emph{top-$K$}, used on multiple choice, can leave whole bins with no token, while \emph{bin coverage}, used on transcription, keeps one token from each of $K_1$ equal time bins.}
\label{fig:pipeline}
\end{figure}

Triage uses one prior per model and a label-free calibration, and leaves the encoder and the language model frozen (Fig.~\ref{fig:pipeline}). Stage~1 cuts the $N$ audio tokens to a shortlist of $K_1$ with the prior of \S\ref{sec:stage1}, before the language model runs. Stage~2 cuts the shortlist to $K$ at layer~2, correcting the prior with the attention observed there. The task decides how the cut is made. Transcription needs all of the speech, so Stage~1 covers the whole timeline and makes the only cut. A question needs only the parts it asks about, so Stage~1 keeps the prior's top tokens and Stage~2 makes the final cut.

\paragraph{Stage~1.} Stage~1 scores each audio token with the prior, $s_i = \mathbf{w}^\top\mathbf{e}_i + b$, on the encoder output the model has already computed, so it needs no extra forward pass. It keeps $K_1 = \max(2,\lfloor r_1 N \rfloor)$ tokens in their original time order. The language model receives only this shorter sequence, with contiguous positions. On multiple choice it keeps the $K_1$ highest-scoring tokens (\emph{top-$K$}). On transcription it keeps the highest-scoring token in each of $K_1$ equal time bins (\emph{bin coverage}), allocating tokens as segmentwise pruning~\cite{segwise} does. Transcription uses bin coverage alone ($K_1{=}K$): at seven of the eight Qwen operating points on LibriSpeech and FLEURS, it has lower WER than both stages together. We keep this choice unchanged for TEDLIUM, Voxtral and Phi-4.

\paragraph{Stage~2.} At layer~2 the prior is still ahead of the attention observed there (\S\ref{sec:stage1}), so Stage~2 uses that attention to correct the prior's ranking, not to replace it. It runs the language model over $[\,$prompt prefix; shortlist; prompt suffix$\,]$ for two layers and, for each head $h$, sums over both layers the attention $c_{hi}$ that shortlist token $i$ receives from every prompt-suffix position. The prompt suffix holds the question, which the prior never sees, so the question first affects the cut here. Prior and observation are put on one scale as percentile ranks over the shortlist, $r^\text{pred}_i$ of $s_i$ and $r^\text{obs}_i$ of $\sum_h c_{hi}$, and combined per token as
\begin{equation}
\label{eq:triage}
\gamma_i \;=\; \alpha_i\, r^\text{obs}_i \;+\; (1-\alpha_i)\, r^\text{pred}_i ,
\end{equation}
where the head-agreement weight $\alpha_i\in[0,1]$ is the percentile rank of $\operatorname{mean}_h c_{hi}/\operatorname{std}_h c_{hi}$. We call Eq.~\ref{eq:triage} \emph{precision fusion} because $\alpha_i$ acts like a precision: where the heads disagree, the observed score is noisy and the prior keeps more weight, and where they agree, the observed attention, which has seen the question, gets more weight. Stage~2 keeps the top $K = \max(1,\lfloor r_2 N \rfloor)$ tokens by $\gamma$, and survivors keep their position IDs, which are already in the KV cache.

\paragraph{Calibration.} For each benchmark and budget, the calibration sets $r_1$ and $r_2$ by self-consistency with the model's own full-audio output. On unlabelled clips it decodes once from the full audio and once per candidate $(r_1,r_2)$, and measures the \emph{drop}: WER against the full-audio transcript, or on multiple choice the fall in the probability of the full-audio answer. Step~1 takes the most compressive $r_2$ whose drop stays within the budget, $.08$ (conservative) or $.20$ (aggressive), on both a calibration and a held-out split. Step~2 then sets the shortlist fraction $r_1{>}r_2$.

\subsection{Setup}
\label{sec:exp-setup}\label{sec:setup}

We run Qwen 3B, Qwen 30B, Voxtral and Phi-4 on six benchmarks at both budgets, giving $24$ \emph{cases} (model, benchmark and budget) per task. Transcription uses LibriSpeech~\cite{librispeech}, FLEURS~\cite{fleurs} and TEDLIUM~\cite{tedlium3}, scored by corpus WER. Multiple choice uses MMSU~\cite{mmsu}, DREAM~\cite{dream,audiobench} in AudioBench's text-to-speech version, and AudioMarathon-RACE~\cite{audiomarathon}, scored by accuracy. The prior and the calibration see only unlabelled audio.

Baselines are grouped by where they cut, which sets their cost. \emph{Encoder-side} baselines select from the encoder output alone: uniform pooling~\cite{ibm}, energy-based voice activity detection (VAD), random selection, and DART~\cite{dart}, a vision-token method whose criterion needs only the token sequence. On multiple choice we add SpeechPrune~\cite{speechprune}, which also cuts before the language model but ranks tokens by their similarity to the question and by an approximation of the first layer's attention. \emph{Mid-prefill} baselines, FastV$_N$~\cite{fastv} and HeadRouter$_N$~\cite{headrouter}, rank all $N$ audio tokens at layer~2 and then cut to the same $K$. HeadRouter's released head weights cover only Qwen 3B, and elsewhere its heads are weighted uniformly. On Qwen 3B, Triage leads HeadRouter$_N$ in all six aggressive cases, three per task. FastV$_N$ scores tokens by the attention from the trailing prompt rows, as FastV is usually implemented. Triage also matches or beats an all-rows variant in eleven of the twelve Qwen multiple-choice cases. We compare against DART, the strongest encoder-side baseline on average, and against the best baseline in each case. Intervals are paired bootstraps over clips.

\subsection{Main Results}
\label{sec:main-results}\label{sec:multimodel}

\begin{table}[t]
\centering
\caption{\textbf{Both tasks at both budgets} (full results in App.~\ref{app:tabnotes}). Every selector in a column keeps the same number of tokens, $K{=}N/c$, where $c$ is the compression in the first row of each budget. \textbf{(a)}~Transcription, corpus WER (lower is better; $n{=}300$ per case, $100$ on Qwen 30B, seven talks on TEDLIUM). Ours is bin coverage. \textbf{(b)}~Multiple choice, accuracy (higher is better; $n{=}400$ per case; RACE is AudioMarathon-RACE). Ours is two-stage Triage with precision fusion. Bin coverage is the same one-stage cut as ours in (a) and never sees the question. $^{\dagger}$Ahead of DART in a one-sided exact McNemar test, Holm-corrected over the twelve cases. \textbf{Bold}: the best selector at the aggressive budget.}
\label{tab:asrmain}
\footnotesize
\setlength{\tabcolsep}{2pt}
\begin{tabular}{@{}l ccc c ccc c ccc c ccc@{}}
\toprule
& \multicolumn{3}{c}{Qwen2.5-Omni-3B} && \multicolumn{3}{c}{Qwen3-Omni-30B} && \multicolumn{3}{c}{Voxtral} && \multicolumn{3}{c}{Phi-4} \\
\cmidrule(lr){2-4}\cmidrule(lr){6-8}\cmidrule(lr){10-12}\cmidrule(l){14-16}
\textbf{(a)} WER & {\scriptsize Libri} & {\scriptsize FLEURS} & {\scriptsize TED} && {\scriptsize Libri} & {\scriptsize FLEURS} & {\scriptsize TED} && {\scriptsize Libri} & {\scriptsize FLEURS} & {\scriptsize TED} && {\scriptsize Libri} & {\scriptsize FLEURS} & {\scriptsize TED} \\
\midrule
full audio & .083 & .153 & .215 && .011 & .037 & .020 && .020 & .043 & .032 && .017 & .045 & .085 \\
\hdashline[2pt/1.5pt]
\addlinespace[1pt]
\emph{conservative} ($\times$) & 1.54 & 1.11 & 1.25 && 1.54 & 1.54 & 1.67 && 1.54 & 1.54 & 2.00 && 1.54 & 1.54 & 1.54 \\
\textbf{ours} & .099 & .177 & .197 && .025 & .050 & .049 && .025 & .043 & .071 && .045 & .064 & .102 \\
\hdashline[2pt/1.5pt]
\addlinespace[1pt]
\emph{aggressive} ($\times$) & 2.00 & 1.18 & 1.67 && 2.00 & 2.00 & 2.00 && 2.86 & 2.86 & 2.38 && 2.86 & 2.86 & 2.86 \\
uniform pooling & .348 & .325 & .403 && .160 & .140 & .123 && .398 & .385 & .241 && .383 & .368 & .488 \\
VAD & .327 & .355 & .496 && .199 & .138 & .166 && .665 & .719 & .317 && .472 & .402 & .531 \\
random & .496 & .236 & .409 && .429 & .385 & .378 && .627 & .646 & .521 && .597 & .583 & .646 \\
DART & .160 & .221 & .290 && .134 & .119 & .123 && .139 & .119 & .441 && .612 & .529 & .697 \\
\addlinespace[1pt]
FastV$_N$ & .395 & .200 & .364 && .354 & .386 & .381 && .753 & .782 & .304 && .456 & .426 & .422 \\
HeadRouter$_N$ & .173 & .215 & .288 && .305 & .210 & .258 && .190 & .109 & .170 && .550 & .472 & .441 \\
\textbf{ours} & \textbf{.116} & \textbf{.174} & \textbf{.195} && \textbf{.060} & \textbf{.081} & \textbf{.083} && \textbf{.088} & \textbf{.066} & \textbf{.152} && \textbf{.312} & \textbf{.285} & \textbf{.397} \\
\midrule
\textbf{(b)} accuracy & {\scriptsize MMSU} & {\scriptsize DREAM} & {\scriptsize RACE} && {\scriptsize MMSU} & {\scriptsize DREAM} & {\scriptsize RACE} && {\scriptsize MMSU} & {\scriptsize DREAM} & {\scriptsize RACE} && {\scriptsize MMSU} & {\scriptsize DREAM} & {\scriptsize RACE} \\
\midrule
full audio & .603 & .892 & .818 && .715 & .922 & .870 && .552 & .882 & .765 && .550 & .843 & .688 \\
\hdashline[2pt/1.5pt]
\addlinespace[1pt]
\emph{conservative} ($\times$) & 1.54 & 2.22 & 1.54 && 1.82 & 1.54 & 1.82 && 2.86 & 1.54 & 1.54 && 1.54 & 1.54 & 1.54 \\
\textbf{ours} & .600 & .865 & .823 && .690 & .940 & .863 && .535 & .885 & .748 && .540 & .812 & .685 \\
\hdashline[2pt/1.5pt]
\addlinespace[1pt]
\emph{aggressive} ($\times$) & 4.00 & 5.00 & 2.86 && 2.86 & 2.22 & 2.86 && 5.00 & 2.86 & 2.86 && 2.86 & 2.86 & 2.86 \\
uniform pooling & .517 & .530 & .705 && .588 & .848 & .755 && .443 & .735 & .675 && .507 & .740 & .647 \\
VAD & .552 & .620 & .718 && .652 & .877 & .740 && .405 & .603 & .610 && .505 & .690 & .630 \\
random & .505 & .569 & .677 && .573 & .759 & .723 && .415 & .596 & .608 && .486 & .604 & .600 \\
DART & .547 & .757 & .805 && .620 & .902 & .807 && \textbf{.532} & .800 & .632 && .460 & .667 & .610 \\
SpeechPrune & .485 & .530 & .608 && .593 & .843 & .720 && \textbf{.532} & .818 & .562 && .445 & .560 & .585 \\
bin coverage & \textbf{.595} & .767 & .818 && .642 & .915 & .802 && .520 & .835 & .657 && \textbf{.510} & \textbf{.767} & .627 \\
\addlinespace[1pt]
FastV$_N$ & .557 & .688 & .730 && .578 & .818 & .740 && .502 & .802 & .608 && .458 & .675 & .613 \\
HeadRouter$_N$ & .557 & .740 & .807 && .588 & .805 & .752 && .527 & .860 & \textbf{.690} && .472 & .665 & .635 \\
\textbf{ours} & .585 & \textbf{.863}$^{\dagger}$ & \textbf{.820} && \textbf{.682}$^{\dagger}$ & \textbf{.943}$^{\dagger}$ & \textbf{.828} && .525 & \textbf{.873}$^{\dagger}$ & .655 && .492 & .743$^{\dagger}$ & \textbf{.650} \\
\bottomrule
\end{tabular}
\end{table}

\paragraph{Transcription.} At the conservative budget ($1.11$--$2.00\times$), bin coverage costs a median of $.017$ WER against full audio and has lower WER than every baseline in ten of the twelve cases. The aggressive budget ($1.18$--$2.86\times$) is a stress test: bin coverage has lower WER than every baseline in all twelve cases, the mid-prefill ones included, and is below DART by a median of $.064$ (Tab.~\ref{tab:asrmain}a). Covering the timeline is not enough: uniform pooling covers it too and has higher WER in all $24$ cases. In the eight Qwen LibriSpeech and FLEURS cases, bin coverage also matches or beats a neural VAD, a word-onset selector and decimation, which keeps evenly spaced tokens (WER $.369$ against $.116$ for Qwen 3B on LibriSpeech at the aggressive budget). Segmentwise pruning keeps the same bins but scores the tokens by the encoder's self-attention. It has higher WER in all eight cases, by a median of $.018$, and trails DART on mean WER. Both parts count: the bins cover the timeline, and the prior picks the token within each bin.

\begin{figure}[t]
\centering
\includegraphics[width=\linewidth]{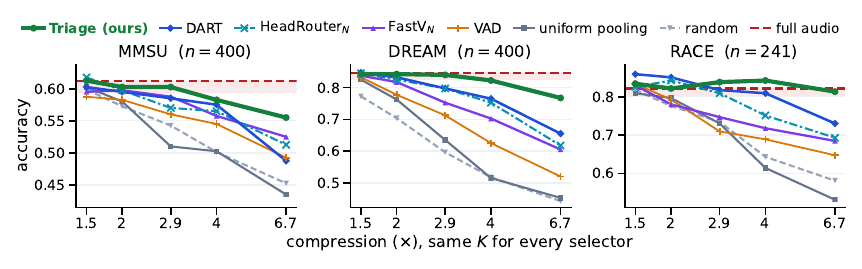}
\caption{\textbf{At the same $K$, Triage stays within two points of full audio to higher compression than every baseline.} Qwen 3B on a second sample of clips ($n{=}400$ per benchmark, $241$ on RACE), disjoint from Tab.~\ref{tab:asrmain}'s clips on MMSU and DREAM. The red dashed line is full audio, and the shaded area is the two-point band. Triage is inside the band to $2.9\times$ on MMSU and DREAM and through $6.7\times$, the highest ratio swept, on RACE. No baseline is inside it past $2\times$ on MMSU and DREAM or past $4\times$ on RACE.}
\label{fig:pareto}
\end{figure}

\paragraph{Multiple choice.} At the aggressive budget ($2.22$--$5.00\times$), precision fusion is ahead of DART by $+.043$ $[+.032,+.053]$ averaged over the twelve cases (Tab.~\ref{tab:asrmain}b). It is ahead of DART in eleven of the twelve cases and on each benchmark on average, by $.024$ to $.073$, and ahead of the best baseline in nine cases. It still leads DART in five of the six Voxtral and Phi-4 cases, and on natural read speech (\S\ref{sec:efficiency}). No baseline is a safe substitute: each trails Triage by at least $.07$ in some Voxtral or Phi-4 case, while Triage is never more than $.035$ behind any. SpeechPrune, which also sees the question, trails Triage in eleven of the twelve cases. Even when DART is applied to the language model's layer-2 states, it trails Triage by at least $.075$ on DREAM. At the conservative budget ($1.54$--$2.86\times$), Triage is within $.02$ of full audio in nine of the twelve cases and never more than $.031$ below. Within two points of full audio, it compresses $1.4$ to $2.0\times$ more than every baseline we swept on Qwen 3B, on every benchmark and in each of two samples (Fig.~\ref{fig:pareto}). Beyond speech, on the sound questions of MMAU-mini and MMAR, the conservative cut stays within $.03$ of full audio on both Qwen models.

\paragraph{What each part adds.} The prior alone already leads: cutting once with bin coverage, before the language model runs and without the question, it is ahead of DART by $.026$ on average and in ten of the twelve aggressive cases (Tab.~\ref{tab:asrmain}b). The full two-stage Triage, whose Stage~2 sees the question, leads DART by $.043$, and is ahead of Stage~1's top-$K$ cut at the same $K$ in eleven of the twelve aggressive cases and tied in the twelfth. The layer-2 cut depends on the prior's shortlist: replacing it with a random shortlist of the same size costs Triage $.047$ to $.147$ accuracy at Qwen 3B's three aggressive points. On the same shortlist, precision fusion is on average $.024$ above cutting on the observed attention alone, and on Qwen 3B, averaged over cuts at layers~2 and~8, it is the most accurate of eight fusion rules, above a supervised learned gate. The prior thus picks the shortlist, and the observed attention, which has seen the question, refines it.

\subsection{Efficiency}
\label{sec:efficiency}

Fewer audio tokens mean faster prefill, more concurrent streams and more audio per context window.

\textbf{Language-model prefill.} Stage~1 cuts before the first layer, so every layer runs on the shorter sequence. At eight deployed operating points (two models, two tasks, two budgets), prefill time falls by $1.2$ to $3.7\times$ on a GH200, on $5$ minutes of audio for Qwen 3B and $40$ minutes for Qwen 30B. The saving grows with length: on an A100, Qwen 3B's aggressive DREAM point ($5\times$) gives $4.5\times$ on $5$ minutes and $6.3\times$ on $20$. The two layers before Stage~2's cut add $3\%$ on average over a prefill of only the $K$ kept tokens.

\textbf{Batched serving.} At the same point on $5$ minutes of audio, one GH200 serves $1{,}024$ concurrent streams, the most we tried and $4\times$ as many as with full audio, at $4.5\times$ the decode throughput. Fewer tokens shrink each stream's KV cache, so more streams fit in GPU memory.

\textbf{With the encoder.} The encoder still runs on all $N$ tokens, and Stage~1 adds only one dot product per token. On the A100, encoder and prefill together speed up $1.28$ to $1.51\times$ for $5$ to $40$ minutes of audio.

\textbf{Context window.} Only a cut before the language model extends the context window: the tokens it keeps take contiguous positions. Qwen 3B's window has room for $21.8$ minutes of audio ($32{,}768$ positions at $1{,}500$ audio tokens a minute). On needle-in-a-haystack QA over $20$ to $45$ minutes of concatenated RACE articles, Stage~1 at the aggressive point cuts the tokens $2.86\times$, which fits about $62$ minutes into the window, and one forward pass matches per-minute window voting, which needs $20$ to $45$ passes ($.722$ against $.706$). On $90$ natural recordings of $20$ to $43$ minutes (QuALITY stories~\cite{quality} read by LibriVox volunteers~\cite{librivox}), only $6$ fit Qwen 3B's window at full length. Here, aggressive Triage reaches $.584$ accuracy: $.046$ above truncating each recording to fit the window, $.026$ above per-minute window voting, which keeps every token, and $.023$ and $.132$ above DART and uniform pooling at the same $K$, with all intervals above zero. Truncation loses each recording's end and voting sees one minute at a time, while Triage keeps the whole recording in one pass.

\section{Conclusion}
\label{sec:conclusion}

Where image pruning cuts, audio tokens are still heavily attended and far from their final ranking, so audio needs a ranking before the language model runs. A linear map on the encoder output, fitted in closed form on unlabelled clips, supplies one on eleven of thirteen LALMs, far better than the signals used by existing methods. At layer~2 it even ranks the tokens better than the language model's own attention. Cutting on it, Triage stays close to full audio at its conservative budget. At its aggressive budget it beats every baseline on transcription and leads DART on multiple choice. Because it cuts before the language model, the context window holds nearly $3\times$ as much audio, about an hour on Qwen 3B. Triage needs no labels and keeps the model frozen.

\paragraph{Limitations and future work.} The linear prior fails on the two Qwen-Audio models (\S\ref{sec:screen}). The screen flags such models from one forward pass, and a nonlinear prior may extend Triage to them. The prior is fitted to attention, a proxy for usefulness, but the same closed-form fit applies to any per-token target, so a closer proxy can replace attention. Combining Triage with KV-cache eviction during decoding is a natural next step.

\ifpreprint\else
\input{tex/body/06-statements}
\fi

\FloatBarrier \clearpage \appendix
\counterwithin{table}{section} \counterwithin{figure}{section}

\section{Where Each Main Claim Is Measured}\label{app:map}

\begin{center}
\captionof{table}{\textbf{Where each main claim of the body is measured}, in body order, with the sample it is measured on. 3B and 30B denote Qwen 3B and Qwen 30B; $n$ counts clips unless the row names another unit.}
\label{tab:map}
\scriptsize
\setlength{\tabcolsep}{4pt}
\begin{tabular}{@{}>{\raggedright\arraybackslash}p{0.055\linewidth} >{\raggedright\arraybackslash}p{0.405\linewidth} >{\raggedright\arraybackslash}p{0.25\linewidth} >{\raggedright\arraybackslash}p{0.23\linewidth}@{}}
\toprule
Body & Claim & Measured in & Sample \\
\midrule
\S\ref{sec:doublebind} & Audio fails both conditions; images and video fail only the ranking one & Fig.~\ref{fig:teaser}, panel~2; Fig.~\ref{fig:massbylayer} & 30B; RACE, MMMU, MVBench; $n{=}100$ each \\
 & The same passage draws more attention as speech than as a rendered page & Tab.~\ref{tab:matched} & 30B, 3B, Phi-4; $40$ passages \\
\midrule
\S\ref{sec:stage1} & Positions that see the question attend mostly to the same tokens & App.~\ref{app:queryindep} & 3B; $50$ RACE passages, $186$ questions \\
 & $g$ is chosen once, on a separate sample, and used unchanged everywhere & Tab.~\ref{tab:targetstudy}; Tab.~\ref{tab:targetmatched} & 3B; DREAM, $n{=}400$; MMSU, DREAM, RACE \\
 & The prior predicts $g$ on all seven models & Tab.~\ref{tab:crossarch}; App.~\ref{app:crossarch} & LibriSpeech; $n_\text{train}{=}120$, $n_\text{test}{=}100$ \\
 & One prior pooled over six benchmarks, including sound and music, at least matches per-benchmark priors & Tab.~\ref{tab:universal} & 3B; six benchmarks, $n{=}200$ each \\
 & The signals used by existing methods predict $g$ weakly & Tab.~\ref{tab:crossarch}; Tab.~\ref{tab:freesignals}; Tab.~\ref{tab:surprise-robust} & as Tab.~\ref{tab:crossarch}; 3B, MMSU, $n{=}120$ \\
 & A regression fitted to the waveform stays well below the prior, and eleven waveform and lexical properties do not explain the prior, even combined linearly & Tab.~\ref{tab:partialout}; App.~\ref{app:keeps} & 3B; LibriSpeech, FLEURS; $n_\text{test}{=}100$ \\
 & The prior predicts attention deep in the model, not just the layer-1 attention, and more than the attention sink & Tab.~\ref{tab:layerwise} & Voxtral, 3B; $n_\text{test}{=}100$ \\
 & Where Stage~2 cuts, the prior ranks the audio tokens better than the language model's own attention & Fig.~\ref{fig:layerconv}; App.~\ref{app:l2probe}; Tab.~\ref{tab:headstart} & 30B, RACE, $n{=}100$; 3B, $n{=}400$ per benchmark \\
\midrule
\S\ref{sec:causal-body} & Deleting the prior's top tokens costs the most & Fig.~\ref{fig:readprops}a; Tab.~\ref{tab:causaldel-ci}; App.~\ref{sec:causal} & Voxtral, Phi-4: LibriSpeech, $n{=}100$; 3B: $150$ per benchmark \\
 & The prior works on eleven of thirteen LALMs, and one unlabelled forward pass separates the ones it fails on & Fig.~\ref{fig:readprops}b; Tab.~\ref{tab:screen}; App.~\ref{app:screen} & $\rho$: LibriSpeech, $n_\text{test}{=}100$; screen: $n{=}20$ per model \\
\midrule
\S\ref{sec:readnotspend} & For targets taken from the language model, the better the prior predicts a target, the more accurate the cut & Tab.~\ref{tab:targetstudy} & 3B; DREAM, $n{=}400$ \\
 & Training a network to predict attention does not make the cut significantly more accurate & App.~\ref{app:priorprobes} & 3B, 30B, Voxtral; $400$ per benchmark; MMAU-mini, $150$ \\
\midrule
\S\ref{sec:method} & On transcription, bin coverage alone has lower WER than both stages together at seven of the eight Qwen points & Tab.~\ref{tab:asrwer} & 3B, 30B; LibriSpeech, FLEURS; $n{=}300$, $100$ on 30B \\
 & Kept tokens get contiguous positions after Stage~1 and keep their position IDs after Stage~2 & App.~\ref{app:splice} & every case \\
 & Calibration needs no labels & App.~\ref{app:calibsweeps} & every case \\
\midrule
\S\ref{sec:setup} & Triage leads HeadRouter$_N$ in all six aggressive Qwen 3B cases and matches or beats an all-rows FastV$_N$ in eleven of twelve Qwen multiple-choice cases & Tab.~\ref{tab:asrmain}; App.~\ref{app:baselinefidelity}; App.~\ref{app:hrname} & as Tab.~\ref{tab:asrmain}; 3B, 30B; $n{=}400$ per case \\
\midrule
\S\ref{sec:main-results} & Bin coverage stays close to full audio at the conservative budget and has lower WER than every baseline at the aggressive one & Tab.~\ref{tab:asrmain}a; Tab.~\ref{tab:asrwer}; Tab.~\ref{tab:asrlong} & $n{=}300$ per case, $100$ on 30B; seven TEDLIUM talks \\
 & Bin coverage matches or beats a neural VAD, a word-onset selector, decimation and segmentwise pruning & Tab.~\ref{tab:newbaselines}; App.~\ref{app:baselinefidelity} & $n{=}300$ per 3B case, $100$ per 30B case \\
 & On multiple choice, Triage leads DART in eleven of twelve aggressive cases and the best baseline in nine & Tab.~\ref{tab:asrmain}b; Tab.~\ref{tab:mcqmcnemar}; Tab.~\ref{tab:ablation} & $n{=}400$ per case \\
 & Within two points of full audio, Triage compresses more than every baseline & Fig.~\ref{fig:pareto}; Tab.~\ref{tab:isoquality}; App.~\ref{app:isoq} & 3B; $n{=}400$, $241$ on RACE; two disjoint samples \\
 & On sound questions, the conservative cut stays close to full audio & Tab.~\ref{tab:cascade} & 3B, 30B; MMAU-mini, MMAR; $n{=}200$ each \\
 & Bin coverage alone, before the language model and without the question, is ahead of DART on average and in ten of twelve aggressive cases & Tab.~\ref{tab:asrmain}b; Tab.~\ref{tab:ablation} & all four models; $n{=}400$ per case \\
 & The prior's shortlist, precision fusion and the layer-2 cut each add accuracy & Tab.~\ref{tab:shortctl}; Tab.~\ref{tab:fusion}; Tab.~\ref{tab:ablation} & shortlist, eight rules: 3B; fusion, layer-2 cut: all four models; $n{=}400$ per case \\
\midrule
\S\ref{sec:efficiency} & Prefill falls at all eight timed points and more on longer audio, the pre-cut layers cost little, and encoder plus prefill still speeds up & Tab.~\ref{tab:eff}; App.~\ref{app:efficiency} & GH200; A100 for audio length and the encoder \\
 & One GPU serves $4\times$ as many streams as with full audio & Tab.~\ref{tab:eff}; Tab.~\ref{tab:stage2eff} & GH200 \\
 & Stage~1 extends the context window, and one compressed pass matches window voting on needle QA & Tab.~\ref{tab:longaudio}; App.~\ref{app:longaudio} & 3B; $n{=}100$ per needle position \\
 & On natural long recordings, aggressive Triage outperforms truncation, window voting, DART and uniform pooling & Tab.~\ref{tab:longnat}; App.~\ref{app:longnat} & 3B; $90$ recordings, $1{,}590$ questions \\
\bottomrule
\end{tabular}
\end{center}
\suppressfloats[t]

\section{Modality Controls}
\label{app:modality}

\begin{figure}[t]
\centering
\includegraphics[width=0.46\linewidth]{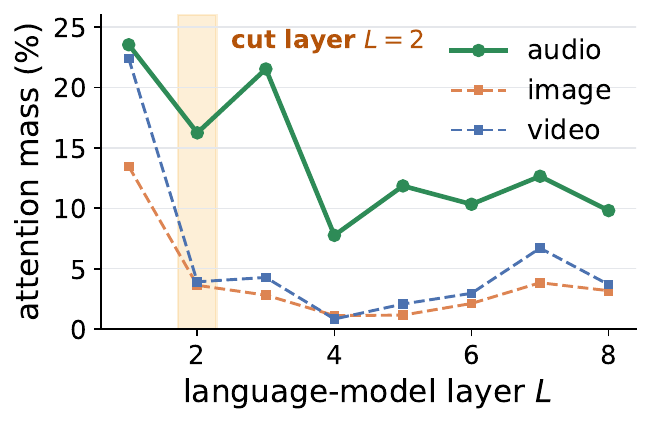}
\caption{\textbf{Attention mass by layer, the panel behind the attention condition of \S\ref{sec:doublebind}.} Each curve is the share of the text positions' attention that goes to one modality, at each of the first eight language-model layers (Qwen 30B, $n{=}100$; audio from RACE, images from MMMU, video from MVBench). Audio draws more attention than \emph{both} visual modalities at every one of these layers. From layer~2 onward, \emph{video} draws much less attention than audio, closer to images, although it has more tokens than audio and is also temporal.}
\label{fig:massbylayer}
\end{figure}

Audio draws more attention than either visual modality at each of the first eight layers in Fig.~\ref{fig:massbylayer}. So a blanket cut at the depth where a mid-prefill method acts would remove tokens that the language model still attends to. Attention mass is a share, so it depends on how many modal tokens each prefill carries. The median is $1{,}170$ tokens for audio ($n{=}100$, with every RACE passage truncated to $90$\,s). On separate draws, it is $286$ for MMMU images ($n{=}120$) and $1{,}728$ for MVBench video ($n{=}20$). Video carries \emph{more} tokens than audio, yet at layer~2 it draws a quarter as much attention, so the gap between audio and video is not a matter of token count. Against images, the matched run below holds the task and the words fixed.

\subsection{The Same Passage, Heard or Read}\label{app:matched}

\S\ref{sec:doublebind} sets audio against MMMU images, a different task on different material. Tab.~\ref{tab:matched} removes that confound. Each item is one AudioMarathon-RACE passage, given in one of three forms: heard, read as a rendered page, or read as text. In each form, the passage is followed by its own multiple-choice question under the same system prompt and chat template. The page shows the words that the text-to-speech system read aloud. The audio is truncated to $90$\,s, and the page and the text are cut to the matching prefix.

\begin{table}[t]
\centering
\caption{\textbf{One passage, three channels, same question.} Mass is the attention mass at layer~2, where Stage~2 cuts, as a mean over passages $\pm$ its standard error. Tokens are the median over passages, and mass\,/\,token is the mean of each passage's own mass per token. Every $\times$ ratio in the text is a per-passage median. Each model answers about as many questions right from the page as from the audio, so both are channels it can use. Phi-4 is not run on text: its two modalities take different LoRA adapters, and plain text takes neither. This run uses its own prompt, so its masses can be compared across its rows but not with the $16.2\%$ of \S\ref{sec:doublebind}.}
\label{tab:matched}
\small
\setlength{\tabcolsep}{6pt}
\begin{tabular}{@{}ll r r r r@{}}
\toprule
Model & the passage & tokens & mass at $L{=}2$ & mass\,/\,token & answered right \\
\midrule
Qwen3-Omni-30B  & heard              & $1{,}170$ & $10.16\pm.14$ & $.0087\%$ & $32/40$ \\
($n{=}40$)      & read, as a page    & $1{,}080$ & $\phantom{0}4.36\pm.06$ & $.0041\%$ & $31/40$ \\
                & read, as text      & $\phantom{00}318$ & $25.87\pm.23$ & $.0806\%$ & $33/40$ \\
\addlinespace
Qwen2.5-Omni-3B & heard              & $2{,}250$ & $23.17\pm.24$ & $.0103\%$ & $29/40$ \\
($n{=}40$)      & read, as a page    & $1{,}426$ & $10.80\pm.13$ & $.0077\%$ & $30/40$ \\
                & read, as text      & $\phantom{00}318$ & $25.03\pm.23$ & $.0780\%$ & $30/40$ \\
\addlinespace
Phi-4-multimodal & heard             & $1{,}125$ & $11.72\pm.21$ & $.0104\%$ & $30/40$ \\
($n{=}40$)       & read, as a page   & $1{,}792$ & $10.63\pm.14$ & $.0059\%$ & $31/40$ \\
                 & read, as text     & --- & --- & --- & --- \\
\bottomrule
\end{tabular}
\end{table}

\textbf{Holding the task and the words fixed leaves audio ahead at the cut layer.} At layer~2, audio draws $2.3\times$ the page's attention on Qwen 30B and $2.1\times$ on Qwen 3B, and it leads on every one of the $40$ passages in both models. On Phi-4 it draws $1.1\times$, and its lead is mainly per token ($1.8\times$). So audio leads on both measures, share and per token, in all three models. The $2.3\times$ understates the gap. The page runs $1{,}080$ tokens on Qwen 30B, against $286$ for a median MMMU image, so this control gives the visual side about four times as many tokens as the cross-benchmark comparison did.

\textbf{Audio spreads the attention it draws over many more tokens than the same words need as text.} Text is the reference here, not a pruning target. Read as text, the passage draws $9.2\times$ as much attention per token as when it is heard on Qwen 30B, and $7.5\times$ as much on Qwen 3B. The mass at layer~2 is why a blanket cut there is unsafe, and this per-token gap is why a selective cut has room.

\section{Generalisation: Across Models, Audio Types, and Lengths}
\label{app:generalisation}

The prior is fitted once per model. We test how well it transfers: to other models, to other audio types and to recordings too long for the uncompressed path. For a new model, one forward pass also predicts in advance whether the prior will work on it. A last subsection fixes the quality and asks how far each selector can compress.

\subsection{Across Models}
\label{app:crossarch}
Every $\rho$ in Tabs.~\ref{tab:crossarch} and~\ref{tab:screen} comes from one protocol: LibriSpeech, $n_\text{train}{=}120$, $n_\text{test}{=}100$, with $g$ summed over all language-model layers and heads as in Eq.~\ref{eq:importance}. Voxtral and Qwen2-Audio~\cite{qwen2audio} are HF-native, so we replicate their audio path (the audio encoder and projector give $\mathbf{e}_i$, and the selected subset is spliced back in), while Phi-4 uses its speech-LoRA adapter. Each model family runs in a separate harness, on an independent $100$-clip sample.

\textbf{The encoder and the language model need not be trained together.} The four models of Tab.~\ref{tab:crossarch} whose encoder and language model were \emph{not} trained together hold its three highest rows and its lowest. These four are Ultravox at $.83$, Phi-4 and Voxtral at $.79$ each, and Granite at $.69$, with the three Qwen-Omni models between them. DiVA, a Whisper encoder attached to a \emph{frozen} Llama-3, reaches the highest $\rho$ in Tab.~\ref{tab:screen}. Its connector emits $448$ learned queries whatever the clip's length, so its tokens carry no time order. The prior holds on the two Qwen2.5-Omni models, which adopt Qwen2-Audio's audio encoder, and on SeaLLMs-Audio-7B and Aero-1-Audio, which attach that encoder to Qwen2.5 language models (Tab.~\ref{tab:screen}). Our hypothesis is that the prior holds when the encoder output already lies in the space the language model takes as input, a property the screen of App.~\ref{app:screen} measures from one forward pass.

\paragraph{The full pipeline.} Voxtral and Phi-4 run Triage in the main harness (Tabs.~\ref{tab:ablation}, \ref{tab:asrwer} and~\ref{tab:asrlong}). On transcription, Triage's bin coverage has a lower WER than every baseline at both budgets on both models. On multiple choice, over both budgets, Triage leads DART in five of Voxtral's six cases and in all six of Phi-4's. The prefill saving holds on Voxtral and Phi-4 as on Qwen 3B: on a matched $7{,}500$ audio tokens, the tightest cut in Tab.~\ref{tab:densityscaling} speeds prefill up $2.8$--$3.0\times$ on all three models.

\paragraph{Phi-4's two modalities.} On Phi-4, the only non-Qwen model here that takes images as well as audio, audio is again the less reliable modality under the ranking condition of \S\ref{sec:doublebind}: on $20$ LibriSpeech streams concatenated to $150$\,s and $20$ MMMU images, its audio tokens' ranking at layer~$2$ agrees with their all-layer ranking at $\rho{=}.44$, against $.68$ for the image tokens. As in \S\ref{sec:doublebind}, the tokens are ranked by the attention the model's response pays them, up to its first $16$ tokens. The response is a verbatim transcript for the audio and an answer to the MMMU question for the image.

\paragraph{The two Qwen-Audio models.} On these two models the prior fails because it is linear. The MLP of App.~\ref{app:probetopk} reaches $\rho{\geq}.90$ on both models, in both prompt formats, so the encoder output carries the signal in a form the linear prior misses. On Qwen2-Audio, which Tab.~\ref{tab:crossarch} leaves out, the prior still leads the other four signals, at $\rho{=}.57$ against at most $|\rho|{=}.32$. The screen of App.~\ref{app:screen} places both models below its threshold without fitting anything.

\paragraph{MLP prior on Qwen2-Audio.} We replace the linear prior inside Stage~1's bin coverage with an MLP fitted to the same $120$ clips as the linear one. The MLP reaches held-out $\rho{=}.94$ and $.95$ on two halves of the test set. On $200$ LibriSpeech clips (full-audio WER $.083$), it lowers WER against the linear prior at every budget, by $.05$ to $.15$, with every $95\%$ interval below zero. Against the best baseline, the difference is not significant at the two milder budgets, and the MLP is significantly ahead at the tightest, $.195$ against DART's $.285$ ($[-.119,-.060]$), where it equals bin coverage on the true attention (Tab.~\ref{tab:q2amlp}).

\begin{table}[h]
\centering\small
\caption{\textbf{On Qwen2-Audio, an MLP in the prior's place gives Stage~1 the lowest WER of any deployable score.} The model is Qwen2-Audio-7B-Instruct on LibriSpeech ($n{=}200$), where full audio has WER $.083$. Each row gives the WER of Stage~1's bin coverage with one score, or of the best of six baselines at each budget (\texttt{pool}, \texttt{vad}, \texttt{random}, \texttt{dart}, \texttt{fastv} and FastV's all-rows variant). \emph{True attention} is bin coverage on the language model's own $g$, an oracle that needs the full prefill. \textbf{Bold} marks the lowest WER without the oracle.}
\label{tab:q2amlp}
\setlength{\tabcolsep}{6pt}
\begin{tabular}{@{}lccc@{}}
\toprule
audio tokens kept & $65\%$ & $50\%$ & $35\%$ \\
\midrule
linear prior & $.139$ & $.186$ & $.348$ \\
\textbf{MLP prior} & $\mathbf{.092}$ & $\mathbf{.126}$ & $\mathbf{.195}$ \\
best baseline & $.101$ (\texttt{vad}) & $.144$ (\texttt{dart}) & $.285$ (\texttt{dart}) \\
\midrule
true attention (oracle) & $.094$ & $.111$ & $.195$ \\
\bottomrule
\end{tabular}
\end{table}

\begin{table}[t]
\centering
\caption{\textbf{At a matched $N{=}7{,}500$ audio tokens, keeping $35\%$ speeds prefill up $2.8$--$3.0\times$ on all three models.} We measure prefill latency (A100-80GB, bf16, \texttt{sdpa} on all three) on synthetic audio of the stated duration, for full audio and for keeping $65/50/35\%$ of the audio tokens. Each latency is a mean over timed passes after two warm-up passes. ``$\times$'' is full$\,\div\,$kept. Voxtral and Phi-4 emit $12.5$\,tok/s, and Qwen 3B emits $25$\,tok/s. \textbf{Bold} marks the entry that keeps $35\%$ in the matched $N{=}7{,}500$ row of each block. Absolute latencies differ with language-model width and depth.}
\label{tab:densityscaling}
\small
\setlength{\tabcolsep}{5pt}
\begin{tabular}{@{}l r r r ccc@{}}
\toprule
Model (tok/s) & audio & $N$ & full (ms) & $.65$ & $.50$ & $.35$ \\
\midrule
\multirow{2}{*}{Qwen2.5-Omni-3B ($25$)}
 & $5$\,min & $7{,}500$ & $315.2$ & $1.61\times$ & $2.06\times$ & $\mathbf{2.92\times}$ \\
 & $10$\,min& $15{,}000$ & $695.1$ & $1.65\times$ & $2.19\times$ & $3.16\times$ \\
\midrule
\multirow{2}{*}{Voxtral-Mini-3B ($12.5$)}
 & $5$\,min & $3{,}750$ & $194.5$ & $1.56\times$ & $2.01\times$ & $2.71\times$ \\
 & $10$\,min& $7{,}500$ & $406.6$ & $1.59\times$ & $2.08\times$ & $\mathbf{2.98\times}$ \\
\midrule
\multirow{2}{*}{Phi-4-multimodal ($12.5$)}
 & $5$\,min & $3{,}750$ & $267.8$ & $1.46\times$ & $1.86\times$ & $2.08\times$ \\
 & $10$\,min& $7{,}500$ & $540.7$ & $1.56\times$ & $2.02\times$ & $\mathbf{2.78\times}$ \\
\bottomrule
\end{tabular}
\end{table}

\subsection{Predicting Where the Prior Works Before Fitting Anything}\label{app:screen}

The screen is one forward pass over $20$ unlabelled clips, spliced into the model's deployment prompt. The pass records how far the audio positions move through the language model's first two layers, $\Delta_2=\lVert \mathbf{h}_2-\mathbf{h}_0\rVert/\lVert \mathbf{h}_0\rVert$, averaged over audio positions and clips. The screen needs no prior, no labels and no downstream run, and it finishes in under two minutes on one A100, model loading included. Layer~$2$ is where Stage~2 already cuts.

\paragraph{The threshold.} The threshold, $0.25$, is the geometric mean of the two class edges, Voxtral's $0.395$ and Qwen-Audio-Chat's $0.159$, and it splits all thirteen models of Tab.~\ref{tab:screen} correctly (Fig.~\ref{fig:screen}). Under leave-one-out on the first nine models screened, it places all nine correctly, and so does a threshold anywhere from $1.1$ to $2.2\times$ the higher of the two Qwen-Audio readings, so the chosen value sits on a plateau. Qwen 30B sits far above the threshold, at $2.100$. SeaLLMs-Audio and Aero attach Qwen2-Audio's encoder to Qwen2.5 language models through a one-layer linear connector, and the prior works on both ($\rho{=}.819$ and $.879$), so the screen does not simply flag that encoder. Two larger checkpoints of families already in the table, granite-speech-3.3-8b and Ultravox-v0.5-Llama-3.1-8B, also fall above the threshold, at $\Delta_2{=}0.685$ and $1.275$, and their priors reach $\rho{=}.646$ and $.748$.

\textbf{Across $79$ measurements of $\Delta_2$, every model stays on the same side of the threshold.} The measurements cover the first nine models screened, four of them under both a bare and a chat prompt, on LibriSpeech, FLEURS and TEDLIUM $60$\,s windows with $6$ to $100$ clips. Six models, including both class edges, were measured on disjoint samples. The two class edges come closest to the threshold, with worst readings of $0.165$ for Qwen-Audio-Chat and $0.340$ for Voxtral. Qwen-Audio-Chat's entry in Tab.~\ref{tab:screen} uses its bare prompt, and in its ChatML format it stays below the threshold ($\Delta_2{=}0.160$, $\rho{=}.255$).

\begin{table}[h]
\centering\small
\caption{\textbf{A threshold of $0.25$ on $\Delta_2$ splits the thirteen models.} $\Delta_2 = \lVert \mathbf{h}_2-\mathbf{h}_0\rVert/\lVert \mathbf{h}_0\rVert$ is the relative distance the audio positions move over the language model's first two layers, from one forward pass over $20$ unlabelled clips. Nine of the thirteen start from a Whisper encoder and fall on both sides, so the encoder family alone does not decide the side. Above the threshold, $\Delta_2$ does not rank $\rho$: it returns a side, not a distance. \textbf{Bold}: the highest $\rho$.}
\label{tab:screen}
\small
\setlength{\tabcolsep}{4pt}
\begin{tabular}{@{}llcc@{}}
\toprule
 & & \emph{before fitting} & \emph{after fitting} \\
\cmidrule(lr){3-3}\cmidrule(l){4-4}
model & audio front end & $\Delta_2$ & $\rho$ \\
\midrule
\multicolumn{4}{@{}l}{\emph{above the $0.25$ threshold: the prior works}}\\
Qwen3-Omni-30B (MoE) & trained together & $2.100$ & $.701$ \\
SeaLLMs-Audio-7B~\cite{seallmsaudio} & Qwen2-Audio $\to$ Qwen2.5 & $1.342$ & $.819$ \\
Aero-1-Audio~\cite{aero} & Qwen2-Audio $\to$ Qwen2.5 & $1.143$ & $.879$ \\
Ultravox-v0.5-Llama-3.2-1B~\cite{ultravox} & Whisper $\to$ Llama-3.2 & $0.941$ & $.829$ \\
Phi-4-multimodal & conformer $\to$ Phi-4 & $0.906$ & $.794$ \\
DiVA-llama-3-v0-8b~\cite{diva} & Whisper $\to$ \emph{frozen} Llama-3 & $0.858$ & $\mathbf{.911}$ \\
MiDashengLM-7B-0804~\cite{midashenglm} & Dasheng $\to$ Qwen2.5-Omni & $0.747$ & $.843$ \\
Qwen2.5-Omni-3B & trained together & $0.739$ & $.750$ \\
Qwen2.5-Omni-7B & trained together & $0.665$ & $.727$ \\
granite-speech-3.3-2b~\cite{granitespeech} & conformer $\to$ Granite & $0.624$ & $.689$ \\
Voxtral-Mini-3B & Whisper $\to$ Mistral & $0.395$ & $.787$ \\
\midrule
\multicolumn{4}{@{}l}{\emph{below it: the prior fails}}\\
Qwen-Audio-Chat~\cite{qwenaudio} & Whisper $\to$ Qwen-7B & $0.159$ & $.288$ \\
Qwen2-Audio-7B-Instruct & Whisper $\to$ Qwen-7B & $0.144$ & $.566$ \\
\bottomrule
\end{tabular}
\end{table}

\subsection{Triage on Qwen 30B}
\label{app:30b}
\textbf{On Qwen 30B's multiple choice, Stage~2 supplies the discrimination that a direct Stage~1 cut lacks.} The prior is weaker on Qwen 30B than on Qwen 3B, so calibration keeps a wider shortlist ($r_1{=}.85$ in five of the six multiple-choice cases). At the aggressive points, precision fusion raises accuracy over top-$K$ on all three benchmarks, most on DREAM ($.877$ to $.943$; Tab.~\ref{tab:ablation}). The prior still guides Stage~2 at this scale: on Qwen 30B it recalls the top-$g$ tokens $15.7$--$17.6$ points better than raw layer-$2$ observation (Tab.~\ref{tab:prior-mechanism}). On transcription, Triage's bin coverage leads DART at five of the six Qwen 30B operating points, by $+.005$ to $+.074$ (Tabs.~\ref{tab:asrwer} and~\ref{tab:asrlong}).

\begin{table}[h]
\centering
\caption{\textbf{The encoder-side prior remains necessary at Stage~2.} Each entry is the token-level top-$g$ recall when keeping $20\%$ of the tokens. The prior retrieves the top tokens by converged $g$ better than raw cumulative layer-2 observation (\emph{obs.}) in all six multiple-choice cases, by $+.067$ to $+.177$, with five of the six gaps above $+.15$. Each benchmark uses up to $150$ clips.}
\label{tab:prior-mechanism}
\small
\setlength{\tabcolsep}{4pt}
\begin{tabular}{@{}llcc@{}}
\toprule
Model & benchmark & prior & obs. \\
\midrule
Qwen 3B & MMSU & .678 & .516 \\
 & DREAM & .644 & .467 \\
 & RACE & .564 & .497 \\
Qwen 30B & MMSU & .574 & .414 \\
 & DREAM & .577 & .401 \\
 & RACE & .548 & .391 \\
\bottomrule
\end{tabular}
\end{table}

\subsection{Across Audio Types}\label{app:universal}
\label{sec:generalization}

\textbf{On Qwen 3B, one prior fit on a pooled mix of six benchmarks predicts language-model attention on each of them at least as well as a prior fit to that benchmark alone} (Tab.~\ref{tab:universal}). The universal prior is ahead on all six, significantly on four, and reaches $\rho{=}.72$ on both general-audio benchmarks, MMAR~\cite{mmar} and MMAU-mini~\cite{mmau}. A benchmark held out of the fit is still predicted well: the zero-shot prior, fitted on the other five, is ahead of the in-domain prior on four of the six benchmarks, by up to $+.065$, and on MMAR reaches $\rho{=}.69$ against $.72$ in-domain. On transcription, the universal prior does about as well as a per-task refit. Across seven calibrated cases, the median $|\Delta|$ in WER is $.0045$, with differences in both directions. For bin coverage, which Triage deploys on transcription, the mean $\Delta$ is $+.0001$. Triage therefore deploys one universal prior per model, for multiple choice, transcription and long audio alike.

\begin{table}[t]
\centering
\caption{\textbf{What draws language-model attention is largely universal, not task-specific} ($n{=}200$ per benchmark). Each entry is the Spearman $\rho$ between predicted and measured language-model attention $g$, for a \emph{per-task} prior (\emph{in-dom.}), one \emph{universal} prior trained on a pooled mix of all six benchmarks, or a \emph{0-shot} prior trained on all \emph{other} benchmarks and tested on the held-out one. \textbf{Bold} marks the better of in-dom./univ.\ where the pair differs ($^{**}$: beyond $2$\,SE and ${\ge}.01$) and \emph{both} where it does not. On Qwen 3B the universal prior leads the per-task one on all six benchmarks, significantly on four, and the 0-shot prior is at most $.027$ below the universal one. On Qwen 30B the per-task prior keeps an edge on MMSU and LibriSpeech, the universal prior has a far larger edge on MMAU-mini, and the 0-shot prior is at most $.021$ below the universal one.}
\label{tab:universal}
\small
\setlength{\tabcolsep}{5pt}
\begin{tabular}{@{}l ccc c ccc@{}}
\toprule
& \multicolumn{3}{c}{\textbf{Qwen2.5-Omni-3B}} && \multicolumn{3}{c}{\textbf{Qwen3-Omni-30B (MoE)}} \\
\cmidrule(lr){2-4}\cmidrule(lr){6-8}
Benchmark & in-dom. & \textbf{univ.} & 0-shot && in-dom. & \textbf{univ.} & 0-shot \\
\midrule
\multicolumn{8}{@{}l}{\emph{speech QA}} \\
MMSU & .698{\scriptsize$\pm$.006} & \textbf{.759}{\scriptsize$\pm$.003}$^{**}$ & .756{\scriptsize$\pm$.004} && \textbf{.653}{\scriptsize$\pm$.007}$^{**}$ & .629{\scriptsize$\pm$.007} & .613{\scriptsize$\pm$.007} \\
AudioM.-RACE & .666{\scriptsize$\pm$.002} & \textbf{.713}{\scriptsize$\pm$.002}$^{**}$ & .690{\scriptsize$\pm$.002} && \textbf{.561}{\scriptsize$\pm$.003} & \textbf{.559}{\scriptsize$\pm$.003} & .549{\scriptsize$\pm$.003} \\
DREAM & .675{\scriptsize$\pm$.004} & \textbf{.745}{\scriptsize$\pm$.003}$^{**}$ & .740{\scriptsize$\pm$.003} && \textbf{.646}{\scriptsize$\pm$.005} & \textbf{.645}{\scriptsize$\pm$.005} & .635{\scriptsize$\pm$.005} \\
\multicolumn{8}{@{}l}{\emph{transcription}} \\
LibriSpeech & .740{\scriptsize$\pm$.004} & \textbf{.767}{\scriptsize$\pm$.003}$^{**}$ & .745{\scriptsize$\pm$.003} && \textbf{.760}{\scriptsize$\pm$.004}$^{**}$ & .713{\scriptsize$\pm$.005} & .692{\scriptsize$\pm$.005} \\
\multicolumn{8}{@{}l}{\emph{general audio (sound, music, speech)}} \\
MMAR & \textbf{.717}{\scriptsize$\pm$.004} & \textbf{.718}{\scriptsize$\pm$.004} & .691{\scriptsize$\pm$.004} && \textbf{.552}{\scriptsize$\pm$.006} & \textbf{.559}{\scriptsize$\pm$.007} & .554{\scriptsize$\pm$.007} \\
MMAU-mini & \textbf{.716}{\scriptsize$\pm$.004} & \textbf{.722}{\scriptsize$\pm$.004} & .702{\scriptsize$\pm$.005} && .486{\scriptsize$\pm$.007} & \textbf{.640}{\scriptsize$\pm$.007}$^{**}$ & .640{\scriptsize$\pm$.006} \\
\bottomrule
\end{tabular}
\end{table}

\subsection{Long Audio: Needle in a Haystack}\label{app:longaudio}

Each needle stream concatenates RACE articles into $20$ to $45$ minutes of audio, and one of the articles answers the question (Tab.~\ref{tab:longaudio}). Past ${\approx}21.8$ minutes, full audio no longer fits in Qwen 3B's context. The default single call runs at every duration but silently truncates: the processor's feature extractor keeps only the first five minutes of any stream. Every other condition, Triage and the baselines alike, encodes the stream one five-minute window at a time and concatenates the outputs, so its tokens cover the whole stream before any cut. We encode the natural recordings of App.~\ref{app:longnat} in the same way. Per-minute window voting is the one baseline that both runs and answers the question asked. Triage runs as deployed, with the universal prior and both stages, at the two operating points the label-free calibration picks at $20$ minutes (App.~\ref{app:needlecal}).

\textbf{One forward pass over the compressed audio matches per-minute window voting, which needs $20$ to $45$ passes.} The conservative point keeps $35\%$ of the tokens and leads voting, $.731$ against $.706$. The lead's paired $95\%$ interval is $[+.003,+.047]$. The aggressive point keeps $20\%$ and scores $.722$, and its interval against voting, $[-.005,+.037]$, covers zero. On the $1{,}626$ questions whose answer-bearing article lies outside the prior's fit set, both points still match voting.

\begin{table}[t]
\centering
\caption{\textbf{Long audio: what runs, and at what cost} (Qwen 3B, needle-in-a-haystack QA over concatenated RACE articles, accuracy $\uparrow$). Each entry averages $100$ questions at each of three needle positions, and the six durations hold $1{,}800$ questions. The positions are $15$, $50$ and $85\%$ of the stream. $D$ is the nominal duration in minutes and $N$ the encoder tokens, about $1{,}500D$ and fewer where a stream runs short of $D$. Two rows are failure modes. Chunked full audio overflows the context on every question past ${\approx}21.8$ minutes, and the single call, today's default, \emph{silently truncates}. The informative baseline is per-minute voting, the only other condition that both runs and answers the question asked. Triage runs at the two operating points the label-free calibration picked at $20$ minutes, kept unchanged at every duration. Its shortlists are $.50N$ and $.35N$. A shortlist wider than the context window is clamped to the positions the prompt leaves free, which at $45$ minutes trims the conservative one on $282$ of $300$ streams. The last row uses a prior fit on $40$ RACE clips from outside the article pool the streams are built from, and is the selector run at $80$ minutes. \textbf{Bold} marks the best method at each duration, not counting the oracle. The mean column is not bolded.}
\label{tab:longaudio}
\small
\setlength{\tabcolsep}{5pt}
\begin{tabular}{@{}l cc cccccc c@{}}
\toprule
Condition & fwd & tokens & 20 & 25 & 30 & 35 & 40 & 45\,min & mean \\
\midrule
oracle span (ceiling) & 1 & --- & .847 & .857 & .873 & .803 & .800 & .843 & .837 \\
\midrule
chunked, full audio & 1 & $N$ & .730 & \multicolumn{5}{c}{\emph{fails: every question over context}} & --- \\
single call (today's default) & 1 & $N$ & .513 & .500 & .497 & .410 & .470 & .500 & .482 \\
per-minute window voting & $D$ & $N$ & .700 & .717 & .740 & \textbf{.667} & .677 & .733 & .706 \\
\midrule
Triage, conserv.\ (.50/.35) & 1 & $.35N$ & \textbf{.787} & \textbf{.773} & .780 & .630 & \textbf{.700} & .713 & .731 \\
Triage, aggr.\ (.35/.20) & 1 & $.20N$ & .733 & .763 & .787 & .607 & .697 & \textbf{.743} & .722 \\
\midrule
\multicolumn{10}{@{}l}{\emph{Stage~1 alone}} \\
\quad universal prior & 1 & $.35N$ & .737 & .757 & \textbf{.793} & .590 & .680 & .740 & .716 \\
\quad universal prior & 1 & $.25N$ & .717 & .717 & .723 & .597 & .673 & .740 & .694 \\
\quad universal prior & 1 & $.20N$ & .670 & .677 & .723 & .593 & .677 & .700 & .673 \\
\quad prior fit on $40$ RACE clips & 1 & $.25N$ & .720 & .730 & .733 & .600 & .693 & .710 & .698 \\
\bottomrule
\end{tabular}
\end{table}

\paragraph{Several needles.} Tab.~\ref{tab:multineedle} spreads $k$ answer-bearing articles through one $20$-minute stream and counts an item as right only when all $k$ answers are right, so keeping one region of the stream no longer suffices. Both deployed points keep at least $89\%$ of full audio's accuracy at every $k$, and the conservative point never falls below full audio. Uniform pooling and truncation fall to $46\%$ and $31\%$ of full audio's accuracy by $k{=}2$. On streams whose answer-bearing articles all lie outside the prior's fit set, the conservative point still stays at or above full audio at every $k$.

\begin{table}[h]
\centering
\caption{\textbf{$k$ answer-bearing articles spread through one $20$-minute stream} ($n{=}100$ per $k$). An item counts as right only when all $k$ answers are right, and the stream stays below the context limit, so full audio still runs. The tokens column gives how many tokens each condition keeps in the end. Triage runs the two operating points of Tab.~\ref{tab:longaudio}. \textbf{Bold} marks the best \emph{selector} in each column. Full audio is the uncompressed reference and is not one of the selectors.}
\label{tab:multineedle}
\small
\setlength{\tabcolsep}{7pt}
\begin{tabular}{@{}l c cccc@{}}
\toprule
$20$\,min, all $k$ needles required & tokens & $k{=}1$ & $k{=}2$ & $k{=}3$ & $k{=}4$ \\
\midrule
full audio (uncompressed reference) & $N$ & $.71$ & $.52$ & $.40$ & $.38$ \\
\midrule
Triage, conserv.\ (.50/.35) & $.35N$ & $\mathbf{.78}$ & $\mathbf{.62}$ & $.48$ & $\mathbf{.40}$ \\
Triage, aggr.\ (.35/.20) & $.20N$ & $.72$ & $.56$ & $\mathbf{.51}$ & $.34$ \\
\multicolumn{6}{@{}l}{\emph{Stage~1 alone}} \\
\quad universal prior & $.35N$ & $.73$ & $\mathbf{.62}$ & $.49$ & $.37$ \\
\quad universal prior & $.25N$ & $.74$ & $.57$ & $.36$ & $.29$ \\
\quad universal prior & $.20N$ & $.67$ & $.48$ & $.30$ & $.23$ \\
\quad prior fit on $40$ RACE clips & $.25N$ & $.74$ & $.48$ & $.34$ & $.35$ \\
\midrule
uniform pooling & $.25N$ & $.48$ & $.24$ & $.11$ & $.08$ \\
truncation & $.25N$ & $.44$ & $.16$ & $.09$ & $.09$ \\
\bottomrule
\end{tabular}
\end{table}

\paragraph{Past the context limit.} An encoder-side cut moves the context limit in proportion to the budget. For Triage, the shortlist sets the limit: the conservative point's $.50N$ raises the limit to roughly $44$ minutes, and the aggressive point's $.35N$, a $2.86\times$ cut, raises it to roughly $62$ minutes. At $80$ minutes ($112{,}059$ encoder tokens against $32{,}768$ positions), full audio fails on every item. Stage~1 alone leads every other single-pass condition measured there ($100$ questions at each needle position). Here Stage~1 uses a prior fitted on $40$ RACE clips from outside the streams' article pool (Tab.~\ref{tab:longaudio}'s last row), not the universal one. At $4\times$ it scores $.653$, against $.527$ for uniform pooling and $.557$ for truncation. At $6.7\times$ it scores $.573$, against $.533$ and $.480$ for the same two.

\FloatBarrier

\subsection{Natural Recordings of Twenty to Sixty Minutes}\label{app:longnat}

\paragraph{Data and protocol.} The needle grid splices short articles into one stream, so we also test on recordings that are long by nature: stories from QuALITY~\cite{quality}, read aloud by LibriVox volunteers~\cite{librivox} and asked about with QuALITY's four-option multiple-choice questions. The main set holds $90$ stories and $1{,}590$ questions over $47$ hours of audio. Each recording runs $20$ to $43$ minutes. The target, the prior, the fusion rule and each model's operating points (its calibrated RACE points) were fixed before these recordings were first run. Each model runs its one universal prior, with no refitting on these recordings. Qwen 30B answers $815$ of the questions, a fixed subset that covers all $90$ stories, and Qwen 3B answers all $1{,}590$. Every $\Delta$ is paired on questions, with a $95\%$ interval from a bootstrap that resamples stories. The questions were written for readers of the text, so the no-audio row measures how much can be answered without listening.

\begin{table}[t]
\centering
\caption{\textbf{Natural long recordings, main set} (QuALITY stories read by LibriVox volunteers, $20$--$43$ minutes, four-option multiple choice, accuracy $\uparrow$). $\Delta$ is paired against each panel's reference, with a $95\%$ interval from a bootstrap that resamples stories. Qwen 30B answers $815$ questions in $90$ stories, and Qwen 3B answers all $1{,}590$. Qwen 3B's conservative point overflows its context on the longest stories, so that column covers $56$ stories ($999$ questions, on which truncation scores $.527$), and on $40$ of them its shortlist is trimmed to fit. Here \texttt{fastv} ranks all $N$ tokens by the attention that Stage~2 reads and, like FastV, continues the same forward pass after the cut. On Qwen 3B, it needs the whole recording in context, so it is run on the same truncated recording as the reference. Every row under an operating point keeps the same number of tokens.}
\label{tab:longnat}
\footnotesize
\setlength{\tabcolsep}{4.5pt}
\begin{tabular}{@{}l cc cc@{}}
\toprule
& \multicolumn{2}{c}{conservative} & \multicolumn{2}{c}{aggressive} \\
\cmidrule(lr){2-3}\cmidrule(l){4-5}
Method & acc & $\Delta$ [95\% CI] & acc & $\Delta$ [95\% CI] \\
\midrule
\multicolumn{5}{@{}l}{\emph{(a) Qwen3-Omni-30B, $r_1/r_2$ of $.85/.55$ and $.65/.35$, reference full audio}} \\
full audio (reference) & \multicolumn{4}{c}{$.779$} \\
no audio & \multicolumn{4}{c}{$.428$ \quad $-.351$ [$-.390,-.313$]} \\
one-minute window vote & \multicolumn{4}{c}{$.667$ \quad $-.112$ [$-.142,-.084$]} \\
\texttt{pool} & $.708$ & $-.071$ [$-.097,-.044$] & $.476$ & $-.303$ [$-.345,-.261$] \\
\texttt{vad} & $.713$ & $-.066$ [$-.093,-.040$] & $.589$ & $-.190$ [$-.227,-.154$] \\
\texttt{random} & $.620$ & $-.160$ [$-.185,-.134$] & $.507$ & $-.272$ [$-.307,-.237$] \\
\texttt{dart} & $.729$ & $-.050$ [$-.073,-.029$] & $.600$ & $-.179$ [$-.214,-.142$] \\
\texttt{fastv} & $.710$ & $-.069$ [$-.093,-.044$] & $.647$ & $-.133$ [$-.160,-.105$] \\
\multicolumn{5}{@{}l}{Triage, Stage~1 only} \\
\quad top-$K$ & $.688$ & $-.091$ [$-.117,-.065$] & $.573$ & $-.206$ [$-.243,-.170$] \\
\quad bin coverage & $.702$ & $-.077$ [$-.104,-.051$] & $.531$ & $-.248$ [$-.284,-.213$] \\
\multicolumn{5}{@{}l}{Triage, both stages} \\
\quad \textbf{deployed} & $.737$ & $-.042$ [$-.063,-.020$] & $.652$ & $-.128$ [$-.161,-.094$] \\
\quad prior at the cut & $.733$ & $-.047$ [$-.065,-.028$] & $.677$ & $-.102$ [$-.131,-.074$] \\
\quad observed-only at the cut & $.707$ & $-.072$ [$-.095,-.050$] & $.653$ & $-.126$ [$-.156,-.096$] \\
\midrule
\multicolumn{5}{@{}l}{\emph{(b) Qwen2.5-Omni-3B, $r_1/r_2$ of $.85/.65$ and $.50/.35$, reference truncation to the context window}} \\
truncation (reference) & \multicolumn{4}{c}{$.538$} \\
no audio & \multicolumn{4}{c}{$.370$ \quad $-.168$ [$-.199,-.137$]} \\
one-minute window vote & \multicolumn{4}{c}{$.558$ \quad $+.020$ [$-.005,+.046$]} \\
\texttt{pool} & $.544$ & $+.017$ [$-.007,+.041$] & $.452$ & $-.086$ [$-.109,-.064$] \\
\texttt{vad} & $.575$ & $+.048$ [$+.021,+.075$] & $.531$ & $-.007$ [$-.031,+.016$] \\
\texttt{random} & $.526$ & $-.001$ [$-.024,+.023$] & $.467$ & $-.071$ [$-.093,-.049$] \\
\texttt{dart} & $.576$ & $+.049$ [$+.019,+.080$] & $.560$ & $+.023$ [$0,+.045$] \\
\texttt{fastv} (part that fits) & $.526$ & $-.001$ [$-.014,+.012$] & $.524$ & $-.014$ [$-.029,+.001$] \\
\multicolumn{5}{@{}l}{Triage, Stage~1 only} \\
\quad top-$K$ & $.571$ & $+.044$ [$+.020,+.068$] & $.577$ & $+.040$ [$+.017,+.062$] \\
\quad bin coverage & $.571$ & $+.044$ [$+.018,+.070$] & $.580$ & $+.042$ [$+.018,+.066$] \\
\multicolumn{5}{@{}l}{Triage, both stages} \\
\quad \textbf{deployed} & $.557$ & $+.030$ [$+.002,+.058$] & $.584$ & $+.046$ [$+.024,+.067$] \\
\quad prior at the cut & $.562$ & $+.035$ [$+.008,+.062$] & $.583$ & $+.045$ [$+.024,+.067$] \\
\quad observed-only at the cut & $.550$ & $+.023$ [$-.005,+.051$] & $.570$ & $+.032$ [$+.010,+.053$] \\
\bottomrule
\end{tabular}
\end{table}

\textbf{On Qwen 3B, aggressive Triage leads both truncation and the one-minute window vote.} Full audio fits in Qwen 3B's context on only $6$ of the $90$ stories, so its reference is the recording truncated to the context window. At the aggressive point, Triage leads truncation by $+.046$ (Tab.~\ref{tab:longnat}b) and the vote, which keeps every token, by $+.026$ $[+.001,+.050]$. Aggressive Triage also leads every same-budget baseline there, DART by $+.023$ $[+.007,+.040]$ and uniform pooling by $+.132$ $[+.109,+.156]$. FastV$_N$, given only the part that fits, trails Triage by $.060$ [$+.038,+.082$]. On Qwen 30B, where full audio fits and is the reference, aggressive Triage is likewise ahead of every same-budget baseline (Tab.~\ref{tab:longnat}a).

\paragraph{Continuation set.} The longest main-set recording runs $43.1$ minutes. Tab.~\ref{tab:longnatext} therefore extends $27$ of the stories to $45.6$--$60$ minutes with the reader's narration of the rest of the story, and asks the same questions. These $27$ are all the main-set stories whose reading continues past the questioned text, in the same reader's voice, to at least $45$ minutes, so we report them in a separate table. At the aggressive point, Triage on Qwen 30B again leads every same-budget baseline. On Qwen 3B, the context window trims the shortlist on all $27$ stories to barely more than the final cut, so Stage~2 has little left to choose from. On Qwen 30B, Triage answers a question on this set $1.58\times$ faster than full audio at the conservative point and $1.91\times$ at the aggressive one, timed on another A100.

\paragraph{Timing and throughput.} Tab.~\ref{tab:longnattime} times Triage, each model's reference, no audio and window voting on one question per story at batch~$1$, from the waveform in memory to the answer letter, with the encoder rerun for every question and method. On Qwen 30B, aggressive Triage lowers the language-model prefill from $5.54$ to $3.57$ seconds. The whole question takes $4.45$ seconds against $6.39$ with full audio, and Stage~2 adds little to it: Stage~1 alone takes $4.37$. We also served Qwen 30B in batches, on $23$ of the timed questions. At a batch of four, the largest at which full audio fits, one A100 answers $1.57\times$ as many questions per second at the conservative point and $1.97\times$ at the aggressive one. The aggressive point also fits a batch of eight, where it answers $2.26\times$ as many as full audio at its largest batch. On the continuation set, where full audio again fits at most a batch of four, the two points reach $1.77\times$ and $2.39\times$.

\begin{table}[t]
\centering
\caption{\textbf{Natural long recordings, continuation set} ($27$ of the stories extended with the reader's narration of the rest of the story, $45.6$ to $60$ minutes). The questions are the main set's, and none asks about the added narration. Rows are as in Tab.~\ref{tab:longnat}. $\Delta$ is paired against full audio on Qwen 30B and against truncation on Qwen 3B. Qwen 30B answers $246$ questions, and Qwen 3B answers $482$. On Qwen 3B only the aggressive point fits, and \texttt{fastv} does not run. The last block compares the reference and deployed Triage with the main set, on the same questions. Truncation never reaches the added narration, so its difference is zero by construction.}
\label{tab:longnatext}
\scriptsize
\setlength{\tabcolsep}{3pt}
\begin{tabular}{@{}l cc cc cc@{}}
\toprule
& \multicolumn{2}{c}{Qwen 30B, conservative} & \multicolumn{2}{c}{Qwen 30B, aggressive} & \multicolumn{2}{c}{Qwen 3B, aggressive} \\
\cmidrule(lr){2-3}\cmidrule(lr){4-5}\cmidrule(l){6-7}
Method & acc & $\Delta$ [95\% CI] & acc & $\Delta$ [95\% CI] & acc & $\Delta$ [95\% CI] \\
\midrule
reference & \multicolumn{4}{c}{$.703$} & \multicolumn{2}{c}{$.560$} \\
no audio & \multicolumn{4}{c}{$.484$ \quad $-.220$ [$-.293,-.141$]} & $.363$ & $-.197$ [$-.264,-.130$] \\
one-minute window vote & \multicolumn{4}{c}{$.703$ \quad $0$ [$-.058,+.056$]} & $.591$ & $+.031$ [$-.021,+.082$] \\
\texttt{pool} & $.679$ & $-.024$ [$-.070,+.021$] & $.508$ & $-.195$ [$-.265,-.127$] & $.456$ & $-.104$ [$-.147,-.062$] \\
\texttt{vad} & $.715$ & $+.012$ [$-.045,+.072$] & $.598$ & $-.106$ [$-.173,-.041$] & $.519$ & $-.041$ [$-.092,+.008$] \\
\texttt{random} & $.622$ & $-.081$ [$-.138,-.028$] & $.514$ & $-.189$ [$-.255,-.125$] & $.459$ & $-.102$ [$-.142,-.063$] \\
\texttt{dart} & $.707$ & $+.004$ [$-.048,+.049$] & $.606$ & $-.098$ [$-.160,-.036$] & $.548$ & $-.012$ [$-.069,+.043$] \\
\texttt{fastv} & $.654$ & $-.049$ [$-.095,0$] & $.565$ & $-.138$ [$-.206,-.073$] & --- & --- \\
\multicolumn{7}{@{}l}{Triage, Stage~1 only} \\
\quad top-$K$ & $.679$ & $-.024$ [$-.066,+.016$] & $.606$ & $-.098$ [$-.164,-.033$] & $.581$ & $+.021$ [$-.027,+.066$] \\
\quad bin coverage & $.654$ & $-.049$ [$-.094,-.004$] & $.545$ & $-.159$ [$-.226,-.089$] & $.558$ & $-.002$ [$-.049,+.043$] \\
\multicolumn{7}{@{}l}{Triage, both stages} \\
\quad \textbf{deployed} & $.691$ & $-.012$ [$-.049,+.025$] & $.626$ & $-.077$ [$-.126,-.029$] & $.577$ & $+.017$ [$-.034,+.066$] \\
\quad prior at the cut & $.687$ & $-.016$ [$-.055,+.024$] & $.642$ & $-.061$ [$-.107,-.016$] & $.566$ & $+.006$ [$-.047,+.057$] \\
\quad observed-only at the cut & $.675$ & $-.028$ [$-.080,+.024$] & $.585$ & $-.118$ [$-.172,-.069$] & $.575$ & $+.015$ [$-.039,+.067$] \\
\midrule
\multicolumn{7}{@{}l}{\emph{Continuation minus main set, same questions (accuracy $\Delta$ [95\% CI])}} \\
reference & \multicolumn{4}{c}{$-.057$ [$-.101,-.012$]} & \multicolumn{2}{c}{$0$ [$0,0$]} \\
\textbf{Triage, both stages (deployed)} & \multicolumn{2}{c}{$-.033$ [$-.080,+.016$]} & \multicolumn{2}{c}{$-.033$ [$-.090,+.028$]} & \multicolumn{2}{c}{$-.031$ [$-.063,+.004$]} \\
\bottomrule
\end{tabular}
\end{table}

\begin{table}[t]
\centering
\caption{\textbf{Where the time goes on the main set} (median seconds per question on one A100, timed \emph{cold}: nothing is cached across questions or methods, though the weights stay loaded and discarded runs warm the GPU first; batch~$1$, except that window voting batches its windows). Each column is a median taken separately over the timed questions, so the columns need not add up. Qwen 3B's conservative rows cover the stories where that point fits.}
\label{tab:longnattime}
\footnotesize
\setlength{\tabcolsep}{6pt}
\begin{tabular}{@{}l l ccc@{}}
\toprule
Model & Method & encoder & language-model prefill & end to end \\
\midrule
Qwen 30B & no audio & --- & $2.26$ & $2.26$ \\
 & full audio (reference) & $0.85$ & $5.54$ & $6.39$ \\
 & one-minute window vote & $0.85$ & $7.43$ & $8.31$ \\
 & Triage, Stage~1 only: top-$K$, conservative & $0.85$ & $4.02$ & $4.84$ \\
 & Triage, both stages, conservative & $0.84$ & $4.09$ & $4.96$ \\
 & Triage, Stage~1 only: top-$K$, aggressive & $0.85$ & $3.51$ & $4.37$ \\
 & Triage, both stages, aggressive & $0.83$ & $3.57$ & $4.45$ \\
\midrule
Qwen 3B & no audio & --- & $0.06$ & $0.06$ \\
 & truncation (reference) & $4.45$ & $1.96$ & $6.41$ \\
 & one-minute window vote & $5.65$ & $1.85$ & $7.51$ \\
 & Triage, Stage~1 only: top-$K$, conservative & $5.09$ & $1.58$ & $6.68$ \\
 & Triage, both stages, conservative & $5.12$ & $1.63$ & $6.73$ \\
 & Triage, Stage~1 only: top-$K$, aggressive & $5.65$ & $0.80$ & $6.44$ \\
 & Triage, both stages, aggressive & $5.63$ & $0.84$ & $6.47$ \\
\bottomrule
\end{tabular}
\end{table}

\subsection{How Far Each Selector Can Compress}\label{app:isoq}

We sweep the eight selectors that appear at every budget over five ratios, $1.5\times$ to $6.7\times$, and record the largest compression each sustains within two points of full audio (Tab.~\ref{tab:isoquality}; selector names as in App.~\ref{app:tabnotes}).

\textbf{On Qwen 3B, Triage sustains $1.4$ to $2.0\times$ more compression than any baseline at the same quality, on every benchmark and in both of two disjoint samples.} Every baseline stays at or below $2\times$ on MMSU and DREAM, while Triage reaches $4\times$ on both in the first sample and $2.9\times$ in the second. On RACE the baselines are stronger, with HeadRouter$_N$ at $2.9\times$ and DART at $4\times$ in the second sample. Triage reaches $4\times$ there in the first sample and ${\ge}6.7\times$ in the second. The ordering also holds away from the threshold: in the second sample, which Fig.~\ref{fig:pareto} plots, Triage at $4\times$ scores $.823$ on DREAM against DART's $.765$. At $6.7\times$ it scores $.555$ on MMSU and $.813$ on RACE, against at most $.525$ and $.730$ for any baseline.

\paragraph{Qwen 30B.} On Qwen 30B, whose full audio scores higher on a separate iso-quality sample ($.950$ on DREAM, $.713$ on MMSU), the same band is stricter. There, one of our two selectors sustains $2\times$ on each benchmark, against $1.5\times$ for DART on MMSU. Two selectors tie at $2\times$ on DREAM and three on RACE. Past the tie, the ordering favours our selector in every case.

\FloatBarrier

\section{Alternatives to the Prior, Measured on the Same Footing}
\label{app:alternatives}

Each alternative below is scored against the prior on the same clips, and none displaces it. Selector names are defined in App.~\ref{app:tabnotes}.

\subsection{Free Signals Against the Language Model's Attention}
\label{app:surprise}

\textbf{No free signal tracks language-model attention both strongly and with a stable sign.} On Qwen 3B (Tab.~\ref{tab:freesignals}), acoustic energy keeps its sign across benchmarks but stays weak, self-information changes sign with the benchmark, and the state norm, the only signal far from noise, is negative: high-norm frames are less attended. Across the seven models of Tab.~\ref{tab:crossarch} the state norm runs from $-.54$ to $+.12$. In text compression, LLMLingua-2~\cite{llmlingua2} likewise replaces self-information with a learned classifier.

\begin{table}[h]
\centering
\caption{\textbf{Three free signals are scored against language-model attention on each benchmark.} Each entry is the Spearman $\rho$ of a signal against $g$ on Qwen 3B. Self-information is the residual of a linear autoregressor that predicts each token's encoder output from the $k$ before it. The last column is a second MMSU sample ($120$ clips, $23{,}951$ tokens), which Tab.~\ref{tab:surprise-robust} and App.~\ref{app:replace} also use.}
\label{tab:freesignals}
\small
\begin{tabular}{@{}lcccc@{}}
\toprule
Signal & MMSU & DREAM & RACE & MMSU, 2nd sample \\
\midrule
Acoustic energy      & $+.236$ & $+.216$ & $+.115$ & $+.276$ \\
Self-information ($k{=}4$) & $+.059$ & $-.080$ & $-.150$ & $-.191$ \\
State norm $\|\mathbf{e}_i\|_2$ & $-.346$ & $-.514$ & $-.538$ & $-.370$ \\
\bottomrule
\end{tabular}
\end{table}

Tab.~\ref{tab:surprise-robust} recomputes self-information on the raw waveform, both as a log-mel autoregressor residual and as per-frame spectral entropy. Self-information is weakly negative on the encoder states and weakly positive on the waveform. Nothing in that table except the prior ($+\PRIORRHOMMSU$) exceeds $|\rho|{=}.28$. One-step novelty $\|\mathbf{e}_i-\mathbf{e}_{i-1}\|$ and local redundancy, close relatives of self-information, give $-.05$ and $-.13$ on Qwen 3B. On Qwen 30B they give $+.07$ and $+.04$ (both models on MMSU, $n{=}200$).

\begin{table}[h]
\centering
\caption{\textbf{However self-information is measured, it stays far below the prior.} Each entry is the Spearman $\rho$ of a signal against $g$ per clip, on the second MMSU sample of Tab.~\ref{tab:freesignals}. Here self-information is the residual of a linear autoregressor fitted on training clips and applied to test clips. The last column of Tab.~\ref{tab:freesignals} fits one autoregressor per clip, which gives $-.191$.}
\label{tab:surprise-robust}
\small
\begin{tabular}{@{}llc@{}}
\toprule
Signal & Representation & $\rho$ vs.\ $g$ \\
\midrule
The prior (ours) & encoder $\mathbf{e}_i$ & $\mathbf{+.67}$ \\
\midrule
Self-information ($k{=}4$) & encoder $\mathbf{e}_i$ & $-.16$ \\
Self-information ($k{=}4$) & raw waveform & $+.24$ \\
Self-information ($k{=}16$) & raw waveform & $+.11$ \\
Spectral entropy & raw waveform & $+.01$ \\
Acoustic energy & raw waveform & $+.28$ \\
\bottomrule
\end{tabular}
\end{table}

We run LLMLingua's rule~\cite{llmlingua}, adapted to audio, as a Stage-1 selector: it keeps the top-$K$ audio tokens by self-information ($k{=}4$ over $\mathbf{e}_i$). On DREAM it falls far below random ($.582$ against $.747$ at the conservative budget, Tab.~\ref{tab:linguasel}). The inverted rule, which keeps the most predictable tokens, scores above random on DREAM. Top-$K$ on the prior still leads the better of the two rules in every case, by $.018$ to $.222$. The lead exceeds the ${\pm}.021$ SE in five of the six cases. LLMLingua's iterative pass (ITPC), which re-estimates self-information on the already-compressed prefix, lowers accuracy further on RACE, to $.677$ at the conservative budget and $.585$ at the aggressive one (a separate run, $n{=}400$).

\begin{table}[h]
\centering
\caption{\textbf{Prompt compression's rule, adapted to audio, is at or below random in five of six cases.} Each entry is gold accuracy on Qwen 3B at the two deployed operating points ($n{=}400$). The clips are those of Tab.~\ref{tab:asrmain}b, except that DREAM uses a separate sample, on which full audio scores $.894$ (against $.892$ on the grid). \emph{self-info} keeps the top-$K$ tokens by self-information, and \emph{inverse} keeps the bottom-$K$. \textbf{Bold} marks our selector, which is ahead of every baseline in five of the six cases. In four of them the lead exceeds the ${\pm}.021$ SE, and the largest lead is $.184$.}
\label{tab:linguasel}
\small
\setlength{\tabcolsep}{5pt}
\begin{tabular}{@{}ll c c cc cc@{}}
\toprule
Bench & Operating point & full & \textbf{top-$K$ (ours)} & self-info & inverse & VAD & random \\
\midrule
\multirow{2}{*}{MMSU} & conserv.\ & \multirow{2}{*}{.603} & \textbf{.603} & .583 & .585 & .595 & .608 \\
 & aggressive & & \textbf{.583} & .515 & .520 & .552 & .505 \\
\midrule
\multirow{2}{*}{DREAM} & conserv.\ & \multirow{2}{*}{.894} & \textbf{.861} & .582 & .801 & .831 & .747 \\
 & aggressive & & \textbf{.806} & .489 & .584 & .622 & .554 \\
\midrule
\multirow{2}{*}{RACE} & conserv.\ & \multirow{2}{*}{.818} & \textbf{.835} & .738 & .777 & .820 & .782 \\
 & aggressive & & \textbf{.820} & .615 & .688 & .718 & .677 \\
\bottomrule
\end{tabular}
\end{table}

\subsection{The Same Signals Fitted}
\label{app:replace}

\textbf{Fitting the free signals does not rescue them.} This control fits a regression like the prior's on the four unfitted signals of panel~3 of Fig.~\ref{fig:xtower} (encoder self-attention, self-information, acoustic energy and the state norm). The control uses Qwen 3B's second MMSU sample of Tab.~\ref{tab:freesignals}, while the figure uses Qwen 30B. The regression uses the four signals jointly and is cross-validated five-fold on $4{,}000$ of that sample's tokens. The fitted regression predicts language-model attention at $\rho{=}.42$, against $.66$ for the prior on the same tokens ($.67$ on the whole sample, Tab.~\ref{tab:surprise-robust}). From the signals that can be computed from the waveform alone, it reaches $.33$.

\subsection{Three Baselines an ASR Practitioner Reaches for First}
\label{app:newbaselines}

The grid of \S\ref{sec:setup} has no decimation, neural VAD or word-onset selector, so Tab.~\ref{tab:newbaselines} adds all three. Bin coverage has the lowest WER of all the selectors in seven of the eight cases. The two speech-specific selectors keep the $K$ most speech-like frames, which spends the budget on the clearest speech and drops whole stretches of the utterance. On Qwen 3B even decimation therefore has lower WER than both. A transcript needs coverage more than clarity, which is also why our transcription selector is bin coverage rather than top-$K$.

\begin{table}[t]
\centering
\caption{\textbf{Bin coverage has lower WER than decimation, neural VAD and word onsets in every case.} Each entry is corpus WER (\emph{lower is better}) on the clips of Tab.~\ref{tab:asrwer}, with $n{=}300$ per Qwen 3B case and $100$ per Qwen 30B case. Selectors shared with that table score the same here, and each model runs at its own operating points for the two budgets. \texttt{stride} keeps $K$ evenly spaced frames and discards the rest. \texttt{vadnn} is the released Silero VAD~\cite{silero}, and \texttt{wordonset} keeps the frames nearest a word onset in a greedy CTC decode by wav2vec~2.0 (\texttt{base-960h})~\cite{wav2vec2}, so neither uses a reference. \textbf{Bold} marks the best selector in each column.}
\label{tab:newbaselines}
\small
\setlength{\tabcolsep}{7pt}
\begin{tabular}{@{}l cc cc@{}}
\toprule
& \multicolumn{2}{c}{LibriSpeech} & \multicolumn{2}{c}{FLEURS} \\
\cmidrule(lr){2-3}\cmidrule(lr){4-5}
Selector & aggressive & conservative & aggressive & conservative \\
\midrule
\multicolumn{5}{@{}l}{\textbf{Qwen2.5-Omni-3B}} \\
\emph{kept} & $.50$ & $.65$ & $.85$ & $.90$ \\
full audio (ceiling) & $.0827$ & $.0827$ & $.1528$ & $.1528$ \\
\textbf{bin coverage (ours)} & $\mathbf{.1160}$ & $\mathbf{.0989}$ & $\mathbf{.1738}$ & $\mathbf{.1774}$ \\
\texttt{stride} (decimation) & $.3693$ & $.2195$ & $.1976$ & $.1778$ \\
\texttt{vadnn} (Silero) & $.5990$ & $.4736$ & $.4462$ & $.3133$ \\
\texttt{wordonset} (CTC onsets) & $.4166$ & $.2531$ & $.4720$ & $.3397$ \\
\texttt{random} (control) & $.4964$ & $.2966$ & $.2355$ & $.2416$ \\
\texttt{pool} & $.3480$ & $.3085$ & $.3247$ & $.2969$ \\
\texttt{vad} (energy) & $.3269$ & $.2648$ & $.3549$ & $.2683$ \\
\midrule
\multicolumn{5}{@{}l}{\textbf{Qwen3-Omni-30B (MoE)}} \\
\emph{kept} & $.50$ & $.65$ & $.50$ & $.65$ \\
full audio (ceiling) & $.0111$ & $.0111$ & $.0373$ & $.0373$ \\
\textbf{bin coverage (ours)} & $\mathbf{.0605}$ & $\mathbf{.0252}$ & $\mathbf{.0812}$ & $.0500$ \\
\texttt{stride} (decimation) & $.2190$ & $.0847$ & $.1803$ & $.0849$ \\
\texttt{vadnn} (Silero) & $.3086$ & $.1507$ & $.2322$ & $.1085$ \\
\texttt{wordonset} (CTC onsets) & $.2112$ & $.0667$ & $.1548$ & $.0656$ \\
\texttt{random} (control) & $.4289$ & $.2135$ & $.3851$ & $.2166$ \\
\texttt{pool} & $.1602$ & $.1255$ & $.1397$ & $.1274$ \\
\texttt{vad} (energy) & $.1991$ & $.0510$ & $.1383$ & $\mathbf{.0491}$ \\
\bottomrule
\end{tabular}
\end{table}

\section{What the Prior Keeps, What Each Stage Removes, and Whether the Model Needs It}\label{app:keeps}

\emph{Every result in this section is on Qwen 3B unless it names another model.} We map each retained audio token back to its time window and label it by acoustics (librosa~\cite{librosa} VAD, RMS energy) and by lexicon (wav2vec2~\cite{wav2vec2} CTC word offsets, content against function words). At a kept fraction of $10\%$, the prior's retained set carries less silence, fewer function words and more energy than attention top-$K$ or random (Tab.~\ref{tab:mechanism}a). On spontaneous TEDLIUM speech ($n{=}50$), $1.8\%$ of the retained tokens are silent, against the corpus's $15.4\%$. On LibriSpeech ($n{=}100$) the function-word share is $14.4\%$, against random's $29.1\%$, and content-word recall is higher, $.132$ against $.096$. The prior is more than a silence detector: deleting the silent tokens and re-scoring on speech alone raises its $\rho$ on LibriSpeech from $.755$ to $.770$, while acoustic energy's falls from $.255$ to $.228$.

\begin{table}[t]
\caption{\textbf{The prior keeps the least silence and the fewest function words, and each stage drops a different kind of token.} \textbf{(a)}~Composition of the retained set at a kept fraction of $10\%$, against attention top-$K$ and random. Attention top-$K$ keeps the tokens the language model itself attends to most, summed over its layers and heads. \textbf{Bold} marks the best value on each metric. \textbf{(b)}~Every audio token, grouped by the stage that drops it ($z$-scored within clip; RACE, $n{=}60$). In this panel Stage~2 reads only the last eight prompt rows. Stage~1 drops low-prior tokens. Stage~2 drops tokens that the prior does not rank low but that receive low observed attention (\textbf{bold}). The kept set ranks highest in every column, including the language model's all-layer attention. Stage~1 never uses that attention, and Stage~2 sees only its first two layers.}
\label{tab:mechanism}
\centering
\small
\begin{tabular}{@{}lccc@{}}
\toprule
\multicolumn{4}{@{}l}{\emph{(a) Retained set, $10\%$ kept}}\\
Metric & Ours & attn top-$K$ & Random \\
\midrule
Silence fraction (TEDLIUM) & \textbf{.018} & .088 & .154 \\
Function-word share (LibriSpeech) & \textbf{.144} & .190 & .291 \\
Mean RMS energy & \textbf{.091} & .074 & .060 \\
\midrule
\multicolumn{4}{@{}l}{\emph{(b) What each stage drops ($z$-scored)}}\\
Removed at & prior & observed & all-layer \\
\midrule
Stage~1 (query-agnostic) & $-.92$ & $-.22$ & $-.11$ \\
Stage~2 (observed refinement) & $+.12$ & $\mathbf{-.35}$ & $-.06$ \\
Kept (final) & $+.94$ & $+.66$ & $+.19$ \\
\bottomrule
\end{tabular}
\end{table}

\paragraph{The prior keeps its $\rho$ when surface properties are controlled for.} We take seven per-token waveform descriptors: position in the clip, RMS, spectral centroid, flatness, rolloff, zero-crossing rate and onset strength. The same ridge regression, refit on them alone, reaches only $\rho{=}.494$, against the prior's $.755$. Controlling for all of them lowers the prior's $\rho$ only to $.741$ ($150$ clips to fit, $100$ to test). Position, the most informative descriptor, correlates far more with $g$ ($-.485$) than with the prior ($-.226$), so it is not what the prior's weights encode. When four lexical descriptors are controlled for as well, $\rho$ stays at $.735$ on LibriSpeech and $.733$ on FLEURS, $97$ and $95\%$ of its uncontrolled value (Tab.~\ref{tab:partialout}).

\begin{table}[h]
\centering
\caption{\textbf{The prior keeps nearly all of its $\rho$ when waveform and lexical descriptors are controlled for} ($\rho$ between the prior and $g$, so \emph{higher is better}; $150$ clips to fit and $100$ to test). Each row removes from both the prior and $g$ everything that the listed descriptors explain. Nothing is removed from the audio. If the prior were a linear readout of loudness and position, the \textbf{bold} row, with both families removed at once, would be zero. The last row gives the bold row as a share of the uncontrolled $\rho$.}
\label{tab:partialout}
\small
\setlength{\tabcolsep}{8pt}
\begin{tabular}{@{}l cc@{}}
\toprule
$\rho$ between the prior and $g$ & on LibriSpeech & on FLEURS \\
\midrule
no controls & $.755$ & $.770$ \\[3pt]
\quad controlling for waveform descriptors & $.741$ & $.741$ \\
\quad {\scriptsize (position, RMS, spectral centroid, flatness, rolloff, zero-crossing rate, onset strength)} & & \\[3pt]
\quad controlling for lexical descriptors & $.740$ & $.747$ \\
\quad {\scriptsize (in-word, content-word, distance to a word boundary, word duration)} & & \\[3pt]
\quad \textbf{controlling for both} & $\mathbf{.735}$ & $\mathbf{.733}$ \\
\midrule
\emph{share of the prior that survives} & $97\%$ & $95\%$ \\
\bottomrule
\end{tabular}
\end{table}

\paragraph{The prior also predicts attention in the deep layers.}\label{app:layerwise} At layer~1 the attention between audio positions is a function of $\mathbf{e}$ by construction, so the prior could be recovering that term rather than anything the deeper layers add. When the prior is refit against $g$ restricted to a range of layers, it does worst on layer~1 alone, on both Voxtral and Qwen 3B. Over the deepest quarter of layers it still reaches $.784$ on Voxtral and $.710$ on Qwen 3B (Tab.~\ref{tab:layerwise}). On Voxtral and Qwen 3B, the first two layers carry only $6.6\%$ and $11.7\%$ of $g$. Dropping the attention sink, the audio token that receives the most attention, raises $\rho$.

\begin{table}[h]
\centering\small
\caption{\textbf{The prior predicts attention in the deepest layers too, and does worst on layer~1 alone} ($\rho$ against $g$ restricted to a layer range, with the prior refit to each range; $120$ clips to fit, $100$ to test, in a run separate from Tab.~\ref{tab:crossarch}'s). If the prior only recovered layer~1's bilinear form from $\mathbf{e}$, the row for layer~1 alone would have the highest $\rho$ here and the deep ranges would collapse. Neither happens. \textbf{Bold}: the all-layer target the prior is fit against.}
\label{tab:layerwise}
\setlength{\tabcolsep}{6pt}
\begin{tabular}{@{}l cc cc@{}}
\toprule
 & \multicolumn{2}{c}{Voxtral-Mini-3B ($30$ layers)} & \multicolumn{2}{c}{Qwen2.5-Omni-3B ($36$ layers)} \\
\cmidrule(lr){2-3}\cmidrule(l){4-5}
$g$ summed over & share of $g$ & $\rho$ & share of $g$ & $\rho$ \\
\midrule
every layer & $100\%$ & $\mathbf{.821}$ & $100\%$ & $\mathbf{.764}$ \\
\quad layer $3$ and deeper & $93\%$ & $.816$ & $88\%$ & $.759$ \\
\quad layer $9$ and deeper & $72\%$ & $.794$ & $74\%$ & $.751$ \\
\quad the deepest quarter & $21\%$ & $.784$ & $22\%$ & $.710$ \\
\midrule
layer $1$ alone & $3\%$ & $.474$ & $6\%$ & $.491$ \\
every layer, sink column dropped & --- & $.822$ & --- & $.772$ \\
\bottomrule
\end{tabular}
\end{table}

\paragraph{At the cut layer the prior ranks the audio tokens better than the model's own attention.}\label{app:l2probe} On the $100$ RACE clips of Fig.~\ref{fig:layerconv} (Qwen 30B), the deployed prior, fitted to $g$ rather than to the figure's target, reaches $\rho{=}.457$ (s.e.\ $.005$). The prior is ahead of the model's attention through layer~2 ($.386$) on $74$ of the clips. Averaging the two rankings gives $.511$ (s.e.\ $.007$), ahead of the deployed prior on $84$ clips and of the layer-2 attention on all $100$.

\subsection{What Each Stage Removes}
\label{app:mechanism}

\paragraph{The two stages remove different kinds of token.} Stage~1 removes what the audio itself makes unimportant. When Stage~1 keeps the prior's top half ($r_1{=}.5$) on $120$ LibriSpeech clips, $41\%$ of the tokens it drops fall outside every forced-aligned word, against $33\%$ of all tokens. On $120$ clips outside the prior's fit set, the two shares are $46\%$ and $37\%$. Stage~2 drops tokens the prior does not rank low (prior $z{=}{+}.12$, essentially the clip mean) but that receive low observed attention ($z{=}{-}.35$; Tab.~\ref{tab:mechanism}b), though they are acoustically ordinary speech. Relative to the prior's own top-$K$, Stage~2 re-selects $23\%$ of the kept set (Jaccard $.62$). The observed attention it uses is only partly correlated with the prior ($\rho{=}.37$), so Stage~2 recovers tokens that the language model attends to but the prior ranks too low.

\paragraph{Attention depends little on the question.}\label{app:queryindep} On AudioMarathon-RACE passages with ${\ge}3$ distinct questions, the kept sets of attention top-$K$ overlap at Jaccard $.71$ across a passage's questions ($50$ passages, $25\%$ of the tokens kept). Two random sets of that size would overlap at about $.14$, and $1.0$ would mean that the question changes nothing. The attention is read from the last eight prompt rows, which come after the question and so attend to it. An MLP given the encoder output plus the pooled question predicts that attention no better than the query-blind MLP ($\rho{=}{+}.62$ against ${+}.66$), so Stage~1 can ignore the question at little cost.

\subsection{Other Predictors and Inputs}\label{app:priorprobes}

\paragraph{The MLP fits $g$ better, mostly among the tokens a cut discards.}\label{app:probetopk} The $1.3$M-parameter MLP fits $g$ better than the prior, which has $2{,}049$ weights ($\rho{=}.72$ against $\PRIORRHOMMSU$ on the $120$-clip MMSU sample of Tab.~\ref{tab:freesignals}). The MLP's advantage lies among the low-ranked tokens that a cut discards: over the bottom $75\%$ of the true ranking, it scores $.493$ against the prior's $.392$. Over the true top $10\%$ and top $25\%$, neither predictor is significantly ahead. As selectors, the two stay within $.025$ of each other in Recall@$K$ at every kept fraction from $5$ to $90\%$. So the MLP's better fit does not make the cut more accurate. On Qwen 3B, over nine paired settings (MMSU, DREAM and RACE at three operating points each, $400$ clips per benchmark), the prior is ahead in $5$ (sign test, $p{=}1.00$), and no difference is significant. The median difference, prior minus MLP, is $+.005$, with a $96\%$ interval of $[-.005,{+}.010]$ (the order-statistic interval for nine settings). Fitted the same way on Qwen 30B and Voxtral, the MLP and the prior differ by at most $.03$ at both deployed operating points, never significantly (MMSU, DREAM and RACE on each model). The MLP is therefore a control, not a competing method: it shows that the signal in the encoder output does not depend on one kind of predictor. We deploy the prior because it needs only one closed-form solve per model, which takes seconds on a CPU.

\textbf{A trained resampler is not significantly more accurate than the prior in the deployed range, and is less accurate below that range.}\label{app:resampler} On the speech category of MMAU-mini~\cite{mmau}, not the sound clips of Tab.~\ref{tab:cascade}, we train an $8.5$M-parameter Perceiver~\cite{perceiver} connector, self-distilled against the frozen model's own answer distribution and anchored on uniform pooling, and fit the prior on the same $120$ training clips. Both are scored with paired $95\%$ bootstrap intervals on $150$ held-out clips, where full audio scores $.527$. The deployed range starts at a kept fraction of $20\%$, the smallest final kept fraction across the paper's $48$ operating points, so $25\%$ and $50\%$ lie inside it and $10\%$ below it. At a kept fraction of $25\%$ the connector leads the prior by $+.040$ $[-.027,+.107]$. At $50\%$ it leads by $+.027$ $[-.027,+.080]$, so neither lead is significant. At $10\%$, below the range, the prior leads by $+.173$ $[+.100,+.247]$: the connector's $.360$ there (one training run) against the prior's $.533$. Uniform pooling, the connector's starting point, scores $.313$, $.420$ and $.487$ from the smallest kept fraction to the largest, below both methods at every kept fraction. Trained on MMSU and scored on MMAU-mini, or the reverse, the two do not differ significantly at any kept fraction. The connector has $4{,}166\times$ the prior's parameters and needs $4{,}000$ gradient steps through the frozen model, where the prior needs one closed-form solve. The connector also has to be trained for each model. Because it outputs new vectors instead of keeping tokens, it gives Stage~2 no per-token score to correct.

\subsection{The Deletion Test: the Prior Marks Tokens the Model Needs}
\label{sec:causal}

\begin{table}[h]
\centering\small
\caption{\textbf{Deleting the prior's top tokens costs more than every control, on Voxtral and on Qwen 3B.} Each cell is prior-top deletion \emph{minus} control deletion, with a $95\%$ bootstrap interval over the same clips. Top: the margins of Fig.~\ref{fig:causaldel}, on Voxtral-Mini-3B and LibriSpeech ($n{=}100$), in mean per-clip WER. Voxtral's prior is fit for this intervention on $120$ other LibriSpeech clips. Seventeen of its eighteen margins exclude zero. The exception is self-information at $35\%$, where both deletions have saturated ($.935$ against $.927$ WER, full audio $.070$). Bottom: Qwen 3B on MMSU, DREAM and RACE ($150$ clips each), in gold-answer margin (the gold option's logit minus the best other option's), entered as its drop from full audio. This run uses the deployed prior unchanged ($\rho{=}.71$ to $.75$ on its clips). Sixteen of its eighteen margins exclude zero, and both exceptions are against the highest-energy frames. \textbf{Bold}: the energy-matched control, the one the claim depends on.}
\label{tab:causaldel-ci}
\resizebox{\linewidth}{!}{%
\begin{tabular}{@{}lccc@{}}
\toprule
Control (same \#tokens removed) & delete $10\%$ & delete $20\%$ & delete $35\%$ \\
\midrule
\multicolumn{4}{@{}l}{\emph{Voxtral-Mini-3B, WER}} \\
at random & $+.510$ [$+.440,+.580$] & $+.738$ [$+.666,+.813$] & $+.659$ [$+.541,+.741$] \\
the prior's own bottom & $+.544$ [$+.471,+.617$] & $+.851$ [$+.788,+.918$] & $+.911$ [$+.879,+.941$] \\
the highest-energy frames & $+.461$ [$+.391,+.532$] & $+.632$ [$+.533,+.719$] & $+.542$ [$+.445,+.608$] \\
matched to its position & $+.421$ [$+.347,+.497$] & $+.754$ [$+.693,+.817$] & $+.860$ [$+.819,+.901$] \\
self-information & $+.391$ [$+.318,+.465$] & $+.259$ [$+.050,+.427$] & $+.007$ [$-.075,+.082$] \\
\textbf{matched to its energy} & $\mathbf{+.528}$ [$+.457,+.598$] & $\mathbf{+.776}$ [$+.666,+.866$] & $\mathbf{+.875}$ [$+.832,+.914$] \\
\midrule
\multicolumn{4}{@{}l}{\emph{Qwen2.5-Omni-3B, gold-answer margin}} \\
at random & $+.293$ [$+.184,+.400$] & $+.643$ [$+.509,+.784$] & $+1.059$ [$+.858,+1.274$] \\
the prior's own bottom & $+.383$ [$+.286,+.485$] & $+.788$ [$+.644,+.929$] & $+1.457$ [$+1.234,+1.672$] \\
the highest-energy frames & $+.127$ [$-.0001,+.246$] & $+.225$ [$+.085,+.364$] & $+.158$ [$-.011,+.326$] \\
matched to its position & $+.361$ [$+.260,+.468$] & $+.759$ [$+.614,+.903$] & $+1.459$ [$+1.239,+1.682$] \\
self-information & $+.351$ [$+.252,+.453$] & $+.716$ [$+.572,+.860$] & $+1.251$ [$+1.043,+1.463$] \\
\textbf{matched to its energy} & $\mathbf{+.362}$ [$+.258,+.472$] & $\mathbf{+.782}$ [$+.639,+.926$] & $\mathbf{+1.413}$ [$+1.190,+1.642$] \\
\bottomrule
\end{tabular}}
\end{table}

A high $\rho$ shows that the prior tracks attention, not that the model needs the prior's top tokens. We therefore delete a fixed fraction of the tokens the prior ranks highest and compare the damage with six controls that delete the \emph{same number} of tokens by other rules: the prior's own bottom, random, the highest-energy frames, self-information, a set matched to the prior-top set's distribution over position, and a set matched to its energy profile. If prior-top deletion costs more than an equally large energy-matched deletion, the prior tracks what the model uses rather than loudness. Of the two energy controls, deleting the highest-energy frames is the less informative: the loudest frames are speech rather than silence, so deleting them also removes words the model needs, and the prior itself favours energetic tokens (Tab.~\ref{tab:mechanism}a). The energy-matched set holds the energy profile fixed and takes its tokens only from those the prior did not rank highest, so its margin is what the prior adds beyond loudness.

\paragraph{On Voxtral, prior-top deletion raises WER more than every control at every rate.} On LibriSpeech ($n{=}100$; the selection uses no reference transcript), seventeen of the eighteen $95\%$ intervals exclude zero (Tab.~\ref{tab:causaldel-ci}, top), and the margin over the energy-matched set, $+.53$ to $+.88$ WER, is as large as the margin over random, in a model family the prior was not designed on. Self-information comes closest: the prior's margin over it is $+.391$ at $10\%$ and $+.259$ at $20\%$, and its interval spans zero only at $35\%$, where both deletions have destroyed the transcript. Both matched controls cost less at $35\%$ than at $20\%$ in Fig.~\ref{fig:readprops}a because a matched set is drawn from the tokens the prior did not rank highest, a pool that shrinks as the rate grows. The paired margins, each taken within a rate, are unaffected.

\paragraph{The contrast also holds on Qwen 3B and on Phi-4.} On Qwen 3B we use the deployed prior unchanged, $150$ clips from each of MMSU, DREAM and RACE, and the gold-answer margin. Prior-top deletion costs $+.36$ more than the energy-matched set at $10\%$, and the gap widens with the rate (Tab.~\ref{tab:causaldel-ci}, bottom). All fifteen intervals against the five controls other than the highest-energy frames exclude zero. Against the same five controls, the contrast is also significant at every rate on KL to the full-audio answer distribution. On accuracy it is significant against the energy-matched set at $20\%$ and $35\%$, reaching $+.11$ at $35\%$. On Phi-4 we follow the same protocol on LibriSpeech ($n{=}100$), with a prior fit for this intervention on $100$ other clips ($\rho{=}.80$), and run four of the six controls at the three rates: the prior's own bottom, random, the highest-energy frames and the energy-matched set. Prior-top deletion is worse in all twelve comparisons. Ten of the twelve intervals exclude zero, and the two exceptions are the highest-energy frames at $20$ and $35\%$. Its margins over the energy-matched set are $+.076$, $+.143$ and $+.291$ WER at $10$, $20$ and $35\%$ deletion. The test thus covers three model families: two with Whisper-style encoders and one with a conformer encoder attached by a speech adapter.

\paragraph{A prior fitted to the trailing-row target fails the same test.} On Phi-4, with the same clips, controls and rates, a prior fitted to the trailing-row target instead of $g$ ($\rho{=}.59$) inverts the sign: deleting its top tokens degrades the model \emph{less} than deleting its bottom tokens ($-.028$, $-.048$, $-.067$ WER). Ten of its twelve comparisons are negative, seven of them significantly, while all twelve are positive for the prior fitted to $g$. On Qwen 3B's DREAM target-study sample, cutting on either target itself gives accuracies within $.03$ of each other (Tab.~\ref{tab:targetstudy}). So a direct cut does not tell the two targets apart, but the deletion test does.

\section{What Each Design Choice Contributes, and How the Budgets Are Set}\label{app:design}

\paragraph{Ranking at layer~2.} At layer~2, where our pipeline cuts, the prior alone is ahead of FastV's trailing-row signal in every setting of Tab.~\ref{tab:headstart}, by $.020$ to $.118$. The best selector in every row is one of ours, precision fusion in six and the prior alone in the other three. Precision fusion's largest gain over the prior alone is $+.035$, on DREAM when keeping $10\%$ of the tokens.

\begin{table}[t]
\centering
\caption{\textbf{At layer~2 the prior selects better than the attention observed there, in eight of the nine settings.} Each entry is gold accuracy on Qwen 3B after a cut at layer~2 that keeps the fraction $r_2$ of audio tokens ranked highest by each signal ($n{=}400$ per benchmark). The DREAM and RACE clips differ from those of the multiple-choice grid. Every selector chooses from the full $N$, so only the ranking signal differs, and the prior alone is computed before any language-model layer runs. The obs$_{\rm tail}$ column is FastV's signal as usually implemented, the attention from the trailing prompt positions. The obs column takes the same attention from every prompt-suffix position, as Eq.~\ref{eq:triage} does. \textbf{Bold} marks the best selector in each row, which is always the prior or precision fusion.}
\label{tab:headstart}
\small
\setlength{\tabcolsep}{6pt}
\begin{tabular}{@{}ll c cccc@{}}
\toprule
Bench & kept $r_2$ & full & obs$_{\rm tail}$ & obs & prior & precision fusion \\
\midrule
\multirow{3}{*}{MMSU}  & $35\%$ & \multirow{3}{*}{$.603$} & $.560$ & $.570$ & $.580$ & $\mathbf{.585}$ \\
                       & $20\%$ &                         & $.535$ & $.573$ & $\mathbf{.583}$ & $.580$ \\
                       & $10\%$ &                         & $.507$ & $.525$ & $.545$ & $\mathbf{.550}$ \\
\midrule
\multirow{3}{*}{DREAM} & $35\%$ & \multirow{3}{*}{$.877$} & $.792$ & $.828$ & $.830$ & $\mathbf{.835}$ \\
                       & $20\%$ &                         & $.723$ & $.745$ & $\mathbf{.797}$ & $.777$ \\
                       & $10\%$ &                         & $.600$ & $.613$ & $.630$ & $\mathbf{.665}$ \\
\midrule
\multirow{3}{*}{RACE}  & $35\%$ & \multirow{3}{*}{$.825$} & $.730$ & $.792$ & $\mathbf{.830}$ & $.823$ \\
                       & $20\%$ &                         & $.652$ & $.745$ & $.770$ & $\mathbf{.772}$ \\
                       & $10\%$ &                         & $.625$ & $.675$ & $.660$ & $\mathbf{.682}$ \\
\bottomrule
\end{tabular}
\end{table}

\begin{figure}[t]
\centering
\includegraphics[width=0.86\linewidth]{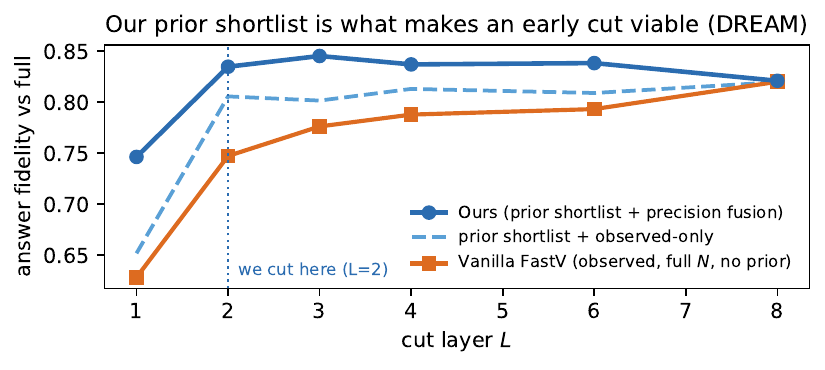}\\[1pt]
{\small (a) only the prior makes an early cut accurate}\\[6pt]
\includegraphics[width=0.86\linewidth]{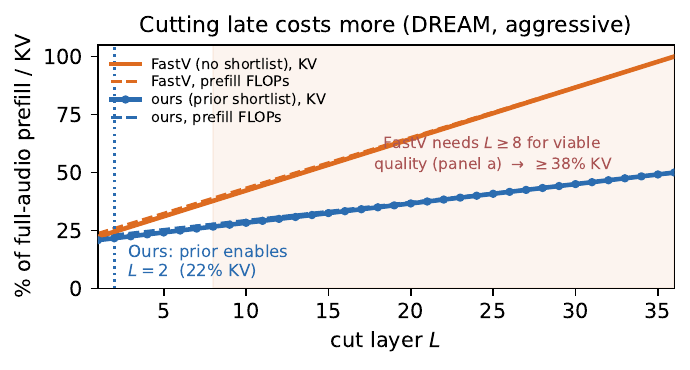}\\[1pt]
{\small (b) an early cut retains less KV}
\caption{\textbf{The prior's shortlist makes a cut at layer~2 accurate, and a cut that early retains little KV.} (a)~plots answer fidelity against cut layer $L$ on DREAM at the aggressive point ($r_1{=}.5$, $r_2{=}.2$, $n{=}100$). Answer fidelity is the pruned model's probability for the option that the full-audio pass picks. FastV, which observes before it cuts and so prefills all $N$, needs about eight layers to reach what our shortlist gives at layer~2, and precision fusion adds to it. (b)~plots retained audio KV and prefill against $L$ at the same point. Our cut at layer~2 retains about $22\%$ of the audio KV. FastV's cut at layer~8, where it first matches that fidelity, retains $38\%$, and later cuts retain more.}
\label{fig:earlycut}
\end{figure}

\paragraph{Cut depth.} FastV observes before it cuts, so it prefills all $N$ tokens, and on DREAM it needs about eight layers to reach the answer fidelity our shortlist gives at layer~2 (Fig.~\ref{fig:earlycut}a). Cutting that late also retains more of the audio KV and of the prefill (panel~b).

\paragraph{Shortlist and rule.} The shortlist does most of the work, and precision fusion adds to it. The control of Tab.~\ref{tab:shortctl} crosses two shortlists (the prior's, or a random one of the same size) with two survivor rules (observed-only, or precision fusion), at each benchmark's deployed operating points. The shortlist changes accuracy more than the rule in five of the six cases. Under precision fusion, which is the deployed rule, a random shortlist costs $.047$ to $.147$ at the aggressive point, and on all three benchmarks it costs more there than at the conservative point. On a random shortlist, precision fusion adds $.038$ at the aggressive point on both DREAM and RACE. On the full grid of Tab.~\ref{tab:ablation}, precision fusion leads observed-only on the same shortlist in $17$ of the $24$ cases. At the aggressive budget, its lead averaged over the four models is $+.024$. On transcription, bin coverage alone has lower WER than two-stage precision fusion at seven of the eight short-form Qwen operating points of Tab.~\ref{tab:asrwer}, so a single cut is deployed there.

\begin{table}[t]
\centering
\caption{\textbf{Replacing the prior's shortlist with a random one costs accuracy, most at the aggressive point.} Each entry is gold accuracy on Qwen 3B with the universal prior and a cut at layer~2 ($n{=}400$ clips per benchmark), at each benchmark's deployed operating points ($r_1$/$r_2$ in parentheses). Columns cross two shortlists (the prior's, or a random one of the same size $K_1$) with two survivor rules (observed-only, which is FastV on that shortlist, or precision fusion). $\Delta_{\rm shortlist}$ is prior minus random under the observed-only rule, and $\Delta_{\rm rule}$ is precision fusion minus observed-only on the prior shortlist, both computed before rounding. The three largest shortlist effects, on DREAM at both points and on RACE at the aggressive one, are $3.4$ to $4.6$ times their unpaired standard error. This table comes from a separate run, and its margins are taken within that run. Its entries differ from Tab.~\ref{tab:ablation} only in three DREAM cells, each by $.003$.}
\label{tab:shortctl}
\small
\setlength{\tabcolsep}{4.5pt}
\begin{tabular}{@{}ll c cc cc cc@{}}
\toprule
& & & \multicolumn{2}{c}{prior shortlist} & \multicolumn{2}{c}{random shortlist} & \multicolumn{2}{c}{effect} \\
\cmidrule(lr){4-5}\cmidrule(lr){6-7}\cmidrule(lr){8-9}
Bench & Operating point & full & \makecell{observed-\\only} & \makecell{precision\\fusion} & \makecell{observed-\\only} & \makecell{precision\\fusion} & $\Delta_{\rm shortlist}$ & $\Delta_{\rm rule}$ \\
\midrule
\multirow{2}{*}{MMSU} & conserv.\ (.85/.65) & \multirow{2}{*}{.603} & .600 & .600 & .595 & .593 & $+.005$ & $.000$ \\
 & aggr.\ (.35/.25) & & .580 & .585 & .548 & .538 & $+.033$ & $+.005$ \\
\midrule
\multirow{2}{*}{DREAM} & conserv.\ (.65/.45) & \multirow{2}{*}{.892} & .878 & .865 & .788 & .805 & $+.090$ & $-.013$ \\
 & aggr.\ (.50/.20) & & .815 & .860 & .675 & .713 & $+.140$ & $+.045$ \\
\midrule
\multirow{2}{*}{RACE} & conserv.\ (.85/.65) & \multirow{2}{*}{.818} & .815 & .823 & .808 & .813 & $+.008$ & $+.008$ \\
 & aggr.\ (.50/.35) & & .833 & .820 & .710 & .748 & $+.123$ & $-.013$ \\
\bottomrule
\end{tabular}
\end{table}

\paragraph{Precision fusion.}\label{app:method} Precision fusion fits no parameters, yet on a fixed shortlist it is the most accurate of the eight survivor rules in Tab.~\ref{tab:fusion}, including a supervised learned gate. Percentile ranks put the prior and the observed attention on a common scale, since one is a linear score on the encoder output and the other a sum of attention mass (Eq.~\ref{eq:triage}). The weight $\alpha_i$ is set per token because head agreement varies from token to token. Across the twelve matched-shortlist comparisons of Tab.~\ref{tab:shortctl}, precision fusion is ahead of observed-only by more than the unpaired standard error in three, and it is never behind by more than that error. The cut is also batching-safe at fixed $L$ and $K$, since every request in a batch keeps the same number of tokens at the same layer.

\begin{table}[t]
\centering
\caption{\textbf{Precision fusion is the most accurate of eight survivor rules on a fixed shortlist.} Each entry is the mean accuracy of Qwen 3B on the multiple-choice grid ($n{=}400$ per benchmark), averaged over benchmarks, budgets and cut layers $L{\in}\{2,8\}$. The shortlist, $L$ and $K$ are the same in every row, so only the rule changes. The $\alpha{=}1$ row ranks by the observed attention alone, which is FastV on that shortlist. Precision fusion fits no parameters, yet it is above five unlearned fusions and a supervised learned gate. \textbf{Bold} marks our rule.}
\label{tab:fusion}
\small
\setlength{\tabcolsep}{8pt}
\begin{tabular}{@{}l c@{}}
\toprule
Design choice & acc \\
\midrule
\multicolumn{2}{@{}l}{\emph{Stage~2, fusion rule on a fixed shortlist} (prior only: $.750$)}\\
observed-only ($\alpha{=}1$) & .762 \\
product & .765 \\
uniform blend ($\alpha{=}.5$) & .766 \\
max-rank & .767 \\
cascade (multi-layer) & .767 \\
learned gate (supervised) & .767 \\
bayes (precision, smooth) & .771 \\
\textbf{precision fusion (per-token, ours)} & \textbf{.777} \\
\bottomrule
\end{tabular}
\end{table}

\paragraph{Needle-task calibration.}\label{app:needlecal} The needle task of App.~\ref{app:longaudio} is calibrated by the same label-free rule on separate streams at $10$ and $20$ minutes, where full audio still fits and supplies the reference answer. These streams are built from the same RACE article pool as the test streams. Past the context limit no full-audio answer exists, so the operating points chosen at $20$ minutes are used unchanged at every longer duration. In the larger grid of Tab.~\ref{tab:longaudio}, both stages lead Stage~1 alone at the same final $K$ by $.015$ at the conservative point and by $.049$ at the aggressive one, on average over durations.

\subsection{Label-Free Calibration}\label{app:calibsweeps}
\paragraph{Procedure.} The calibration of \S\ref{sec:deployment} scores every candidate $(r_1,r_2)$ against the model's full-audio output, so no label enters at any point. Each candidate is judged by the larger of its drops on a calibration split and on a held-out split. Step~1 fixes the final kept fraction $r_2$ as the smallest $r_2$ whose drop stays within the budget. On multiple choice the drop is taken at a full shortlist, and on transcription $r_2$ qualifies if some shortlist $r_2{<}r_1{<}1$ keeps the drop within the budget. Step~2 then fixes the shortlist $r_1{>}r_2$, again taking the smallest value that qualifies. On multiple choice, $r_1$ qualifies if it is not clearly worse than the full shortlist. On transcription, it qualifies if its drop is within $.01$ of the best in-budget shortlist at that $r_2$. The final pair must also keep the drop within the budget. The deployed points are printed in the operating-point columns of Tabs.~\ref{tab:ablation}, \ref{tab:asrwer} and~\ref{tab:asrlong}.

\paragraph{Samples and scoring.} On both Qwen models, each multiple-choice case is calibrated on $100$ calibration and $100$ held-out clips, and each transcription case on $50$ and $50$ ($25$ each on Qwen 30B). Long-form cases use a coarser grid of candidates, scored on $60$\,s TEDLIUM windows. On transcription the drop is the WER against the full-audio transcript, averaged over clips.

\paragraph{Overlap with evaluation.} Calibration clips overlap the evaluation clips, but the calibration never sees a reference transcript, a gold answer or any other label. On the evaluation clips that the calibration did not use, our selector still leads every baseline of Tab.~\ref{tab:asrmain} at the aggressive budget: in all four short-form Qwen transcription cases by at least $.034$ WER, and on MMSU and DREAM for both Qwen models.

\FloatBarrier

\subsection{Every Candidate Target}
\label{app:targetfull}

The candidates in this section are the acoustic-energy control and sixteen language-model targets. Tab.~\ref{tab:targetstudy} prints thirteen of the sixteen for Qwen 3B on DREAM ($n{=}400$). Nine of them are attention variants, and four need a backward pass (gradient saliency and three attention--gradient mixtures). The other three, added for Tab.~\ref{tab:targetmatched}, are trailing-row attention weighted by value norm, input-times-gradient saliency and a fourth mixture (all rows RoPE-free $+$ gradient).

\paragraph{Predictability and accuracy.} Over the thirteen targets of Tab.~\ref{tab:targetstudy}, the rank correlation between $\rho$ and achieved accuracy is $+.66$ when $25\%$ of the tokens are kept. With $35\%$ of the tokens kept, the same rank correlation is $+.91$. At $25\%$ kept, the targets within one per-case SE of the best are exactly the eight that the prior predicts best.

\paragraph{Dispersion test.} Tab.~\ref{tab:targetmatched} compares two spreads across the sixteen language-model targets. The \emph{direct cut} keeps the tokens that a target itself ranks highest, and the \emph{achieved} accuracy is that of the cut on the prior fitted to the target. The achieved spread is significantly wider than the direct-cut spread in six of the seven rows, all but MMSU at $25\%$ kept. On DREAM the SD ratio grows as the budget tightens, from $1.63$ to $1.99$ to $2.25$, and RACE shows the same growth at larger ratios. The exact $p$ comes from a swap permutation test, which centres both series and enumerates all $2^{16}$ ways of swapping or keeping each target's pair of deviations. At $25\%$ kept, the deployed target has the highest achieved accuracy of all the candidates on every benchmark.

\begin{table}[h]
\centering
\caption{\textbf{On DREAM and RACE the SD ratio grows as the budget tightens, and only MMSU at $25\%$ kept shows no effect.} The test covers the sixteen language-model targets of App.~\ref{app:targetfull} on Qwen 3B. MMSU uses the clips of the multiple-choice grid, and DREAM and RACE use those of the target study. The \emph{direct cut} keeps the tokens that each target itself ranks highest, and \emph{achieved} is the accuracy of cutting on the prior fitted to that target. SD\,/\,SE puts the spread of direct-cut accuracy across targets in units of the binomial error of a test set that size, and $\chi^2$ $p$ tests it against a flat direct cut. The SD ratio divides the achieved spread by the direct-cut spread, with an exact $p$ from the swap permutation test. \textbf{Bold} marks the largest SD ratio on each of DREAM and RACE.}
\label{tab:targetmatched}
\small
\begin{tabular}{@{}ll r cc cc@{}}
\toprule
& & & \multicolumn{2}{c}{direct cut} & \multicolumn{2}{c}{achieved vs direct cut} \\
\cmidrule(lr){4-5}\cmidrule(lr){6-7}
Benchmark & kept & $n$ & SD\,/\,SE & $\chi^2$ $p$ & SD ratio & exact $p$ \\
\midrule
DREAM & $35\%$ & $400$ & $1.05$ & $.34$ & $1.63$ & $<.0001$ \\
DREAM & $25\%$ & $400$ & $1.48$ & $.005$ & $1.99$ & $.0005$ \\
DREAM & $15\%$ & $399$ & $1.54$ & $.002$ & $\mathbf{2.25}$ & $<.0001$ \\
\midrule
RACE & $35\%$ & $317$ & $0.45$ & $1.00$ & $2.59$ & $.042$ \\
RACE & $25\%$ & $316$ & $0.75$ & $.90$ & $\mathbf{2.87}$ & $.0004$ \\
\midrule
MMSU & $35\%$ & $400$ & $0.57$ & $.99$ & $1.31$ & $.043$ \\
MMSU & $25\%$ & $400$ & $0.77$ & $.88$ & $0.96$ & $.67$ \\
\bottomrule
\end{tabular}
\end{table}

\subsection{Long-Form Transcription}
\label{app:longformnotes}

\paragraph{Protocol.} The long-form results in the TEDLIUM columns of Tab.~\ref{tab:asrmain}a and in Tab.~\ref{tab:asrlong} use the universal prior. Following the standard long-form protocol~\cite{whisper}, the hypotheses of each talk's $60$\,s windows are concatenated and scored against the gold transcript of the whole talk. The test split holds seven whole TEDLIUM talks ($92$ minutes of audio), and the bootstrap resamples talks. Selection never uses a reference, and the generation cap is a fixed $400$ tokens per window.

\paragraph{Results.} On Qwen 3B, bin coverage leads the nearest baseline significantly at both budgets, and it is ahead on six of the seven talks at each. At $1.67\times$ the margin over DART is $.095$, wider than at any of Qwen 3B's short-form points (at most $.047$). VAD, an energy-based heuristic, is the worst selector at both Qwen 3B TEDLIUM points. On each of two disjoint halves of the $60$\,s chunks, bin coverage leads every encoder-side baseline.

\section{Protocol, Data, and Provenance}
\label{app:provenance}

This appendix records how the measurements were made: the hardware, the evaluation settings, decoding and scoring, the sample behind each $\rho$, how each baseline was reproduced, and how a cut token is removed.

\subsection{Hardware}
\label{app:hardware}

The accuracy tables run on a GH200 ($96$\,GB, aarch64; PyTorch~2.7.1+cu128, Transformers~4.57, bf16), as do Tab.~\ref{tab:eff} and the prefill and serving measurements behind it. The target-definition study (Tab.~\ref{tab:targetstudy}), the baseline re-runs, the prefill-scaling table across models (Tab.~\ref{tab:densityscaling}) and the encoder-plus-prefill accounting of App.~\ref{app:efficiency} run on an A100-80GB (x86; PyTorch~2.6, Transformers~4.57, bf16). Every speedup is a ratio taken within one platform. Qwen 30B's all-layer attention capture does not fit in $80$\,GB, so Qwen 30B's accuracies come from the GH200, except those of SpeechPrune, layer-2 DART and the MLP, which need no such capture and run on the A100. Its encoder-plus-prefill figure comes from one A100 run on $40$\,min of audio at its aggressive MMSU point ($r_1{=}.85$, $r_2{=}.35$), since the GH200 encoder runs stop at $10$\,min. There, the encoder takes $1.44$\,s, and the language-model prefill takes $6.72$\,s on full audio and $3.96$\,s on ours, which gives the $1.51\times$.

\subsection{The Nine Evaluation Settings} \label{app:benchmarks}

The deployed grids use six of the nine settings in Tab.~\ref{tab:benchmarks}, and the other three are breadth checks. Multiple choice is scored on $400$ clips per benchmark, drawn after the prior's fit clips are set aside, and transcription on $300$ clips each of LibriSpeech and FLEURS. Each list serves all four models, but Qwen 30B takes only the first $100$ clips of each transcription list. Transcription is also scored on seven TEDLIUM talks per model. MMAU-mini is the AudioBench~\cite{audiobench} copy of MMAU test-mini (\texttt{AudioLLMs/MMAU-mini-do-not-use} on Hugging Face, revision \texttt{db5fcbf6}), released before MMAU's v05.15.25 revision.

\begin{table}[h]
\centering
\caption{\textbf{The nine settings, what each one tests, and where it is scored.} The three breadth checks sit outside the deployed grids. Architecture figures quoted in the text are Qwen 3B's: a frozen audio encoder ($32$ layers, ${\approx}25$ tokens per second of audio) feeding a $36$-layer language model.}
\label{tab:benchmarks}
\small
\begin{tabular}{@{}llll@{}}
\toprule
Benchmark & What it tests & Metric & Scored in \\
\midrule
\multicolumn{4}{@{}l}{\emph{Multiple-choice QA}} \\
MMSU~\cite{mmsu} & targeted listening, ${\sim}6$\,s & gold accuracy & Tab.~\ref{tab:ablation} \\
DREAM~\cite{dream,audiobench} & dialogue comprehension, ${\sim}45$\,s & gold accuracy & Tab.~\ref{tab:ablation} \\
AudioMarathon-RACE~\cite{audiomarathon,race} & reading comprehension, ${\sim}2.5$\,min & gold accuracy & Tab.~\ref{tab:ablation} \\
\addlinespace[2pt]
\multicolumn{4}{@{}l}{\emph{Transcription}} \\
LibriSpeech~\cite{librispeech} & read-aloud speech, test-clean & corpus WER & Tab.~\ref{tab:asrmain} \\
FLEURS~\cite{fleurs} & read-aloud speech, English split & corpus WER & Tab.~\ref{tab:asrmain} \\
TEDLIUM~\cite{tedlium3} & spontaneous talks, $6$--$20$\,min & corpus WER & Tab.~\ref{tab:asrmain} \\
\addlinespace[2pt]
\multicolumn{4}{@{}l}{\emph{Breadth check: beyond the words}} \\
MMAR~\cite{mmar} & sound & gold accuracy, $\rho$ & Tabs.~\ref{tab:cascade},~\ref{tab:universal} \\
MMAU-mini~\cite{mmau} & sound and music & gold accuracy, $\rho$ & Tabs.~\ref{tab:cascade},~\ref{tab:universal} \\
IEMOCAP~\cite{iemocap} & emotion & gold accuracy & Tab.~\ref{tab:cascade} \\
\bottomrule
\end{tabular}
\end{table}

\subsection{Decoding, Scoring and How the Prior Is Fitted}
\label{app:decoding}

Transcription uses greedy decoding, with no sampling, beam search or rescoring, so selectors differ only in which tokens they keep. It is scored by \emph{corpus} WER (total edit distance over total reference words) after Whisper's~\cite{whisper} spelling normaliser. On a few clips, the model given full audio answers with commentary rather than a transcript: six of Qwen 3B's $300$ LibriSpeech clips, two of its FLEURS clips, and none on the other models. These clips are dropped for every selector. A compressed selector that answers with commentary is scored as it stands. If we instead drop every clip on which any selector answers with commentary, our lead at Qwen 3B's aggressive LibriSpeech budget widens: our WER goes from $.116$ to $.090$, and DART's from $.160$ to $.145$. Multiple choice is scored from the option-letter logits, so no answer is lost to parsing.

Unless the text names another sample, \emph{the prior's $\rho$} is the per-clip Spearman correlation on one split: $120$ unlabelled LibriSpeech clips to fit and $100$ held out to test. This is the protocol of Tab.~\ref{tab:crossarch} and of the screen in \S\ref{sec:screen}. Other $\rho$ values in the paper are labelled with what they measure (pooled over tokens, leave-one-out, after controls, or per audio type) and are compared only with values measured the same way.

One universal prior per model serves every run and every task, with three exceptions that fit their own prior: the Voxtral and Phi-4 deletion tests of App.~\ref{sec:causal}; the cascade runs of Tab.~\ref{tab:cascade}, fitted on $40$ clips of each benchmark; and the long-audio rows labelled ``prior fit on $40$ RACE clips'', whose prior also serves the $80$-minute runs (App.~\ref{app:longaudio}). Studies that vary the prior itself, such as the target study of Tab.~\ref{tab:targetstudy}, refit it for each candidate. Two runs over the same test sample, on the same GPU type and batch size, agree to the third decimal in every column in which nothing is fitted (full audio, uniform pooling, VAD, random, DART, FastV$_N$ and HeadRouter$_N$). Every margin is taken inside one table (App.~\ref{app:tabconv}). Within a case, every selector keeps the same $K$, and all selectors with the same kind of cut remove tokens the same way (App.~\ref{app:splice}).

\paragraph{What the prior and the target were chosen on.} The deployed prior is fitted the same way as the one behind \emph{the prior's $\rho$}, but on a larger pool of clips from several benchmarks (\S\ref{sec:stage1}). Its fit clips are set aside from the multiple-choice evaluation samples, and the fit sees only audio under a generic prompt, never a transcript, an answer or a label. The target was chosen once, on Qwen 3B's DREAM target-study sample (Tab.~\ref{tab:targetstudy}). Transcription's one-stage rule was chosen on the Qwen models' LibriSpeech and FLEURS points and then kept unchanged for TEDLIUM, Voxtral and Phi-4. On the two Qwen models, its WER is below that of both two-stage variants in eleven of the twelve cases (Tabs.~\ref{tab:asrwer} and~\ref{tab:asrlong}). On Voxtral and Phi-4, a two-stage variant is lower in eight of the twelve, but we kept the rule fixed rather than tune it per model. The two-stage design was chosen for multiple choice on the evaluation runs. It still sustains more compression than Stage~1 alone on all three benchmarks of Qwen 3B's second iso-quality sample (sample B of Tab.~\ref{tab:isoquality}), which is disjoint from the grid on MMSU and DREAM. The target, the prior, the fusion rule and each model's operating points were all fixed before the natural long recordings of App.~\ref{app:longnat} were first run, so those recordings test the whole method on data it was never chosen on.

\subsection{Baseline Fidelity}
\label{app:baselinefidelity}

We checked each baseline against its source paper. Unless stated otherwise, figures in this subsection come from a separate fidelity sample on Qwen 3B and Qwen 30B, with $n{=}400$ per case.

\paragraph{HeadRouter and FastV.} HeadRouter$_N$ follows its source on Qwen 3B and uses HeadRouter's own head weights only there, since the released weights are for Qwen 3B (App.~\ref{app:hrname}). FastV's paper defines its score as ``the average attention-score one token received from all other tokens'', whereas HeadRouter describes FastV as using the attention from the last generated text token, so we run both readings. FastV$_N$ scores tokens by the attention from the trailing prompt rows, as FastV is usually implemented, and is the reading Tab.~\ref{tab:ablation} reports. FastV$_N^{\text{all}}$ takes the attention from all rows. On the runs behind Tab.~\ref{tab:ablation}, Triage is at or above FastV$_N^{\text{all}}$ in eleven of the twelve Qwen multiple-choice cases.

\paragraph{DART and pooling.} DART~\cite{dart} is a vision-token method, but its criterion uses only the token sequence, so we port it. The port cuts once, on the encoder output before the language model. Like DART, it keeps $8$ pivot tokens and fills the rest of the budget with the tokens whose highest cosine similarity to any pivot is lowest. DART takes as pivots the tokens with the largest key norm inside the language model. Before the language model there are no keys, so the port takes the $8$ encoder outputs with the largest $\ell_1$ norm. On Qwen 3B's RACE fidelity sample at the aggressive point, DART scores $.800$ with these pivots and $.745$ with $8$ random ones. Uniform pooling averages $K$ contiguous segments, as published.

\paragraph{DART on layer-2 states.} DART reads keys inside the language model, so we also run its rule where FastV$_N$ cuts: on the audio tokens' hidden states at layer~$2$, with the $8$ pivots of largest norm there. We run it on all $N$ tokens (\emph{layer~2}) and, like Triage, on Stage~1's shortlist of $K_1$ tokens (\emph{shortlist}), on the clips of Tab.~\ref{tab:ablation}. Neither variant closes the gap on DREAM, where Triage leads both by at least $.075$ on all three models (Tab.~\ref{tab:dartl2}). On MMSU and RACE, no variant is significantly ahead of Triage.

\begin{table}[h]
\centering
\caption{\textbf{Moving DART into the language model does not close the gap on DREAM} (accuracy on the clips and at the aggressive operating points of Tab.~\ref{tab:ablation}, $n{=}400$ per case). \emph{layer~2} runs DART's rule on the layer-$2$ hidden states over all $N$ tokens, and \emph{shortlist} does so over Stage~1's $K_1$ tokens. On Qwen 30B, all three DART columns come from an A100 run (App.~\ref{app:hardware}), so they can differ slightly from Tab.~\ref{tab:ablation}. \textbf{Bold}: the best selector in each row.}
\label{tab:dartl2}
\small
\begin{tabular}{@{}ll c ccc@{}}
\toprule
Model & Bench & \textbf{Triage} & DART & layer~2 & shortlist \\
\midrule
\multirow{3}{*}{Qwen 3B} & MMSU & \textbf{.585} & .547 & .550 & .565 \\
 & DREAM & \textbf{.863} & .757 & .738 & .778 \\
 & RACE & \textbf{.820} & .805 & .812 & .818 \\
\midrule
\multirow{2}{*}{Qwen 30B} & MMSU & \textbf{.682} & .620 & .652 & .618 \\
 & DREAM & \textbf{.943} & .907 & .853 & .843 \\
\midrule
\multirow{3}{*}{Voxtral} & MMSU & .525 & .532 & \textbf{.537} & .448 \\
 & DREAM & \textbf{.873} & .800 & .797 & .775 \\
 & RACE & \textbf{.655} & .632 & .647 & .632 \\
\bottomrule
\end{tabular}
\end{table}

\paragraph{Segmentwise pruning.} Segmentwise pruning~\cite{segwise} scores each audio token by the attention it receives inside the audio encoder, which we sum over layers and heads, and keeps the best-scoring token in each equal time segment, so it differs from bin coverage only in the score. On the clips and operating points of Tab.~\ref{tab:asrwer}'s Qwen blocks, where this run reproduces the full-audio column clip for clip, bin coverage has the lower WER in all eight cases, by a median of $.018$, and five of the eight paired bootstrap intervals exclude zero. Segmentwise pruning trails DART on mean WER, $.134$ against $.129$, so DART remains the strongest encoder-side baseline on average. Tab.~\ref{tab:ablation} does not print segmentwise pruning, but on the clips and operating points of its six Qwen 3B multiple-choice cases, segmentwise pruning trails DART in all six. Triage leads it in five, four of them with a paired bootstrap interval that excludes zero.

\paragraph{SpeechPrune.} SpeechPrune~\cite{speechprune} releases no code, so we implement it from its equations. Its first phase scores each audio token by its cosine similarity to the text of the question and splits the budget across one-second frames by their total similarity. Its second phase keeps the tokens that receive the most attention in a binarized copy of the language model's first layer, leaving out biases, normalization and RoPE as its equations do. The first phase keeps $K_1$ tokens, as many as Triage's shortlist, and the second keeps the same $K$ as every other selector. On Phi-4 the second phase uses the first layer's speech-adapted projection. Tab.~\ref{tab:ablation} prints it as \texttt{sp}. Our implementation is competitive: at the conservative budget on Voxtral DREAM, it is the best encoder-side selector, $.025$ above top-$K$ with a paired bootstrap interval that excludes zero.

\subsection{How a Dropped Token Is Dropped}\label{app:splice}

An \emph{encoder-side} cut hands the language model a genuinely shorter sequence, so positions run $0$ to $K{-}1$. A \emph{mid-prefill} cut cannot shorten the sequence, because the positions are already in the KV cache, so the kept set indexes into \texttt{position\_ids}, \texttt{cache\_position} and the RoPE $(\cos,\sin)$ alike. Within each kind of cut, the splice is identical for every selector, so the gap between selectors comes from selection, not bookkeeping. At Qwen 3B's conservative LibriSpeech budget, for example, random selection scores $.297$ WER and our selector $.099$.

\section{Reading the Main Tables}\label{app:tabnotes}

This appendix holds the full results behind Tab.~\ref{tab:asrmain}. It gives the table conventions, the counts that the body cites, the full grids and notes on individual tables.

\subsection{Conventions}
\label{app:tabconv}

\textbf{Cases and protocol.} Each task has $24$ cases: four models, three benchmarks and two budgets. Every multiple-choice cut at layer~$2$ is \emph{faithful}: the $K_1$ shortlist tokens pass through layers $1$--$2$ together, and only the $K$ survivors continue to the later layers. On transcription, selection runs on a forward pass over the prompt alone, so no reference transcript enters it. Comparisons stay within one sample. A model's full-audio score on a benchmark is the same in Tabs.~\ref{tab:asrmain}, \ref{tab:ablation}, \ref{tab:asrwer} and~\ref{tab:asrlong}, and each short-clip benchmark has one clip list for all four models, of which Qwen 30B's transcription cases use the first $100$.

\textbf{Comparator.} $\Delta$ compares two fixed selectors: ours (bin coverage on transcription, precision fusion on multiple choice) and DART, the strongest encoder-side baseline on average. $\Delta$ is ours minus DART on accuracy and DART minus ours on WER, so a positive $\Delta$ favours us. Full audio is the uncompressed reference, not a competitor.

\textbf{Columns.} The grids use short code names, given here with the names used in the body. Bold names are ours, and the group headings show where each group of selectors cuts, which sets its cost. \emph{Encoder-side} selectors cut before the language model, and all but \texttt{sp} choose from the encoder output alone: \texttt{pool} (uniform pooling), \texttt{vad} (energy-based VAD), \texttt{random} (a random-selection control), \texttt{dart} (DART~\cite{dart}, which uses no attention), \texttt{sp} (SpeechPrune~\cite{speechprune}, which also reads the question and so runs on multiple choice only), top-$K$ (keeps the tokens that the prior scores highest) and bin coverage (keeps the highest-scoring token in each of $K$ equal time bins~\cite{segwise}). \emph{Mid-prefill} selectors, \texttt{fastv} (FastV$_N$) and \texttt{headrouter} (HeadRouter$_N$), rank all $N$ tokens at layer~$2$, so they save no prefill before the cut. \emph{Shortlist} selectors cut twice and differ only in the Stage-2 rule: raw observed attention (\emph{observed-only}) or \emph{precision fusion} (Eq.~\ref{eq:triage}). The reference column, \texttt{full@N}, is full audio and keeps every token.

\paragraph{The \texttt{headrouter} column.}\label{app:hrname} HeadRouter~\cite{headrouter} ranks tokens by RoPE-free query--key attention and weights the heads by a released profile for Qwen 3B's $16$ heads, so only Qwen 3B uses the released head weights. On Qwen 30B and Voxtral ($32$ query heads), and on Phi-4 after slicing its fused $QKV$ projection, HeadRouter$_N$ uses RoPE-free query--key attention with uniform head weights.

\begin{table}[t]
\centering
\caption{\textbf{Iso-quality compression: the largest ratio each selector sustains within two points of full audio.} Each selector is swept over five compression ratios, $1.5\times$ to $6.7\times$, with $n{=}400$ per case unless a block header says otherwise. On Qwen 3B, sample A is the grid sample of Tab.~\ref{tab:ablation} and sample B is the next $400$ clips of the same pool, so the two are disjoint. RACE's pool holds only $483$ clips, so there A is the first $241$ clips of the grid sample and B the next $241$. ``---'' means the selector is outside the band even at the smallest ratio, and ``$\ge 6.7\times$'' means it is still inside at the largest. \textbf{Bold} marks the largest ratio in a column, ties included. On Qwen 3B, precision fusion is one to two ratio steps above DART in all six samples. On Qwen 30B, precision fusion or top-$K$ sustains $2\times$ on each benchmark, and DART does so on DREAM and RACE.}
\label{tab:isoquality}
\small
\setlength{\tabcolsep}{8pt}
\begin{tabular}{@{}l ccc@{}}
\toprule
Selector & MMSU & DREAM & RACE \\
\midrule
\multicolumn{4}{@{}l}{\textbf{Qwen2.5-Omni-3B}, two disjoint samples, A\,/\,B ($n{=}400$; RACE $n{=}241$ each)} \\
\textbf{precision fusion (ours)} & $\mathbf{4.0}$\,/\,$\mathbf{2.9\times}$ & $\mathbf{4.0}$\,/\,$\mathbf{2.9\times}$ & $\mathbf{4.0}$\,/\,$\mathbf{\ge 6.7\times}$ \\
top-$K$ (prior only, ours) & $\mathbf{4.0}^{\dagger}$\,/\,$2.0\times$ & $1.5$\,/\,$1.5\times$ & $2.9$\,/\,$4.0\times$ \\
\midrule
DART & $2.0$\,/\,$2.0\times$ & $2.0$\,/\,$2.0\times$ & $2.0$\,/\,$4.0\times$ \\
HeadRouter$_N$ & $2.0$\,/\,$2.0\times$ & $2.0$\,/\,$2.0\times$ & $2.9$\,/\,$2.9\times$ \\
FastV$_N$ & $2.0$\,/\,$2.0\times$ & $1.5$\,/\,$1.5\times$ & $1.5$\,/\,$1.5\times$ \\
\midrule
uniform pooling & $2.0$\,/\,$1.5\times$ & ---\,/\,$1.5\times$ & $2.0$\,/\,$1.5\times$ \\
energy VAD & $2.0$\,/\,--- & $1.5$\,/\,$1.5\times$ & $1.5$\,/\,$1.5\times$ \\
random & $1.5$\,/\,$1.5\times$ & ---\,/\,--- & ---\,/\,$1.5\times$ \\
\midrule
\multicolumn{4}{@{}l}{\textbf{Qwen3-Omni-30B (MoE)}, $n{=}400$; DREAM $n{=}397$, leaving out $3$ clips too long for this run} \\
\textbf{precision fusion (ours)} & $\mathbf{2.0\times}$ & $\mathbf{2.0\times}$ & $1.5\times$ \\
top-$K$ (prior only, ours) & $\mathbf{2.0\times}$ & $1.5\times$ & $\mathbf{2.0\times}$ \\
DART & $1.5\times$ & $\mathbf{2.0\times}$ & $\mathbf{2.0\times}$ \\
uniform pooling & --- & $1.5\times$ & $\mathbf{2.0\times}$ \\
\bottomrule
\end{tabular}
\vspace{0.2em}

\raggedright\footnotesize{$^{\dagger}$This entry is the largest ratio inside the band: top-$K$ alone is inside it at $4\times$ ($.583$) but just outside it at $2.9\times$ ($.580$).}
\end{table}

\subsection{Case Counts and the Aggregate}
\label{app:tabcounts}

\textbf{Transcription.} Bin coverage has lower WER than DART in $23$ of the $24$ transcription cases. The exception is Qwen 30B on FLEURS at the conservative budget, where DART is ahead by $.002$. In the four conservative LibriSpeech and FLEURS cases on Voxtral and Phi-4, bin coverage leads DART by $+.024$ to $+.149$. At the conservative budget, Tab.~\ref{tab:asrmain}a prints only bin coverage, and Tabs.~\ref{tab:asrwer} and~\ref{tab:asrlong} print every baseline. Over the twelve conservative cases, the median margin over DART is $+.031$.

\textbf{Multiple choice.} Precision fusion is ahead of DART in $11$ of the $12$ aggressive cases and $19$ of the $24$ overall, and ahead of the best baseline in nine of the twelve aggressive cases. Over those twelve, its mean margin over DART is $+.043$ $[+.032,+.053]$ under a stratified paired bootstrap over clips. On average it also leads HeadRouter$_N$ by $+.047$, FastV$_N$ by $+.074$ and uniform pooling by $+.081$. Even at $n{=}400$ per case, five of the twelve cases are significant on their own after Holm correction (Tab.~\ref{tab:mcqmcnemar}). At fixed quality, within two points of full audio, precision fusion sustains the largest compression in all six Qwen 3B samples (Tab.~\ref{tab:isoquality}).

\begin{table}[h]
\centering\small
\caption{\textbf{Precision fusion leads DART in eleven of the twelve aggressive multiple-choice cases.} Ours is precision fusion, and each case has $n{=}400$ items. $b$ counts the items that precision fusion answers correctly and DART does not, and $c$ counts the reverse, so $\Delta=(b-c)/n$, computed before rounding. Significance comes from an exact McNemar test~\cite{mcnemar} on the $b+c$ items where the two differ, and Holm's method corrects the one-sided $p$-values across the twelve cases. The Holm column marks the five cases that are significant on their own after this correction. Pooled over all $4{,}800$ items, the margin is $+.043$.}
\label{tab:mcqmcnemar}
\setlength{\tabcolsep}{6pt}
\begin{tabular}{@{}ll ccc rr c@{}}
\toprule
model & bench & ours & \texttt{dart} & $\Delta$ & $b$ & $c$ & Holm \\
\midrule
Qwen2.5-Omni-3B & MMSU & $.585$ & $.547$ & $+.037$ & $30$ & $15$ & \\
 & DREAM & $.863$ & $.757$ & $+.105$ & $51$ & $9$ & \checkmark \\
 & RACE & $.820$ & $.805$ & $+.015$ & $13$ & $7$ & \\
\midrule
Qwen3-Omni-30B & MMSU & $.682$ & $.620$ & $+.062$ & $45$ & $20$ & \checkmark \\
 & DREAM & $.943$ & $.902$ & $+.040$ & $22$ & $6$ & \checkmark \\
 & RACE & $.828$ & $.807$ & $+.020$ & $27$ & $19$ & \\
\midrule
Voxtral-Mini-3B & MMSU & $.525$ & $.532$ & $-.007$ & $16$ & $19$ & \\
 & DREAM & $.873$ & $.800$ & $+.072$ & $43$ & $14$ & \checkmark \\
 & RACE & $.655$ & $.632$ & $+.022$ & $42$ & $33$ & \\
\midrule
Phi-4-multimodal & MMSU & $.492$ & $.460$ & $+.032$ & $39$ & $26$ & \\
 & DREAM & $.743$ & $.667$ & $+.075$ & $58$ & $28$ & \checkmark \\
 & RACE & $.650$ & $.610$ & $+.040$ & $54$ & $38$ & \\
\bottomrule
\end{tabular}
\end{table}

\begin{sidewaystable}
\centering
\caption{\textbf{The full multiple-choice grid: full audio and all eleven selectors at both budgets.} Scores are accuracy against the gold answers ($n{=}400$ per case), and every cut at layer~$2$ is faithful. Selectors are grouped by where they cut, which sets their cost. \textbf{Bold} marks the best selector within each group, because the groups cost different amounts of prefill. The \texttt{random} control and the reference column, which keeps every token, are not bolded. $^{\circ}$ marks a row whose non-reference columns all lie within $2$ per-case SE of each other. Significance tests are in App.~\ref{app:tabcounts}, not in this table. App.~\ref{app:tabconv} defines the columns and the faithful cut.}
\label{tab:ablation}
\footnotesize
\setlength{\tabcolsep}{2.0pt}
\begin{tabular}{@{}ll c<{\hspace{2pt}} : >{\hspace{2pt}}c cccccc<{\hspace{2pt}} : >{\hspace{2pt}}c c<{\hspace{2pt}} : >{\hspace{2pt}}c c@{}}
\toprule
& & reference & \multicolumn{7}{c}{\makecell{encoder-side cut\\[-1pt]{\scriptsize $N\xrightarrow{\text{\tiny cut}}K\xrightarrow{\text{\tiny prefill}}K$}}} & \multicolumn{2}{c}{\makecell{mid-prefill cut\\[-1pt]{\scriptsize $N\xrightarrow{\text{\tiny prefill}}N\xrightarrow{\text{\tiny cut}}K$}}} & \multicolumn{2}{c}{\makecell{shortlist cut\\[-1pt]{\scriptsize $N\xrightarrow{\text{\tiny top-$K$}}K_1\xrightarrow{\text{\tiny prefill}}K$}}} \\
\cmidrule(lr){3-3}\cmidrule(lr){4-10}\cmidrule(lr){11-12}\cmidrule(lr){13-14}
Bench & Operating pt.\ ($r_1/r_2$) & \texttt{full@N} & \texttt{pool} & \texttt{vad} & \texttt{random} & \texttt{dart} & \texttt{sp} & \textbf{top-$K$} & \makecell[b]{\textbf{bin}\\\textbf{coverage}} & \texttt{fastv} & \texttt{headrouter} & \makecell[b]{\textbf{observed-}\\\textbf{only}} & \makecell[b]{\textbf{precision}\\\textbf{fusion}} \\
\midrule
\multicolumn{14}{@{}l}{\textbf{Qwen2.5-Omni-3B}} \\
MMSU & conserv.\ .85/.65 ($1.54\times$)$^{\circ}$ & \multirow{2}{*}{.603} & .600 & .595 & .608 & .598 & .585 & \textbf{.603} & .595 & .598 & \textbf{.603} & \textbf{.600} & \textbf{.600} \\
 & aggr.\ .35/.25 ($4.00\times$) & & .517 & .552 & .505 & .547 & .485 & .583 & \textbf{.595} & \textbf{.557} & \textbf{.557} & .580 & \textbf{.585} \\
DREAM & conserv.\ .65/.45 ($2.22\times$) & \multirow{2}{*}{.892} & .812 & .833 & .733 & .868 & .645 & .858 & \textbf{.870} & .833 & \textbf{.868} & \textbf{.875} & .865 \\
 & aggr.\ .50/.20 ($5.00\times$) & & .530 & .620 & .569 & .757 & .530 & \textbf{.805} & .767 & .688 & \textbf{.740} & .818 & \textbf{.863} \\
RACE & conserv.\ .85/.65 ($1.54\times$) & \multirow{2}{*}{.818} & .815 & .820 & .782 & .828 & .782 & \textbf{.835} & .823 & \textbf{.820} & .810 & .815 & \textbf{.823} \\
 & aggr.\ .50/.35 ($2.86\times$) & & .705 & .718 & .677 & .805 & .608 & \textbf{.820} & .818 & .730 & \textbf{.807} & \textbf{.833} & .820 \\
\midrule
\multicolumn{14}{@{}l}{\textbf{Qwen3-Omni-30B (MoE)}} \\
MMSU & conserv.\ .85/.55 ($1.82\times$) & \multirow{2}{*}{.715} & .662 & .688 & .652 & \textbf{.698} & .635 & .690 & .695 & \textbf{.672} & .662 & .685 & \textbf{.690} \\
 & aggr.\ .85/.35 ($2.86\times$) & & .588 & \textbf{.652} & .573 & .620 & .593 & .637 & .642 & .578 & \textbf{.588} & .632 & \textbf{.682} \\
DREAM & conserv.\ .85/.65 ($1.54\times$) & \multirow{2}{*}{.922} & .915 & .927 & .873 & .930 & .912 & .922 & \textbf{.952} & .900 & \textbf{.917} & .935 & \textbf{.940} \\
 & aggr.\ .85/.45 ($2.22\times$) & & .848 & .877 & .759 & .902 & .843 & .877 & \textbf{.915} & \textbf{.818} & .805 & .887 & \textbf{.943} \\
RACE & conserv.\ .85/.55 ($1.82\times$) & \multirow{2}{*}{.870} & .858 & .848 & .804 & \textbf{.868} & .843 & .850 & .865 & .828 & \textbf{.833} & .840 & \textbf{.863} \\
 & aggr.\ .65/.35 ($2.86\times$) & & .755 & .740 & .723 & \textbf{.807} & .720 & .790 & .802 & .740 & \textbf{.752} & .797 & \textbf{.828} \\
\midrule
\multicolumn{14}{@{}l}{\textbf{Voxtral-Mini-3B}} \\
MMSU & conserv.\ .65/.35 ($2.86\times$) & \multirow{2}{*}{.552} & .512 & .420 & .460 & .527 & .527 & .527 & \textbf{.535} & .542 & \textbf{.552} & .532 & \textbf{.535} \\
 & aggr.\ .35/.20 ($5.00\times$) & & .443 & .405 & .415 & \textbf{.532} & \textbf{.532} & .520 & .520 & .502 & \textbf{.527} & \textbf{.525} & \textbf{.525} \\
DREAM & conserv.\ .85/.65 ($1.54\times$) & \multirow{2}{*}{.882} & .865 & .743 & .800 & .863 & \textbf{.895} & .870 & .875 & .880 & \textbf{.887} & .882 & \textbf{.885} \\
 & aggr.\ .50/.35 ($2.86\times$) & & .735 & .603 & .596 & .800 & .818 & \textbf{.850} & .835 & .802 & \textbf{.860} & .850 & \textbf{.873} \\
RACE & conserv.\ .85/.65 ($1.54\times$) & \multirow{2}{*}{.765} & .735 & .693 & .716 & .725 & .740 & .740 & \textbf{.743} & .713 & \textbf{.757} & .708 & \textbf{.748} \\
 & aggr.\ .50/.35 ($2.86\times$) & & \textbf{.675} & .610 & .608 & .632 & .562 & .625 & .657 & .608 & \textbf{.690} & .627 & \textbf{.655} \\
\midrule
\multicolumn{14}{@{}l}{\textbf{Phi-4-multimodal}} \\
MMSU & conserv.\ .85/.65 ($1.54\times$)$^{\circ}$ & \multirow{2}{*}{.550} & \textbf{.547} & \textbf{.547} & .536 & .515 & .535 & .527 & .530 & \textbf{.537} & .530 & .530 & \textbf{.540} \\
 & aggr.\ .50/.35 ($2.86\times$) & & .507 & .505 & .486 & .460 & .445 & .487 & \textbf{.510} & .458 & \textbf{.472} & \textbf{.492} & \textbf{.492} \\
DREAM & conserv.\ .85/.65 ($1.54\times$) & \multirow{2}{*}{.843} & \textbf{.830} & .787 & .746 & .772 & .723 & .812 & .828 & \textbf{.833} & .805 & \textbf{.828} & .812 \\
 & aggr.\ .50/.35 ($2.86\times$) & & .740 & .690 & .604 & .667 & .560 & .662 & \textbf{.767} & \textbf{.675} & .665 & .733 & \textbf{.743} \\
RACE & conserv.\ .85/.65 ($1.54\times$) & \multirow{2}{*}{.688} & .688 & .677 & .655 & .655 & .647 & .672 & \textbf{.695} & .682 & \textbf{.708} & \textbf{.698} & .685 \\
 & aggr.\ .50/.35 ($2.86\times$) & & \textbf{.647} & .630 & .600 & .610 & .585 & .620 & .627 & .613 & \textbf{.635} & .598 & \textbf{.650} \\
\bottomrule
\end{tabular}\end{sidewaystable}

\begin{sidewaystable}
\captionsetup{singlelinecheck=false}
\centering
\caption{\textbf{The full transcription grid: full audio and all ten selectors on LibriSpeech and FLEURS.} The LibriSpeech and FLEURS columns of Tab.~\ref{tab:asrmain}a summarize it. Scores are corpus WER, the total edit distance over the total number of reference words (lower is better). \textbf{Bold} marks the best selector within each column group. The operating points come from the label-free calibration of \S\ref{sec:deployment}. Each benchmark has one sample, with $n{=}300$ clips per case on every model except Qwen 30B, which has $n{=}100$. On Qwen 3B, dropping the clips on which full audio answers with commentary instead of a transcript (App.~\ref{app:decoding}) leaves $294$ on LibriSpeech and $298$ on FLEURS. FLEURS uses English clips only. App.~\ref{app:tabconv} defines the columns.}
\label{tab:asrwer}
\footnotesize
\setlength{\tabcolsep}{2.05pt}
\begin{tabular}{@{}ll c<{\hspace{2pt}} : >{\hspace{2pt}}c ccccc<{\hspace{2pt}} : >{\hspace{2pt}}c c<{\hspace{2pt}} : >{\hspace{2pt}}c c@{}}
\toprule
& & reference & \multicolumn{6}{c}{\makecell{encoder-side cut\\[-1pt]{\scriptsize $N\xrightarrow{\text{\tiny cut}}K\xrightarrow{\text{\tiny prefill}}K$}}} & \multicolumn{2}{c}{\makecell{mid-prefill cut\\[-1pt]{\scriptsize $N\xrightarrow{\text{\tiny prefill}}N\xrightarrow{\text{\tiny cut}}K$}}} & \multicolumn{2}{c}{\makecell{shortlist cut\\[-1pt]{\scriptsize $N\xrightarrow{\text{\tiny bin coverage}}K_1\xrightarrow{\text{\tiny prefill}}K$}}} \\
\cmidrule(lr){3-3}\cmidrule(lr){4-9}\cmidrule(lr){10-11}\cmidrule(lr){12-13}
Bench & Operating pt.\ ($r_1/r_2$) & \texttt{full@N} & \texttt{pool} & \texttt{vad} & \texttt{random} & \texttt{dart} & \textbf{top-$K$} & \makecell[b]{\textbf{bin}\\\textbf{coverage}} & \texttt{fastv} & \texttt{headrouter} & \makecell[b]{\textbf{observed-}\\\textbf{only}} & \makecell[b]{\textbf{precision}\\\textbf{fusion}} \\
\midrule
\multicolumn{13}{@{}l}{\textbf{Qwen2.5-Omni-3B}} \\
LibriSpeech\ & conserv.\ .85/.65 ($1.54\times$) & \multirow{2}{*}{.083} & .309 & .265 & .297 & .123 & .106 & \textbf{.099} & .214 & \textbf{.124} & .126 & \textbf{.118} \\
 & aggr.\ .80/.50 ($2.00\times$) & & .348 & .327 & .496 & .160 & .158 & \textbf{.116} & .395 & \textbf{.173} & .172 & \textbf{.159} \\
FLEURS-en & conserv.\ .95/.90 ($1.11\times$) & \multirow{2}{*}{.153} & .297 & .268 & .242 & .199 & .226 & \textbf{.177} & \textbf{.175} & .191 & .220 & \textbf{.217} \\
 & aggr.\ .95/.85 ($1.18\times$) & & .325 & .355 & .236 & .221 & .249 & \textbf{.174} & \textbf{.200} & .215 & .258 & \textbf{.252} \\
\midrule
\multicolumn{13}{@{}l}{\textbf{Qwen3-Omni-30B (MoE)}} \\
LibriSpeech\ & conserv.\ .90/.65 ($1.54\times$) & \multirow{2}{*}{.011} & .126 & .051 & .213 & .030 & .055 & \textbf{.025} & .162 & \textbf{.120} & .056 & \textbf{.047} \\
 & aggr.\ .70/.50 ($2.00\times$) & & .160 & .199 & .429 & .134 & .133 & \textbf{.060} & .354 & \textbf{.305} & .107 & \textbf{.096} \\
FLEURS-en & conserv.\ .80/.65 ($1.54\times$) & \multirow{2}{*}{.037} & .127 & .049 & .217 & \textbf{.048} & .054 & .050 & .181 & \textbf{.080} & .046 & \textbf{.045} \\
 & aggr.\ .70/.50 ($2.00\times$) & & .140 & .138 & .385 & .119 & .109 & \textbf{.081} & .386 & \textbf{.210} & .101 & \textbf{.085} \\
\midrule
\multicolumn{13}{@{}l}{\textbf{Voxtral-Mini-3B}} \\
LibriSpeech\ & conserv.\ .70/.65 ($1.54\times$) & \multirow{2}{*}{.020} & .062 & .373 & .220 & .063 & .062 & \textbf{.025} & .492 & \textbf{.051} & \textbf{.022} & .024 \\
 & aggr.\ .50/.35 ($2.86\times$) & & .398 & .665 & .627 & .139 & .217 & \textbf{.088} & .753 & \textbf{.190} & \textbf{.086} & .108 \\
FLEURS-en & conserv.\ .70/.65 ($1.54\times$) & \multirow{2}{*}{.043} & .080 & .399 & .219 & .067 & .061 & \textbf{.043} & .515 & \textbf{.050} & \textbf{.041} & .043 \\
 & aggr.\ .50/.35 ($2.86\times$) & & .385 & .719 & .646 & .119 & .160 & \textbf{.066} & .782 & \textbf{.109} & \textbf{.059} & .077 \\
\midrule
\multicolumn{13}{@{}l}{\textbf{Phi-4-multimodal}} \\
LibriSpeech\ & conserv.\ .70/.65 ($1.54\times$) & \multirow{2}{*}{.017} & .051 & .093 & .212 & .194 & .145 & \textbf{.045} & \textbf{.075} & .129 & .060 & \textbf{.051} \\
 & aggr.\ .50/.35 ($2.86\times$) & & .383 & .472 & .597 & .612 & .476 & \textbf{.312} & \textbf{.456} & .550 & .334 & \textbf{.320} \\
FLEURS-en & conserv.\ .70/.65 ($1.54\times$) & \multirow{2}{*}{.045} & .068 & .077 & .199 & .163 & .110 & \textbf{.064} & \textbf{.081} & .109 & \textbf{.061} & .063 \\
 & aggr.\ .50/.35 ($2.86\times$) & & .368 & .402 & .583 & .529 & .389 & \textbf{.285} & \textbf{.426} & .472 & .273 & \textbf{.244} \\
\bottomrule
\end{tabular}
\end{sidewaystable}

\begin{sidewaystable}
\centering
\caption{\textbf{The full long-form grid: full audio and all ten selectors on TEDLIUM.} The TEDLIUM columns of Tab.~\ref{tab:asrmain}a summarize it, and it has the same columns as Tab.~\ref{tab:asrwer}. Every model transcribes the same seven talks ($92$ minutes of audio), and scores are corpus WER over whole talks against the gold transcripts. \textbf{Bold} marks the best selector within each column group, and the reference column is not bolded.}
\label{tab:asrlong}
\footnotesize
\setlength{\tabcolsep}{2.3pt}
\begin{tabular}{@{}ll c<{\hspace{2pt}} : >{\hspace{2pt}}c ccccc<{\hspace{2pt}} : >{\hspace{2pt}}c c<{\hspace{2pt}} : >{\hspace{2pt}}c c@{}}
\toprule
& & reference & \multicolumn{6}{c}{\makecell{encoder-side cut\\[-1pt]{\scriptsize $N\xrightarrow{\text{\tiny cut}}K\xrightarrow{\text{\tiny prefill}}K$}}} & \multicolumn{2}{c}{\makecell{mid-prefill cut\\[-1pt]{\scriptsize $N\xrightarrow{\text{\tiny prefill}}N\xrightarrow{\text{\tiny cut}}K$}}} & \multicolumn{2}{c}{\makecell{shortlist cut\\[-1pt]{\scriptsize $N\xrightarrow{\text{\tiny bin coverage}}K_1\xrightarrow{\text{\tiny prefill}}K$}}} \\
\cmidrule(lr){3-3}\cmidrule(lr){4-9}\cmidrule(lr){10-11}\cmidrule(lr){12-13}
& Operating pt.\ ($r_1/r_2$) & \texttt{full@N} & \texttt{pool} & \texttt{vad} & \texttt{random} & \texttt{dart} & \textbf{top-$K$} & \makecell[b]{\textbf{bin}\\\textbf{coverage}} & \texttt{fastv} & \texttt{headrouter} & \makecell[b]{\textbf{observed-}\\\textbf{only}} & \makecell[b]{\textbf{precision}\\\textbf{fusion}} \\
\midrule
\multicolumn{13}{@{}l}{\textbf{Qwen2.5-Omni-3B}} \\
 & conserv.\ .85/.80 ($1.25\times$) & \multirow{2}{*}{.215} & .397 & .473 & .280 & .247 & .208 & \textbf{.197} & \textbf{.245} & .259 & .221 & \textbf{.200} \\
 & aggr.\ .75/.60 ($1.67\times$) & & .403 & .496 & .409 & .290 & .224 & \textbf{.195} & .364 & \textbf{.288} & \textbf{.234} & .235 \\
\midrule
\multicolumn{13}{@{}l}{\textbf{Qwen3-Omni-30B (MoE)}} \\
  & conserv.\ .75/.60 ($1.67\times$) & \multirow{2}{*}{.020} & .102 & .073 & .242 & .059 & .087 & \textbf{.049} & .253 & \textbf{.145} & .124 & \textbf{.085} \\
  & aggr.\ .75/.50 ($2.00\times$) & & .123 & .166 & .378 & .123 & .141 & \textbf{.083} & .381 & \textbf{.258} & .260 & \textbf{.166} \\
\midrule
\multicolumn{13}{@{}l}{\textbf{Phi-4-multimodal}} \\
  & conserv.\ .85/.65 ($1.54\times$) & \multirow{2}{*}{.085} & .109 & .156 & .346 & .351 & .173 & \textbf{.102} & .149 & \textbf{.111} & .149 & \textbf{.119} \\
  & aggr.\ .50/.35 ($2.86\times$) &  & .488 & .531 & .646 & .697 & .480 & \textbf{.397} & \textbf{.422} & .441 & .355 & \textbf{.347} \\
\midrule
\multicolumn{13}{@{}l}{\textbf{Voxtral-Mini-3B}} \\
  & conserv.\ .70/.50 ($2.00\times$) & \multirow{2}{*}{.032} & .076 & .203 & .403 & .329 & .094 & \textbf{.071} & .232 & \textbf{.100} & .169 & \textbf{.100} \\
  & aggr.\ .60/.42 ($2.38\times$) &  & .241 & .317 & .521 & .441 & .154 & \textbf{.152} & .304 & \textbf{.170} & .220 & \textbf{.145} \\
\bottomrule
\end{tabular}\end{sidewaystable}

\begin{table}[t]
\centering
\caption{\textbf{What should the Stage-1 target be?} The study runs Qwen 3B on DREAM ($n{=}400$, a sample of its own rather than the grid's) at two keep rates within the deployed range. Each row is a candidate definition of $g$. \emph{Direct cut} is the accuracy of cutting on the target itself, \emph{$\rho$} is how well the prior predicts the target, and \emph{achieved} is the accuracy of cutting on the prior fitted to that target. Setting the energy control aside, achieved accuracy is more closely rank-correlated with $\rho$ than with the direct cut: $+.91$ against $+.64$ at $35\%$ kept, and $+.66$ against $+.23$ at $25\%$. \textbf{Bold} marks the deployed target. At $25\%$ kept it has the highest achieved accuracy, and the top five lie within $.005$ of each other (per-case SE $\pm.020$ at these accuracies). The trailing-row target has a direct cut close to ours but achieves far less, and on Phi-4 the deletion test of App.~\ref{sec:causal} separates the two as well.}
\label{tab:targetstudy}
\small
\setlength{\tabcolsep}{5pt}
\begin{tabular}{@{}l c cc cc@{}}
\toprule
& & \multicolumn{2}{c}{keep $35\%$} & \multicolumn{2}{c}{keep $25\%$} \\
\cmidrule(lr){3-4}\cmidrule(lr){5-6}
Target $g$: query rows $\mathcal{Q}$ & $\rho$ & direct cut & achieved & direct cut & achieved \\
\midrule
trailing rows & .658 & .817 & .782 & .780 & .695 \\
all text rows & .677 & .825 & .825 & .818 & .770 \\
text rows, template tail dropped & .675 & .827 & .823 & .810 & .772 \\
\textbf{audio rows (ours)} & .707 & .815 & .825 & .798 & .815 \\
all rows & .714 & .815 & .825 & .802 & .812 \\
\midrule
trailing, RoPE-free & .656 & .778 & .742 & .735 & .643 \\
text, RoPE-free & .712 & .833 & .830 & .792 & .810 \\
audio, RoPE-free & .775 & .843 & .825 & .795 & .800 \\
all rows, RoPE-free & .768 & .840 & .830 & .795 & .798 \\
\midrule
audio rows $+$ gradient ($\tfrac12$/$\tfrac12$) & .790 & .833 & .833 & .805 & .805 \\
text RoPE-free $+$ gradient ($\tfrac12$/$\tfrac12$) & .794 & .830 & .837 & .812 & .810 \\
all rows $+$ gradient ($\tfrac12$/$\tfrac12$) & .794 & .840 & .835 & .815 & .810 \\
\midrule
gradient saliency $\|\partial\text{logit}/\partial\mathbf{e}_i\|$ & .617 & .825 & .823 & .808 & .765 \\
\midrule
acoustic energy \emph{(control)} & .881 & .762 & .752 & .677 & .662 \\
\bottomrule
\end{tabular}
\end{table}

\textbf{The lead over DART holds across clips, cases, benchmarks and models.} The bootstrap interval of the aggressive mean margin stays clear of zero whether clips or cases are resampled. The mean margin is also positive on every benchmark ($+.024$ to $+.073$) and for every model (Tab.~\ref{tab:mcqmcnemar}).

\textbf{Distance to full audio.} At the conservative budget, a paired bootstrap over clips finds nine of the twelve cases indistinguishable from full audio. Of the other three, precision fusion is above full audio in one and below it in two, Qwen 3B and Phi-4 on DREAM ($-.027$ and $-.031$).

\subsection{Notes on Individual Tables}
\label{app:tabper}

\paragraph{Tab.~\ref{tab:ablation}, the multiple-choice grid.} Full audio is a reference, not a ceiling: some selector exceeds it in seven of the $24$ rows, and one of ours exceeds it in six of them, on all four models. The per-case SE, binomial at each case's full-audio accuracy, is $\pm.023$--$.025$ on MMSU, $\pm.017$--$.023$ on RACE and $\pm.013$--$.018$ on DREAM.

\paragraph{Operating points.} The label-free calibration of App.~\ref{app:calibsweeps} sets every model's operating points; no label enters the choice of any of them.

\paragraph{Tab.~\ref{tab:targetstudy}, the Stage-1 target.} Over the thirteen language-model targets, at $25\%$ kept, achieved accuracy rank-correlates $+.66$ with $\rho$ but only $+.23$ with the direct cut. What separates the targets is therefore how well the prior predicts them, not the accuracy of their direct cut (App.~\ref{app:targetfull}). The energy control is predicted best ($\rho{=}.88$) but achieves only $.662$ at $25\%$ kept, against $.815$ for our target. Our target also leads the trailing-row target by $+.120\,[+.080,+.163]$ under a paired bootstrap.

\FloatBarrier

\FloatBarrier
\section{Efficiency and Cost}
\label{app:efficiency}
We time language-model prefill and batched serving at the eight deployed operating points on a GH200 (Tab.~\ref{tab:eff}), then encoder plus prefill on an A100, and compare the pipeline with an ASR$\to$text cascade (App.~\ref{app:cascade}).

\begin{table}[t]
\centering
\caption{\textbf{Efficiency at the deployed operating points, against full audio on one $96$\,GB GH200.} Per request, we report the language-model prefill speedup and the share of the audio KV cache that is kept (KV\%). We also report the largest number of concurrent streams that fit in memory and their aggregate decode throughput (tok/s), timed on one decoding step at that batch size. Prefill excludes the audio encoder, which both sides run on every frame and which holds no KV cache. Each row is the most compressible multiple-choice (MCQ) or short-utterance transcription (ASR) point at its budget: DREAM on Qwen 3B and MMSU on Qwen 30B for MCQ, and LibriSpeech for ASR. The ASR rows time both stages at the transcription points. Transcription itself deploys a single cut to $r_2$ before the language model, which passes $K$ rather than $K_1$ tokens through the first two layers, so it saves at least as much. \textbf{Bold} marks the aggressive rows. $^{\ast}$\,The batch search stops at $1024$ streams, and this point reaches it.}\label{tab:eff}
\small
\setlength{\tabcolsep}{4pt}
\begin{tabular}{@{}ll l ccc c l ccc@{}}
\toprule
&& \multicolumn{4}{c}{Qwen2.5-Omni-3B ($5$\,min)} && \multicolumn{4}{c}{Qwen3-Omni-30B (MoE, $40$\,min)} \\
\cmidrule(lr){3-6}\cmidrule(lr){8-11}
Task & Budget & $r_1/r_2$ & prefill\,/\,KV\% & streams & tok/s && $r_1/r_2$ & prefill\,/\,KV\% & streams & tok/s \\
\midrule
MCQ & conserv. & .65/.45 & 2.3$\times$\,/\,$46\%$ & 2$\times$ & 2.1$\times$ && .85/.55 & 1.4$\times$\,/\,$56\%$ & 2$\times$ & 1.9$\times$ \\
MCQ & \textbf{aggr.} & .50/.20 & \textbf{3.7$\times$}\,/\,$22\%$ & 4$\times^\ast$ & \textbf{4.5$\times$} && .85/.35 & \textbf{1.6$\times$}\,/\,$37\%$ & \textbf{3$\times$} & \textbf{2.8$\times$} \\
\midrule
ASR & conserv. & .85/.65 & 1.6$\times$\,/\,$66\%$ & 1.5$\times$ & 1.5$\times$ && .90/.65 & 1.2$\times$\,/\,$66\%$ & 1.5$\times$ & 1.5$\times$ \\
ASR & \textbf{aggr.} & .80/.50 & \textbf{2.0$\times$}\,/\,$52\%$ & \textbf{2$\times$} & \textbf{1.9$\times$} && .70/.50 & \textbf{1.5$\times$}\,/\,$51\%$ & \textbf{2$\times$} & \textbf{2.0$\times$} \\
\bottomrule
\end{tabular}

\end{table}

\textbf{Cutting before the language model shortens its prefill at all eight timed points, and the saving grows with audio length.} On the GH200, prefill falls by up to $3.7\times$ on Qwen 3B and $1.6\times$ on Qwen 30B. On an A100, Qwen 3B's aggressive multiple-choice point gives $4.5\times$ on $5$ minutes of audio. On $20$ minutes it gives $6.3\times$, because the compressed prefill grows far more slowly with the token count than the full one. Qwen 30B gains less for two reasons: a measured ${\sim}2.0$\,s of its MoE prefill is weight-bound and does not fall with the token count, and on multiple choice its Stage~1 keeps more tokens ($r_1{=}.85$, against $.65$ and $.50$ on Qwen 3B).

\paragraph{Batched serving.} Across the eight deployed points the cut shrinks the audio KV cache to $22$--$66\%$ of its full size. A single stream's memory is mostly the frozen weights, so the saving shows in batched serving: there, many concurrent streams fill the GPU, and each stream's KV cache scales with the tokens kept. At the aggressive multiple-choice point one GH200 serves $4\times$ as many concurrent Qwen 3B streams as with full audio, at $4.5\times$ the decode throughput. Qwen 30B's aggressive multiple-choice point reaches $3\times$ the streams and $2.8\times$ the throughput.

\paragraph{Stage attribution.} Serving is set by the decode KV cache and so by $r_2$, because the shortlist $K_1$ occupies only the first two layers (Tab.~\ref{tab:stage2eff}). Cutting to $K$ therefore provides the extra serving capacity on both models. Stage~1 adds prefill saving to the extent that it cuts before the language model, which is substantial on Qwen 3B at $r_1{=}.5$ and small on Qwen 30B at $r_1{\approx}.85$.

Because the cut is at layer~2, the full pipeline comes within a few percent of the lower bound on prefill. That bound is an encoder-only-$K$ prefill, in which no language-model layer ever runs on more than $K$ tokens. Over the deployed points measured on the GH200, the two pre-cut layers add $+0.6\%$ to $+6.9\%$ over it, with a mean of $+3.0\%$.

The timings in Tab.~\ref{tab:eff} use SDPA attention. Stage~2 needs only the attention rows of the prompt suffix in the two layers before the cut, so we recompute just those rows alongside SDPA.

\begin{table}[h]
\centering
\caption{\textbf{Which stage produces the gain depends on the model.} We compare Stage~1 alone (shortlist $K_1$, no layer-2 cut) with the full pipeline (cut to $K$) on one GH200, on $5$ minutes of audio for Qwen 3B and $40$ minutes for Qwen 30B. Each ratio is against the full audio of the same run, and run-to-run noise on Qwen 30B is $\pm1$\,tok/s. \textbf{Bold} marks the full pipeline's serving figures, except the Qwen 3B stream count, which reaches the search ceiling ($^\ast$).}
\label{tab:stage2eff}
\small
\setlength{\tabcolsep}{4pt}
\begin{tabular}{@{}l cccc@{}}
\toprule
Config & prefill & KV\% & streams & tok/s \\
\midrule
\multicolumn{5}{@{}l}{\textbf{Qwen2.5-Omni-3B} (5-min), \emph{aggressive (MCQ)} $.50/.20$}\\
full ($N$) & $120$\,ms & $100\%$ & 256 & 666 \\
Stage-1 only ($K_1$) & $57$\,ms ($2.1\times$) & $50\%$ & 512 & 1326 ($2.0\times$) \\
\;\;+\,Stage-2 ($K$) & $32$\,ms ($3.7\times$) & $22\%$ & 1024$^\ast$ & \textbf{2993} ($4.5\times$) \\
\midrule
\multicolumn{5}{@{}l}{\textbf{Qwen3-Omni-30B} (40-min), \emph{aggressive (MCQ)} $.85/.35$}\\
full ($N$) & $3.6$\,s & $100\%$ & 8 & $\sim$28 \\
Stage-1 only ($K_1$) & $3.2$\,s ($1.1\times$) & $85\%$ & 8 & 30 ($1.1\times$) \\
\;\;+\,Stage-2 ($K$) & $2.3$\,s ($1.6\times$) & $37\%$ & \textbf{24} & \textbf{79} ($2.8\times$) \\
\midrule
\multicolumn{5}{@{}l}{\textbf{Qwen3-Omni-30B} (40-min), \emph{aggressive (ASR)} $.70/.50$}\\
full ($N$) & $3.5$\,s & $100\%$ & 8 & $\sim$29 \\
Stage-1 only ($K_1$) & $2.8$\,s ($1.3\times$) & $70\%$ & 12 & 42 ($1.5\times$) \\
\;\;+\,Stage-2 ($K$) & $2.4$\,s ($1.5\times$) & $51\%$ & \textbf{16} & \textbf{57} ($2.0\times$) \\
\bottomrule
\end{tabular}
\vspace{0.2em}

\raggedright\footnotesize{$^\ast$The Qwen 3B stream count reaches the search ceiling (1024) at $r_2{\le}.25$, so throughput is the cleaner metric for Qwen 3B.}
\end{table}

\paragraph{Encoder plus prefill.} Full audio and ours run the same encoder over all $N$ frames, so it drops out of the ratios above but adds to every request. We time it on the A100 of App.~\ref{app:hardware}, in the same run and on the same clip as the prefill it joins, so all figures in this paragraph come from the A100 and exclude decoding. On Qwen 3B the encoder costs more than the full-audio language-model prefill itself, $1.9$ to $2.6\times$ as much. The prefill speedup on $5$ minutes therefore falls to $1.28\times$ for encoder plus prefill, and on $20$ minutes to $1.41\times$. On $40$ minutes, Qwen 30B's encoder costs only $21\%$ of its full-audio language-model prefill. Its $1.70\times$ prefill speedup there becomes $1.51\times$ for encoder plus prefill. The more a language model costs relative to its encoder, the more of the prefill saving reaches the whole request.

\subsection{LALM against an ASR\texorpdfstring{$\to$}{->}Text Cascade}
\label{app:cascade}
\textbf{Our conservative cut stays close to full audio on sound and ahead of the cascade on emotion.} On the four sound cases it stays within $.03$ of full audio (Tab.~\ref{tab:cascade}). On emotion, the same language model fed a Whisper~\cite{whisper} transcript trails full audio by $.05$ and $.09$, and our conservative cut is ahead of this cascade on both models.

\begin{table}[t]
\centering
\caption{\textbf{LALM against an ASR$\to$text cascade.} The same Qwen language model answers zero-shot from the audio or from a Whisper-small.en transcript ($n{=}200$; $204$ on the lexical control). The benchmarks cover non-speech sound (MMAU-mini, MMAR), acted-speech emotion (IEMOCAP~\cite{iemocap}) and a lexical control drawn from MMSU, whose answer is in the words. Ours runs at each model's MMSU points of Tab.~\ref{tab:ablation}, unchanged on every benchmark here: $.85/.65$ and $.35/.25$ on Qwen 3B, and $.85/.55$ and $.85/.35$ on Qwen 30B. \textbf{Bold} marks audio (full or our conservative cut) that is ahead of the cascade. The control is not bolded.}
\label{tab:cascade}
\small
\setlength{\tabcolsep}{5pt}
\begin{tabular}{@{}ll cc:c c@{}}
\toprule
Benchmark & Model & audio full & ours (conserv.) & ours (aggr.) & Whisper$\to$text \\
\midrule
MMAU-mini sound & Qwen 3B & \textbf{.62} & \textbf{.62} & .59 & .44 \\
 & Qwen 30B & \textbf{.66} & \textbf{.63} & .66 & .51 \\
MMAR sound & Qwen 3B & \textbf{.54} & \textbf{.54} & .48 & .39 \\
 & Qwen 30B & \textbf{.66} & \textbf{.66} & .62 & .42 \\
IEMOCAP emotion & Qwen 3B & \textbf{.62} & \textbf{.62} & .54 & .57 \\
 & Qwen 30B & \textbf{.65} & \textbf{.58} & .53 & .56 \\
\midrule
MMSU lexical \emph{(control)} & Qwen 3B & .90 & .89 & .77 & .84 \\
 & Qwen 30B & .96 & .93 & .84 & .93 \\
\bottomrule
\end{tabular}
\end{table}

\paragraph{Time to first token.} Our pipeline also answers sooner than the cascade (Tab.~\ref{tab:latency}), mostly because it skips transcription: the cascade must transcribe before its language model starts, so its time to first token grows with audio length. Our compressed language-model prefill holds at ${\approx}30$\,ms on Qwen 3B across all three lengths, so our time to first token grows only as fast as the model's own encoder.

\begin{table}[t]
\centering
\caption{\textbf{Time to first token, our pipeline against the cascade.} Both run on the same clip on one GH200, at the aggressive multiple-choice points ($.50/.20$ on Qwen 3B, $.85/.35$ on Qwen 30B). Ours is the audio encoder plus the compressed prefill, and the cascade is Whisper-small.en transcription plus the same language model on the transcript, so both encoders are counted. Speedup is cascade over ours, not over full audio. Qwen 3B's margin falls after one minute because its encoder then takes longer than its whole language-model prefill. \textbf{Bold} marks the largest speedup in each block.}
\label{tab:latency}
\small
\setlength{\tabcolsep}{6pt}
\begin{tabular}{@{}ll ccc@{}}
\toprule
Model & Audio & ours (compressed) & cascade (Whisper$+$text) & speedup \\
\midrule
Qwen 3B & 30\,s & $48$\,ms & $1.07$\,s & $22\times$ \\
& 1\,min & $61$\,ms & $2.40$\,s & $\mathbf{39\times}$ \\
& 5\,min & $497$\,ms & $13.9$\,s & $28\times$ \\ \midrule
Qwen 30B & 30\,s & $1.71$\,s & $2.62$\,s & $1.5\times$ \\
& 5\,min & $1.80$\,s & $16.1$\,s & $8.9\times$ \\
& 10\,min & $1.97$\,s & $30.8$\,s & $\mathbf{16\times}$ \\
\bottomrule
\end{tabular}
\end{table}

\FloatBarrier

\section{Extended Related Work}
\label{app:related}
This section extends \S\ref{sec:related} with vision token reducers, the nearest counterparts we test, an end-to-end test of self-information, and two further methods.

\paragraph{Vision token pruning.} LLaVA-PruMerge merges visual tokens before the language model, choosing them by the vision encoder's CLS attention~\cite{visionprune}, and VisPruner selects them by the encoder's own self-attention~\cite{vispruner}. DynamicViT and ToMe prune or merge inside the vision transformer and never consult the language model~\cite{dynamicvit,tome}. SparseVLM prunes inside the language model by the attention that text tokens give each visual token~\cite{sparsevlm}. Like IVTP~\cite{ivtp} and Lei et al.~\cite{tasktoken}, it must therefore run the model it prunes. For audio, the encoder's own self-attention reaches at most $\rho{=}{+}.31$ against the language model's attention (\S\ref{sec:stage1}). Audio is also the one modality still heavily attended at the depth where a cut inside the language model would act (\S\ref{sec:doublebind}), so audio tokens have to be chosen before the language model runs.

\paragraph{Nearest tested counterparts.} These vision methods do not apply to a frozen LALM as they stand, so for four of their families we test the nearest counterpart. Our uniform-pooling baseline is the nearest counterpart of content-blind pooling and pixel-shuffle. Triage is ahead of uniform pooling in $42$ of the $48$ deployed cases of Tabs.~\ref{tab:ablation}, \ref{tab:asrwer} and~\ref{tab:asrlong}, by a median of $+.056$. Over the $24$ aggressive cases it is ahead in $22$, by a median of $+.090$. Similarity-based merging (ToMe) is nearest to DART, the comparator in our tables. DART keeps the tokens least similar to a small pivot set. Pruning by CLS or encoder self-attention would rely on the $\rho{=}{+}.31$ signal above. Its nearest counterpart, segmentwise pruning, selects by the audio encoder's attention and has a higher WER than bin coverage in all eight Qwen LibriSpeech and FLEURS cases (App.~\ref{app:baselinefidelity}). Trained resamplers such as Q-Former~\cite{blip2} and Perceiver~\cite{perceiver} emit new vectors, so they give Stage~2 no per-token score to correct. We trained one ourselves. An $8.5$M-parameter Perceiver connector does not significantly beat the $2{,}049$-parameter prior at the deployed kept fractions, and it falls behind the prior when only $10\%$ of the tokens are kept (App.~\ref{app:resampler}).

\paragraph{Self-information, tested end to end.} LLMLingua's rule~\cite{llmlingua}, adapted to audio tokens, is at or below random selection in five of the six multiple-choice cases on Qwen 3B (Tab.~\ref{tab:linguasel}). Keeping the $K$ tokens it ranks lowest even scores higher than keeping the $K$ it ranks highest, in all six cases, by $+.002$ to $+.219$.

\paragraph{Further methods.} FastAdaSP merges audio tokens inside a speech language model's decoder layers~\cite{fastadasp}, so every token first enters the model, as in FastV. In text, SpecPrefill ranks a large model's prompt by the attention of a smaller draft model and prefills the large model on the kept tokens alone~\cite{specprefill}. The draft is a second language model run over every token, and its attention is used in place of the large model's, without being fitted to it. The prior, in contrast, is fitted to the served model's own attention, in closed form and without labels. On eleven of thirteen LALMs it reaches $\rho{\ge}.69$, and one unlabelled forward pass separates the two it fails on (\S\ref{sec:screen}).

\end{document}